\documentclass[a4paper,12pt]{article}
\usepackage[centertags]{amsmath}
\usepackage{amssymb}

\usepackage{graphicx}
\usepackage{epsfig}
\usepackage{ulem}
\usepackage[english]{babel}
\usepackage{array}
\usepackage{amsthm}
\usepackage{latexsym}

\usepackage{yfonts}
\usepackage{mathrsfs}
\usepackage[mathcal]{euscript}

\newcommand{\Um}{{\mathscr{U}}}
\newcommand{\Vm}{{\mathscr{V}}}
\newcommand{\Tm}{{\mathscr{T}}}
\newcommand{\Xm}{{\mathscr{X}}}
\pdfoutput=1
\usepackage{epsfig}
 \usepackage{jheppub}
\usepackage{mathdots}
\usepackage{MnSymbol}
\usepackage{multirow}

\definecolor{darkraspberry}{rgb}{0.53, 0.15, 0.34}

\definecolor{darkblue}{rgb}{0., 0, 1}
\newcommand{\IA}{\textcolor{darkblue}}

\definecolor{dgreen}{rgb}{0.,0.6,0.}

\newcommand{\nn}{\nonumber}
\newcommand{\be}{\begin{equation}}
\newcommand{\ee}{\end{equation}}
\newcommand{\bea}{\begin{eqnarray}}
\newcommand{\eea}{\end{eqnarray}}

\newcommand{\e}{\mathrm{e}}

\newcommand{\RR}{\mathbb{R}}
\newcommand{\MM}{\mathbb{M}}

\newcommand{\cM}{{\cal M}}

\newcommand{\cU}{\mathcal{U}}
\newcommand{\cV}{\mathcal{V}}
\newcommand{\cT}{\mathcal{T}}
\newcommand{\cX}{\mathcal{X}}

\newcommand{\fb}{\mathfrak{b}}

\newcommand{\ff}{\mathfrak{f}}

\newcommand{\fs}{\mathfrak{s}}

\newcommand{\fB}{\mathfrak{B}}

\definecolor{dgreen}{rgb}{0.,0.6,0.}
\title{Quantum explosions of black holes and thermal coordinates 
}
\author{Irina Aref'eva  and}
\author{Igor Volovich}
\emailAdd{arefeva@mi-ras.ru, volovich@mi-ras.ru}
\affiliation{Steklov Mathematical Institute, Russian Academy of Sciences,\\Gubkina str. 8, 119991, Moscow, Russia}

\abstract{

The Hawking temperature for  Schwarzschild black hole $T_H=1/8\pi M$ is singular in the limit of vanishing mass $M\to 0$. However, the Schwarzschild metric itself is regular when the black hole mass $M$ tends to zero, it is reduced to the Minkowski metric  and there are no reasons to believe that the temperature becomes  infinite. 
It is  pointed out that this discrepancy may be due to the fact that the Kruskal coordinates are singular in the limit of the vanishing mass of the black hole. 
To improve the situation, new  coordinates for the Schwarzschild metric are introduced, called thermal coordinates, which depend on the black hole mass $ M $ and the parameter $ b $. The thermal coordinates are regular in the limit of vanishing black hole mass $ M $. In this limit, the Schwarzschild metric is reduced to the Minkowski metric, written in coordinates dual  to the Rindler coordinates.

Using the thermal coordinates the Schwarzschild black hole radiation is reconsidered and it is found that the Hawking formula for temperature is  valid only for large black holes while for small black holes the temperature is $T=1/2\pi (4M+b)$. 
The thermal observer in Minkowski space
 sees  radiation with temperature $T=1/2\pi b$, similar to the Unruh effect  with non-constant acceleration.
Now, during evaporation, in the thermal coordinates the black hole mass is  decreasing inverse proportional to time and the black hole life time  is infinite. 
The thermal coordinates for more general spherically symmetric metrics, including the Reissner-Nordstrom, de Sitter and anti-de Sitter are also considered. In these coordinates one sees a Planck distribution with constant temperature. One obtains that the property to have a temperature distribution for quantum fields in classical gravitational background is not restricted to the cases of black holes or constant acceleration,    but is valid for any spherically symmetric metric written in thermal coordinates.
 Implications  for primordial black
 holes  and for the information loss problem are mentioned.

}
\begin{document}
\maketitle

\section{Introduction}
Hawking showed that black holes emit radiation like black bodies with a temperature of $
T _H= 1/8 \pi M $, where $ M $ is the mass of the black hole \cite{Hawking:1974,Hawking:1975}.
Considering the scenario where a black hole that emits its energy ends up completely evaporating, a problem arises.
From Hawking's formula it follows that the energy density of radiation emitted by a black hole according to
the Stefan-Boltzmann formula behaves at small $ M $ as $ M ^ {- 4} $.
  If during evaporation the mass of the black hole disappears, then the black hole releases an infinite amount of energy, which is clearly not physical.
  This can be called the problem of the  big bang of black holes. 
 The  information loss problem \cite{Hawking:1976,Susskind:2005,tHooft:2012} is closely related to this big bang problem, since the radiation entropy diverges for small $ M $ as $ M ^ {- 3} $.
 Considerations in more complicated cases, such as de Sitter Schwarzschild black hole evaporation, do not improve the situation in an essential way.\\

Standard transformation from the Schwarzschild coordinates to the Kruskal ones includes first transformation from the Schwarzschild coordinates 
to Eddincton-Finkelstein coordinates \cite{HE,FN,Wald,Ydri:2017}.
The Schwarzschild metric in the Schwarzschild  coordinates
\be
\label{Scw-Int}
ds^2=-(1-\frac{2M}{r})dt^2+(1-\frac{2M}{r})^{-1}dr^2+ r^2\,d \Omega^2,\qquad r>2M>0,\ee
 obviously  admits the $M\to 0$ limit, that defines the Minkowski space.
The  Kruskal coordinates $(U,V)$ are defined as
\bea\label{UK-ii}
U&=&-e^{-u/4 M},\,\,V=e^{v/4M}
\eea
here $u$ and $v$ are the Eddington-Finkelstein coordinates. The Kruskal coordinates are used to get a maximal analytic extension, but we note, that in the limit of vanished mass $M$  black hole,  even outside of horizon, $r>2M$, the Kruskal coordinates and the metric get a singularity, instead to become the  Minkowski one. 
 This leads also to the singular behaviour of the Hawking temperature $T_H=1/8\pi M$ in the limit $M\to 0$. To improve the situation, E(xponential)-coordinates $\Um $ and $\Vm $ for the Schwarzschild metric are introduced, 
\bea
\Um =-e^{-\frac{u}{4 M+b}},\quad
\Vm=e^{\frac{v}{4 M+b}},  \label{UmVn}
\eea 
which depend on the black hole mass $M$ and a parameter  $ b>0$. The E-coordinates are regular in the limit of the vanishing black hole mass $M$. 
Obviously in this limit the  Schwarzschild metric is reduced to the Minkowski one written in the E-coordinates.

Black hole radiation is considered and it is found that the Hawking formula for temperature is approximately valid only for large black holes while for small black holes for the temperature of black hole the following formula is obtained: 
\be T=\frac{1}{2\pi (4M+b)}.\ee
 As a result, black holes could completely evaporate in terms of classical geometry, but it is shown that this requires the infinite time because the mass is decreasing 
 in the inverse proportional to time,
 \be
 M(t)=\frac{C}{t}, \quad t\to \infty\label{Mt}\ee 
It is noted that E-observer in Minkowski space
will see radiation with the temperature 
\be T=\frac{1}{2\pi b}. \ee
This effect is similar (dual) to the Unruh effect \cite{Unruh:1976,BD} for the Rindler metric \cite{Rindler:1966} but in our case the acceleration is not a constant. \\

 We define   the E(xponential)-coordinates $(\Um,\Vm)$ and L(ogariphmic)-coordinates $(\nu,\vartheta)$ for the arbitrary static metric of the form
 \be
ds^2=-f(r)dt^2 +f(r)^{-1}dr^2+r^2d\Omega^2=-f(r)dudv+r^2d\Omega^2\label{ds-f}
\ee
as
 \be \Um=-e^{-\frac{u}{B}}, \quad\Vm=e^{\frac{v}{B}},\quad B>0,\ee 
\be \vartheta=\frac{1}{a}\log(av), \quad\nu=-\frac{1}{a}\log( - au)\quad a>0.\ee It will be shown that one has the Planck distribution with temperature $T=1/2\pi B$ and $T=a/2\pi$ for quantum  fields in the gravitational background 
\eqref{ds-f} with an arbitrary function $f(r)$ in E- and L-coordinates, respectively.
We have the following general scheme (duality)
\bea
\left(
\begin{array}{c}
  \mbox{E-coord.}  \\
  (\Um,\Vm)  
  \end{array}
\right)
\xleftarrow[\Um=-e^{-\frac{u}{B}}]{\Vm=e^{\frac{v}{B}}} \left(
\begin{array}{c}
  \cM  \\
  (u,v) \\
  ds_2^2=- f(r) dudv
  \end{array}
\right)\xrightarrow[\vartheta=\frac{1}{a}\log(av)]{\nu=-\frac{1}{a}\log( - au)}
\left(
\begin{array}{c}
  \mbox{L-coord.}  \\
  (\nu,\vartheta)  
  \end{array}
\right)\eea
\\

The  physical meaning of above formulae for temperature is that they give temperature of radiation for different observers moving along different trajectories in the same background. The simplest examples of such  special trajectories are ones in the  Minkowski space. The standard Rindler observer moves with constant acceleration and sees radiation from Minkowski vacuum
as it has temperature defined by its acceleration. The E-observer moves along  a hyperbola and  feels the 
temperature. E- and L-coordinates can be called the thermal coordinates since in these coordinates one sees a Planck distribution with constant temperature. Actually, the property to have a temperature is not connected with black holes and horizons but with the usage of the thermal coordinates and the temperature can be obtained for any metric.
Implications for the information loss problem  and primordial black
 holes are mensioned.\\
 
 The paper is organized as follows. We start Sect.\ref{Sect:Kruskal} reminding the standard definition of the Kruskal coordinates and also we discuss problem that arises with them when one considers the limit $M\to 0$. Then in Sect.\ref{quasiK-Sch} we introduce in E-coordinates for Schwarzschild metric, and in 
 Sect.\ref{quasiK-Sch-T}
 temperature of  Schwarzschild  black holes in E-coordinates. In the next Sect.\ref{QK-Min} we discuss E-coordinates in Minkowski space. We show in Sect.\ref{QK-Min-4} that   two-dimensional Minkowski space  can be   represented as a union of 4 disconnected regions, right
 ({\bf R}), future ({\bf F}), left ({\bf L}) and past  ({\bf P}),
and each  of them is isometric to two-dimensional Minkowski space. In \ref{QK-geodesics} we study geodesics in E-coordinates. In Sect.\ref{Acc-QK} we calculate acceleration of a E-observer.
Sect.\ref{qK-Rindler}  is devoted to comparison of E- and Rindler coordinates in Minkowski space. In Sect.\ref{Gen-QK}
we introduce general E-coordinates. In Sect.\ref{Sect:QK-accel} acceleration along special trajectories $\Xm=\Xm_0$ in black hole backgrounds are calculated. Then in next sections we consider some examples, In next Sect.\ref{GenQR} general L-coordinates are introduced. 
In Sect.\ref{Sect: q-Rindler-T} temperature in L-coordinates is calculated and it has been found that it is given by an universal formula
that does not depend on characteristic of the black hole under consideration.  The origin of this phenomena is that the choice of the coordinate system depends in essentially way on the metric itself.
In Sect.\ref{time-evap} we present an estimation of evaporation time as it can been seen by observers in different coordinate systems.
In Sect.\ref{Sec:Disc-Concl} we summarise the obtained results and discuss their physical applications.

\section{Exponential Coordinates}\label{Sect:Kruskal}
\subsection{Kruskal coordinates}
Standard transformation from the Schwarzschild coordinates to the Kruskal ones includes first transformation from the Schwarzschild coordinates 
to Eddincton-Finkelstein coordinates.
The Schwarzschild metric in the Schwarzschild  coordinates  is
\be
\label{Scw}
ds^2=-(1-\frac{2M}{r})dt^2+(1-\frac{2M}{r})^{-1}dr^2+ r^2\,d \Omega^2,\qquad r>2M>0.\ee
where $d\Omega^2=d \theta ^2 +\sin^2\theta  d\varphi^2$. It is obvious that  the exterior Schwarzschild spacetime ($r>r_h=2M$) admits the $M\to 0$ limit, that defines the Minkowski spacetime\footnote{The  Kretschmann invariant 
$ K = \frac{48  M^2}{ r^6}\to 0$ as $M\to 0$ for any fixed  $r>2M$.
}.

One can introduce the  tortoise coordinate $r_*$
\be\label{rstar}
r_*=r+2 M \log \left(\frac{r}{2 M}-1\right), 
\ee
which solves the equation  $dr_*=(1-2M/r)^{-1}dr$. To keep the reality condition one has to assume $r>2M$. Then one defines
\bea\label{uv}
u=t-r_*, \qquad
v=t+r_*.
\eea
(coordinates $u,v$ cover whole $\RR^2$),
and one has
\be
ds_2^{2}=-(1-\frac{2M}{r})dt^2+(1-\frac{2M}{r})^{-1}dr^2=-(1 - \frac{2 M}{r}) du dv\label{ds-uv}\ee

The Kruskal coordinates are 
\bea\label{UK}
U&=&-e^{-\frac{u}{4 M}},\,\,V=e^{v/4M}
\eea
and  the Schwarzschild metric becomes 
\bea\label{mUV}
ds^2=-\frac{32M^3}{r} \,e^{-r/2M}\,dUdV\,+ r^2d \Omega^2,\label{SchKR}
\eea
where  $r$ is defined from  equation
\bea
\left(\frac{r}{2 M}-1\right)e^{\frac{r}{ 2M}}=-UV \label{rUV}
\eea

Note that  the Kruskal coordinates \eqref{UK} and metric \eqref{mUV} are singular in the limit $M\to 0$.

\subsection{E-coordinates for the Schwarzschild metric}\label{quasiK-Sch}

To be able to send the mass $M$ of black hole  to zero we define  coordinates (we call them E-coordinates) as follows 
\bea\label{UmK}
\Um&=&-e^{-\frac{u}{4 M+b}},\,\,
\Vm=e^{\frac{v}{4M+b}},
\eea
here $b$ is a positive constant.
The coordinates run over the region $\Um<0, \Vm >0$. The question of the existence of an extension of the Schwarzschild metric that is analytic not only with respect to space and time variables,  but also with respect to the mass parameter, requires a separate consideration. 

The Schwarzschild metric in the E-coordinates is ($r>2M$)
\bea
ds^2&=&-(1-\frac{2M}{r})dt^2+(1-\frac{2M}{r})^{-1}dr^2+ r^2d \Omega^2\nn\\&=&
(4M+b)^2 (1 - \frac{2 M}{r}) \frac{\,d\Um d\Vm}{\Um \Vm}
+ r^2d \Omega^2,\label{Scw-Q-Kmm}
\eea
here $r$ is derived from the relation 
\be
e^{2r_*/(4M+b)}=-\Um \Vm.
\label{r-UV}
\ee
It is clear that in the limit $b\to 0$ the metric \eqref{Scw-Q-Kmm} 
rewritten as 
\bea
ds^2&=& -16(M+b/4)^2 \left(2 M\right)^{\frac{M}{M+b/4}} \frac{(r-2M)^{\frac{b}{4M+b}}}{r} e^{-\frac{r}{2(M+b/4)}}\cdot  
\,d\Um d\Vm + r^2d \Omega^2
\eea
becomes the Schwarzschild-Kruskal metric
 \eqref{SchKR}.

At the limit $M\to 0$ the metric \eqref{Scw-Q-Kmm} 
becomes
\bea\label{bUV}
 ds^2&=&-b^2e^{-\frac{2r}{b}}d\Um \,d\Vm+r^2d \Omega^2,\label{M0QK}\eea
here $r$ is defined by
\be
-\Um\Vm=e^{\frac{2r}{b}}.\label{r-UV-mm}\ee
Equation \eqref{r-UV-mm} is nothing but the formula  \eqref{r-UV}  rewritten as 
\be
(r-2M)^{\frac{M}{M+b/4}}=(2M)^{M/(M+b/4)}(-\Um\Vm)e^{-\frac{r}{2(M+b/4)}}\label{r-UV-m}\ee
in the limit $M\to 0$.
Explicit check shows that metric \eqref{M0QK} is the Minkowski metric.

\subsection{Temperature of  Schwarzschild  black holes in E-coordinates}\label{quasiK-Sch-T}
We consider the scaler field on the Schwarzschild background in two systems of coordinates,  the Eddington-Finkelstein $(u,v)$  and E-coordinates $( \Um,\Vm)$ related  as
\be
 \Um =-\exp\{-\frac{u}{4M+b}\},\qquad
   \Vm=\exp\{ \frac{v}{4M+b}\},\quad u,v\in \RR,\qquad M>0,
   \quad b>0.\label{Um}
\ee

The two-dimensial parts of  the Schwarzschild metric in these coordinate systems read
\bea
ds_2^2&=&-(1 - \frac{2 M}{r}) du dv=
(4M+b)^2 (1 - \frac{2 M}{r}) \frac{\,d\Um d\Vm}{\Um \Vm},\quad r>2M.
 \label{2Scw-Q-K}
\eea

The wave equations   for the scalar field $\phi(u,v)=\Phi(\Um, \Vm)$  in these coordinate systems are 
\bea
\partial_v\partial_u\phi&=&0,\,\,u,v\in\mathbb{R}\label{plane-weq}\\
\partial_  \Vm\partial_  \Um\,\Phi&=&0,\,\, \Um<0,\, \Vm>0.\label{fUV-weq}\eea
They can be represented as combinations of the left and write modes, $\phi(u,v)=\phi_R (u)+\phi_L (v)$.
For the real right mode (for the left mode all consideration is similar and will be omitted) one has
\bea
\phi _R(u)&=&\int _0^\infty d\omega (f_\omega b_\omega+f_\omega ^*b^+_\omega),\quad 
f_\omega(u)=\frac{1}{\sqrt{4\pi \omega}}e^{-i\omega u},\label{phiR}
\eea
where
\be
[b_\omega,b^+_{\omega^\prime}]=\delta(\omega-\omega^\prime).\label{bb+}
\ee
One  also has representation for $\Phi$-field
$
\Phi( \Um, \cV)=\Phi_R ( \Um)+\Phi_L ( \cV)$
where for right mode(similar for left ones) \bea
\Phi _R( \Um)&=&\int _0^\infty d\mu \Big( \fb_\mu\ff_\mu( \Um)+\fb^+_\mu\ff_\mu ^*( \Um)\Big),\quad 
\ff_\mu( \Um)=\frac{1}{\sqrt{4\pi \mu}}e^{-i\mu  \Um}
\eea
where
\be
[\fb_\mu,\fb_{\mu^\prime}]=\delta(\mu-\mu^\prime)\label{fbfb+}
\ee
Right (and left) modes in different coordinate system are relared
$
\phi _R(u)=\Phi _R( \Um(u))$ and therefore,
\bea
\int _0^\infty d\omega (f_\omega b_\omega+f_\omega ^*b^+_\omega)&=&\int_0^\infty d\mu (\ff_\mu \fb_\mu+\ff_\mu ^*\fb^+_\mu),
\qquad u\in \RR\label{PhiRQM}\
\eea
Multiplying \eqref{PhiRQM} on  $f_{\omega'}(u)$ and integrate the first equation over $\RR$ one  gets \footnote{For $b=0$ we get the standard formula for the  Schwarzschild metric in the Kruskal coordinates }
 \bea b_\omega=\int d\mu\Big(\beta ^*_{\omega\,\mu}\fb^+_{\mu}+\alpha^* _{\omega\,\mu}\fb_{\mu}\Big),\qquad  b^+_\omega&=&\int d\mu\Big(\beta _{\omega\,\mu}\fb_{\mu}+\alpha _{\omega\,\mu}\fb^+_{\mu}\Big),\label{bb}
\eea
where
\bea
\beta _{\omega\,\mu}&=&\int _{\RR} \,\frac{du}{2\pi }\sqrt{\frac{\omega}{ \mu}}e^{-i\omega u}e^{-i\mu \Um},
\label{delta}\\
\alpha _{\omega\,\mu}&=&\int _{\RR} \,\frac{du}{2\pi }\sqrt{\frac{\omega}{ \mu}}e^{-i\omega u}e^{i\mu \Um}.
\label{gamma}
\eea

The E-observer has the E-vacuum
\be
\fb_\omega|0_{E}\rangle=0,\ee
i.e. the state  $|0_{E}\rangle$  does   not contain $\fb$ particles. But it contains the Minkowski $b$ particles:
\begin{eqnarray}
&\,&\langle 0_{E}|N_{\omega}(b)|0_{E}\rangle \equiv\langle 0_{E}|b_{\omega}^+b_{\omega}|0_{E}\rangle
=\int_0^{\infty} d\mu \,|\beta_{\omega\mu}|^2.\label{beta-gamma}
\end{eqnarray}
The Bogoliubov coefficient $\beta_{\omega\nu}$ is given by \eqref{delta} with $\Um$ as in \eqref{Um} and we have
\bea \beta _{\omega\,\mu}=\frac{B}{2\pi  }\sqrt{\frac\omega\mu}\,e^{-\frac{\pi B\omega}{2}}\,\Big(\mu\Big)^{-iB\omega} \Gamma (i B\omega), \qquad B=4M+b\eea
Using the formula $\left|\Gamma\left(ix\right)\right|^2=
\pi/(x\sinh(\pi x))$ we get he Planck distribution
\bea
|\beta _{\omega\,\mu}|^2&=&\frac{B}{2\pi  \mu}\frac{1}{e^{2\pi B \omega }-1}, \label{delta2}\eea
 with the temperature
\be
T=\frac{1}{2\pi B}=\frac{1}{2\pi (4M+b)}
\ee

 \section{E-Coordinates in Minkowski Space }\label{QK-Min}
  \subsection{Minkowski space in terms of new coordinates $ \cU, \cV$ and $ \cT, \cX$} \label{QK-Min-4}
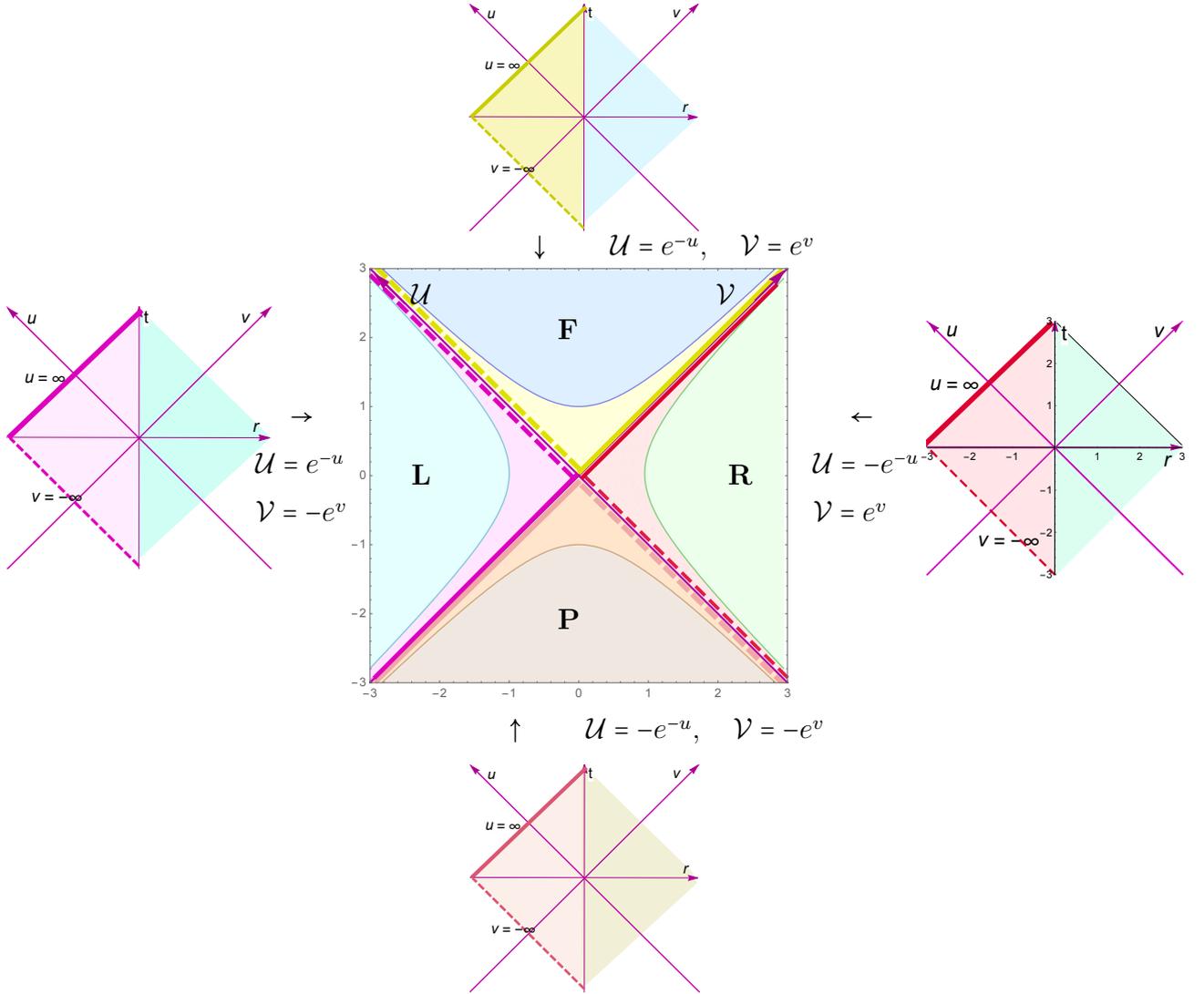
\begin{figure}[h!]
      $$\,$$
$$\,$$
$$\,$$
   $$\,$$
$$\,$$
$$\,$$
  \centering
   \qquad\qquad  \qquad\includegraphics[scale=0.25]{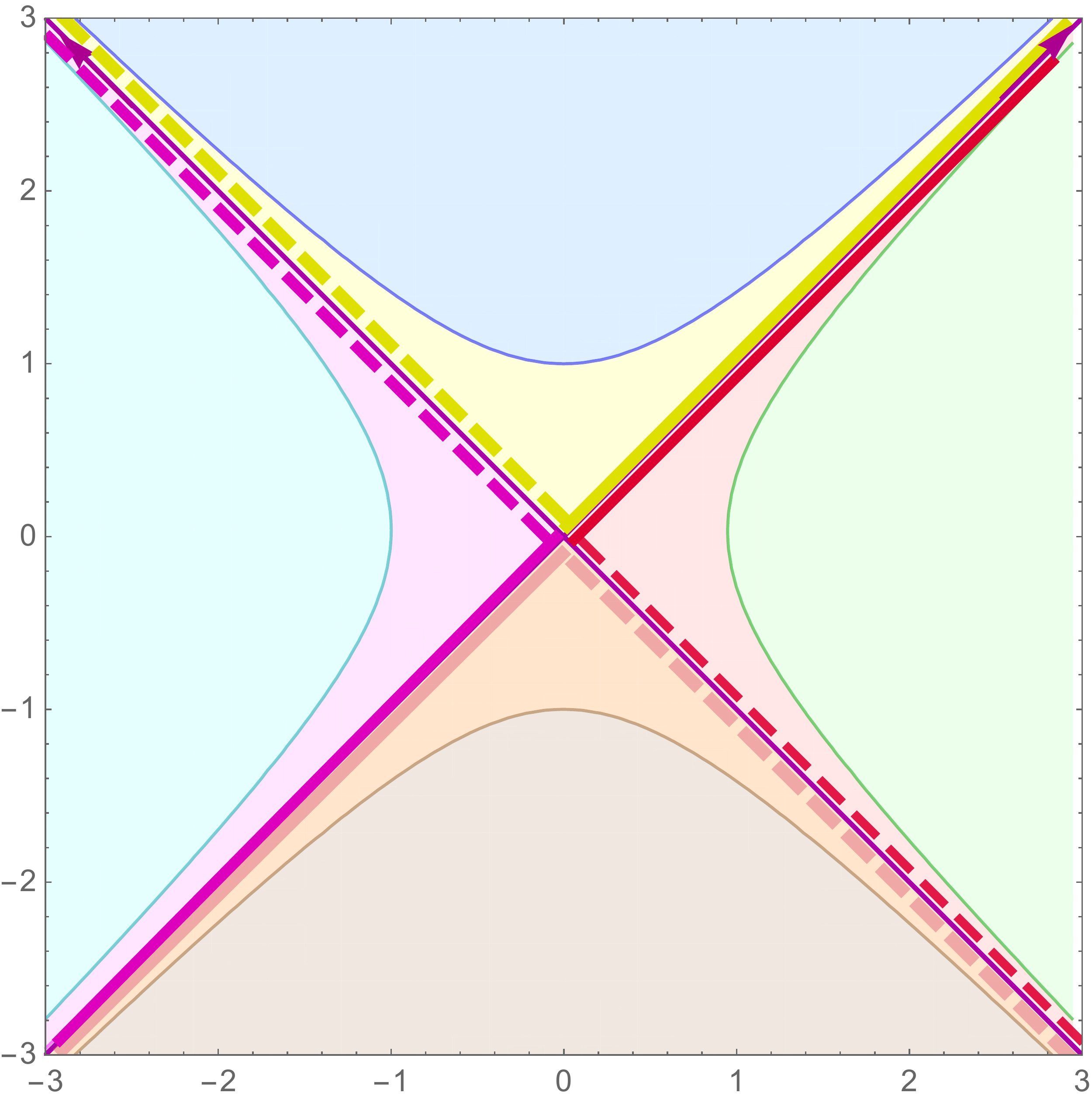}  
        \begin{picture}(50,150)\put(-35,165){$ \cV $}  
     \put(-160,165){$ \cU$}\put(-160,90){{\bf L}}\put(-30,90){{\bf R}}
     \put(-100,150){{\bf F}}
     \put(-100,30){{\bf P}}
          \put(-140,195)  { \includegraphics[scale=0.13]{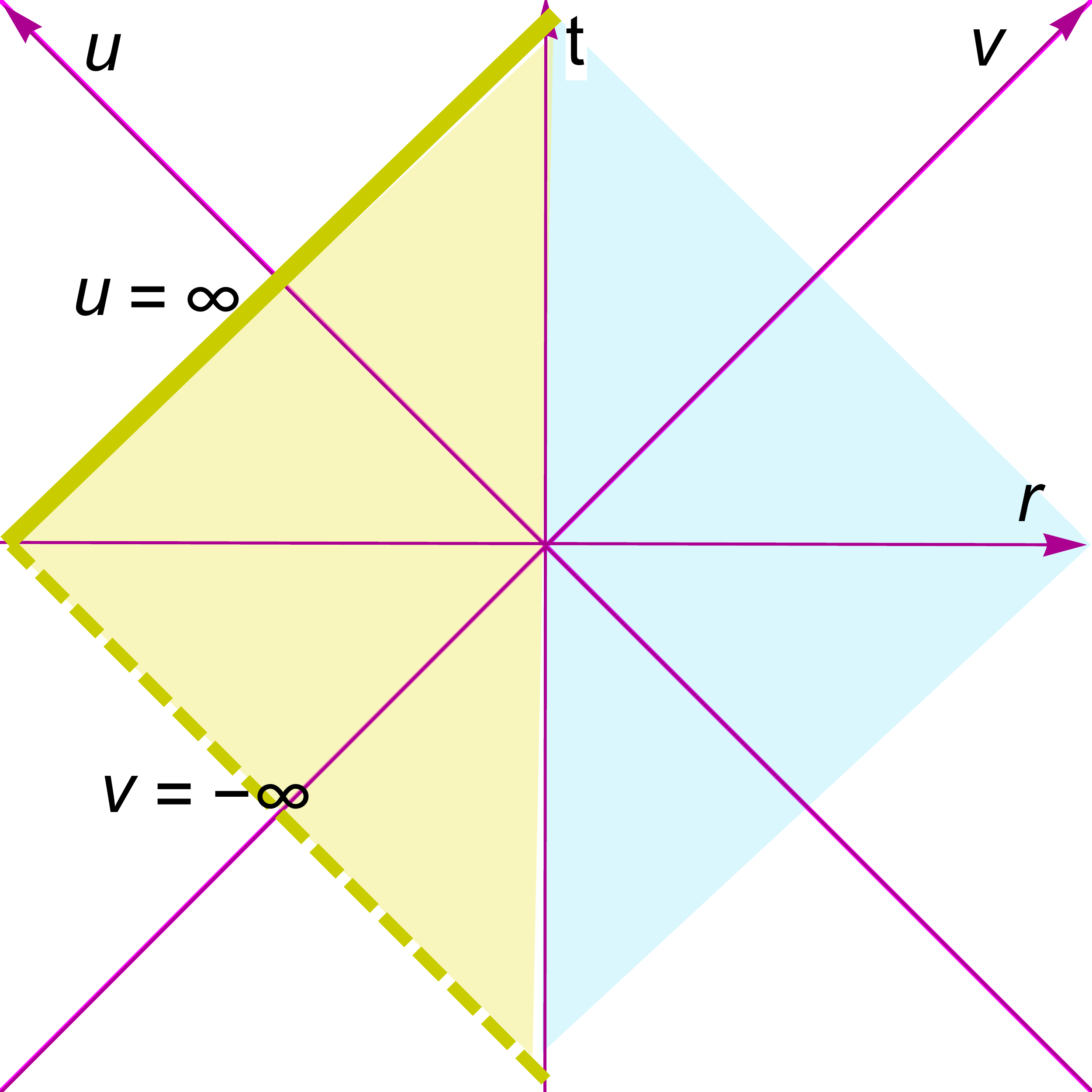} }
                     \put(-140,-120)  { \includegraphics[scale=0.13]{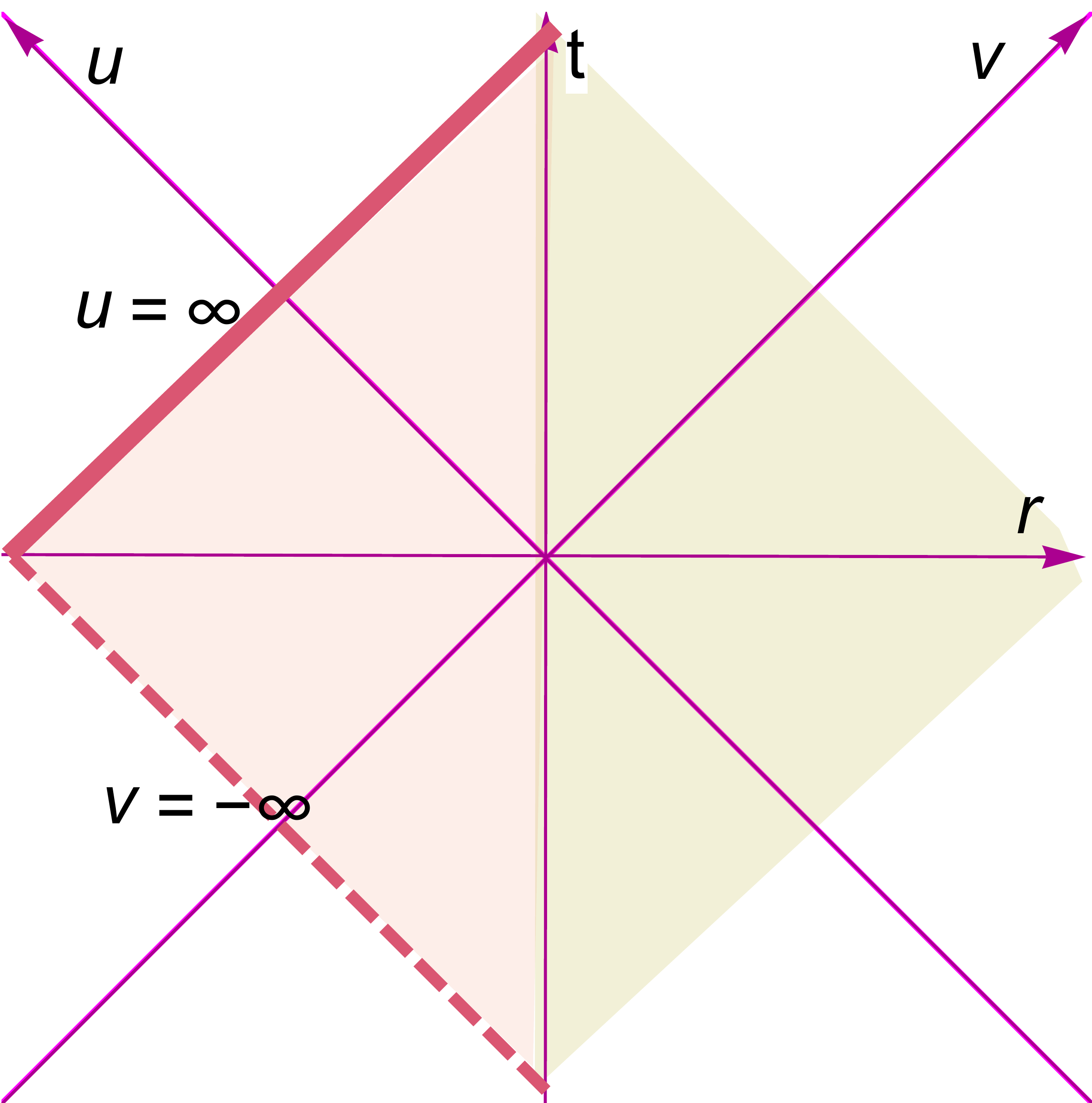} }
   \put(-110,185){$\downarrow\qquad  \cU =e^{-u},
   \quad \cV=e^{ v}$}     
   \put(-120,-15){$ \uparrow\qquad
     \cU =-e^{-u},\quad \cV=-e^{ v}$}	
 \put(-330,55)  { \includegraphics[scale=0.15]{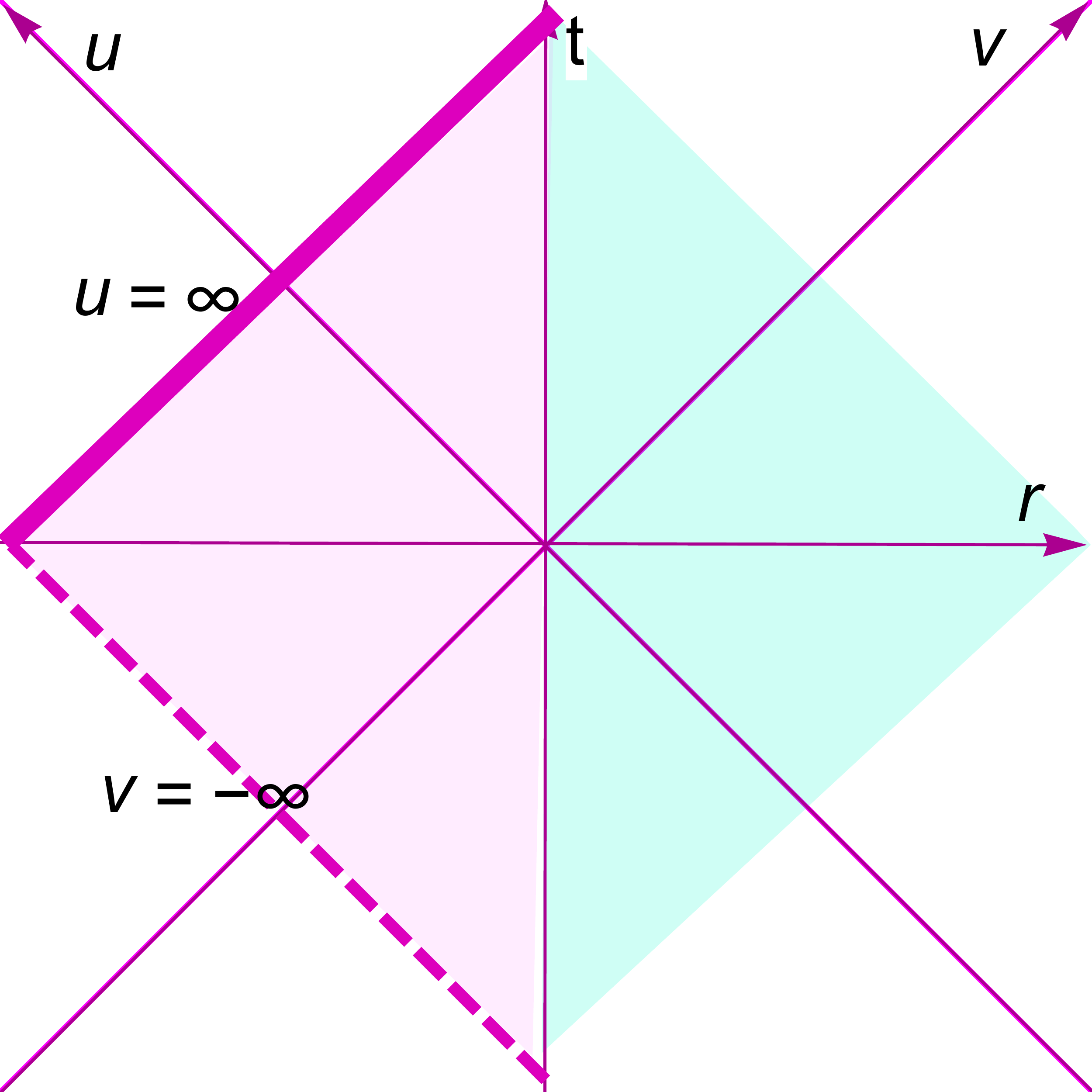}} 
     \put(45,50)  { \includegraphics[scale=0.18]{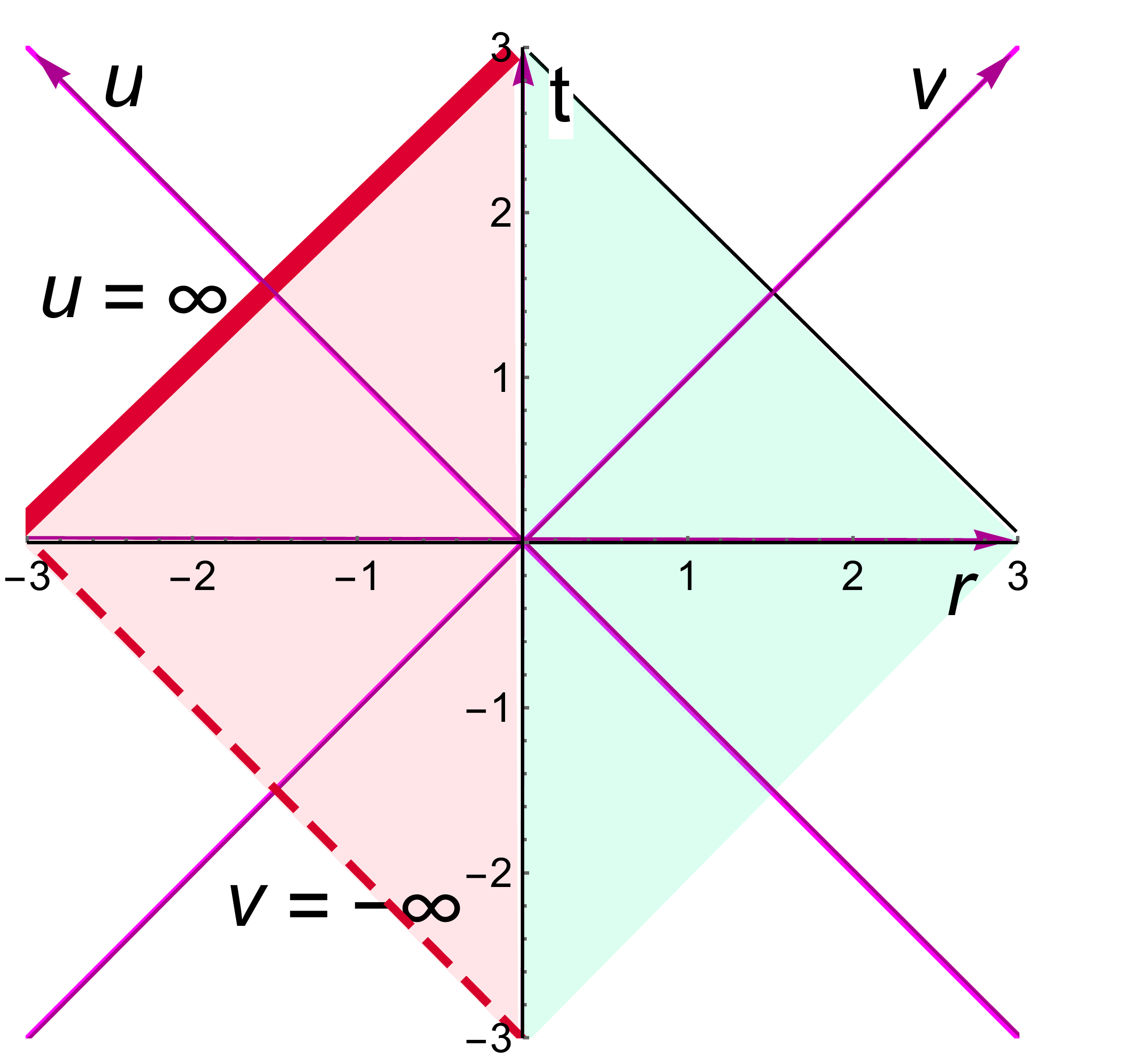} }
\put(-210,115){$\rightarrow$}
\put(-224,95){$ \cU =e^{-u}$}
\put(-224,75){$  \cV=-e^{ v}$}
\put(20,115){$ \boldmath{\leftarrow}$}
\put(5,95){$ \cU =-e^{-u}$}
\put(5,75){$  \cV=e^{ v}$}
\end{picture}
  $$\,$$
$$\,$$
$$\,$$
$$\,$$
 \caption{Map of 4 copies ${\mathbb M}^{1,1}$.  Here $b=1$}
 \label{Fig:QKMXT}
\end{figure}

  Starting from Minkowski coordinates 
 \be\label{Min}
 ds^2=ds_2^2+ r^2d \Omega^2, \qquad ds_2^2=-dt^2+dr^2, \qquad r\in {\mathbb R}_+,\quad t\in {\mathbb R}.
 \ee
 we introduce  the E-ones 
 \bea\label{quasiKU-R}
 \cU&= & \cU^{(R)}(t,r) =-\exp\{\frac{r-t}{b}\}=-\exp\{-\frac{u}{b}\},\qquad u=t-r\\
   \cV&=&  \cV^{(R)}(t,r) =\exp\{ \frac{t+r}{b}\}=\exp\{ \frac{v}{b}\},\qquad v=t+r,\,b>0,
 \eea
 $r>0$ corresponds to $  \cU \cV<-1$.  One has an expression of $r$ in terms of coordinates $\cU,\cV$
 \be
 r=b/2\log(-\cU \cV)\label{r-logUV}\ee
 The 2-dimensional Minkowski  metric in \eqref{Min} after this change becomes 
\bea
 ds^2_2=b^2\frac{d \cU \,d \cV}{ \cU \cV}
 \eea
 One can see that the metric \eqref{bUV} admits an extension to the region $-\cU \cV<0$. In this case by using formula \eqref{r-logUV}
 the coordinate $r$ is extending to the region $r\leq 0$, see Fig.\ref{Fig:QKMXT}.

 We also   define the coordinates
\be
 \cT=\frac{ \cU+ \cV}{2}, \qquad  \cX=\frac{ \cV- \cU}{2}\ee
We cover the  {\bf R}-region 
\bea
\mbox{\bf R}& =&\{(\cT,\cX)\in \RR^2\,| \,\cX^2- \cT^2>0, \cX>0\}\nn\eea
by the map
\bea
R&:&\qquad  \cT=e^{r/b}\sinh\frac{t}{b},\qquad  \cX=e^{r/b}\cosh\frac{t}{b},\qquad (t,x)\in M^{(1,1)}
\eea
The inverse transformation is
\bea
 t&=&b\, \mbox{arctanh}\frac{ \cT}{ \cX}\qquad r=b/2\,\log( \cX^2- \cT^2),\qquad  (\cT,\cX)\in R
 \eea
The metric  here is
\be
ds_2^2=-dt^2+dr^2=\frac{b^2}{ \cX^2- \cT^2}(-d \cT^2+d \cX^2)\label{dsM-QK}\ee
 Let us introduce  the future ({\bf F}) E-coordinates
 \bea
  \cU= \cU^{(F)}(r,t) =\exp\{\frac{r-t}{b}\}=\exp\{-\frac{u}{b}\},\quad
    \cV=  \cV^{(F)}(r,t)=\exp\{ \frac{t+r}{b}\}=\exp\{ \frac{v}{b}\},\nn\\\label{FquasiKU}
 \eea
and they cover the {\bf F}-part of $( \cU,  \cV)$-plane, see the top part of Fig.\ref{Fig:QKMXT},
\be
  \cU  \cV>0\ee
The  Minkowski  metric  \eqref{Min} after \eqref{FquasiKU} becomes 
\bea
 ds_2^2&=&
 =- b^2\frac{d \cU \,d \cV}{ \cU\cV}\label{FUV}
 \eea
 
 One   also introduces  the left ({\bf L})  E-coordinates
 \bea
 \cU= \cU^{(L)}(t,r) =\exp\{\frac{r-t}{b}\}=\exp\{-\frac{u}{b}\},\qquad
  \cV= \cV^{(L)}(t,r)=-\exp\{ \frac{t+r}{b}\}=-\exp\{ \frac{v}{b}\},\nn\\\label{LquasiKU}
 \eea
with the inequality
\be
  \cU \cV<0\ee
and past (P)  E-coordinates
\bea
\cU=  \cU^{(P)}(t,r)  =-\exp\{\frac{r-t}{b}\}=\exp\{-\frac{u}{b}\},\qquad
  \cV=  \cV^{(P)}(t,r) =-\exp\{ \frac{t+r}{b}\}=-\exp\{ \frac{v}{b}\},\nn\\\label{PquasiKU}
 \eea
with the inequality
\be
  \cU  \cV>0\ee
 The metric can be written in the universal way
  \bea
 ds_2^2&=&-dudv=- b^2\frac{d \cU \,d \cV}{| \cU \cV|}, \qquad   \cU \cV\neq 0\label{ds4}
\eea
 These maps are shown in Fig.\ref{Fig:QKMXT}.

 To summarize, we have obtained the 2-dimensional plane divided on 4 disconnected regions with E-coordinates $(\cU,\cV)$
 and the metric  given by \eqref{ds4}.
 Or in other words, we have obtained that two-dimentional  space is represented as a union of 4 disconnected regions 
 {\bf R,F,P} and {\bf L},
 each of which of them is isometric to two-dimensional Minkowski space.

 \subsection{Geodesics in E-coordinates }\label{QK-geodesics}
 The geodesics in E-coordinates  $(\cU,\cV)$ can be obtained simply by changing of variables from geodesics in Minkowski space. 
 However it seems instructive and useful for quantization to investigate  geodesics directly in E-coordinates. 
 Geodesics equations for the metric \eqref{bUV}
are
 \bea
\frac{  \cV'(\fs)^2- \cV(\fs)  \cV''(\fs)}{ \cU(\fs)  \cV(\fs)^2}&=&0,\label{QKgeoV}\\
 \frac{  \cU'(\fs)^2- \cU(\fs)  \cU''(\fs)}{ \cU(\fs)^2  \cV(\fs)}&=&0.\label{QKgeoU}
 \eea
These geodesics equations \eqref{QKgeoV} and \eqref{QKgeoU} can be solved to get
 \be
  \cV(\fs)=c_2 e^{c_1 \fs},\qquad   \cU(\fs)=c_4 e^{c_3 \fs},\label{geod-UV}\ee
 here $-\infty<\fs<\infty$.
 
To guaranty that the geodesics run in one of regions {\bf R,F,L,P}  we have to take
\bea
\mbox{\bf R}: \quad c_2>0,\quad c_4<0, \qquad \mbox{\bf F}: \quad c_2>0,\quad c_4>0\\
\mbox{\bf L}: \quad c_2<0,\quad c_4>0,  \qquad \mbox{\bf P}: \quad c_2<0,\quad c_4<0\eea
here $-\infty<s<\infty$, $c_1$ and $c_3$ can have any signs.

The geodesics \eqref{geod-UV} in $( \cT, \cX)$ coordinates are 
  \bea
   \cX(\fs)=\frac{1}{2} \left(c_4 e^{c_3 \fs}-c_2 e^{c_1\fs}\right),\qquad
    \cT(\fs)=\frac{1}{2} \left(c_2 e^{c_1 \fs}+c_4 e^{c_3 \fs}\right)
  \eea
  
  One can check that these geodesics  after mapping to $t,x$ (Minkowski space-coordinates) are  (as should be) the straight lines
  \bea
   t=t(\fs)&=&b\, \mbox{arctanh}\frac{ \cT(\fs)}{ \cX(\fs)}=\frac{1}{2} \left(\fs (-c_1-c_3))-\log \left(-\frac{c_2}{c_4}\right)\right)\label{geod-ts}\\
r=r(\fs)&=&2b\,\log( \cX(\fs)^2- \cT(\fs)^2)=2b\Big((c_1+c_3)\fs+\log(-c_2c_4)\Big)\label{geod-rs}\eea

We  have 3 types of geodesics:
\begin{itemize}
\item $c_1c_3<0$, in this case both "ends" of geodesics are in infinities, see for example plot Fig. \ref{Fig:geod-XX-TT}.A
\item $c_1c_3>0$, in this case one "end" of geodesics is at  infinity and  the second one  is at zero, see for example plot Fig. \ref{Fig:geod-XX-TT}.B
\item $c_1=0, c_3\neq 0$ or $c_3=0, c_1\neq 0$, in this case geodesics are bounded by characteristics  $ \cU=0$ or $ \cV=0$, see Fig.\ref{Fig:geod-XX-TT}.C
\end{itemize}
From \eqref{geod-UV} follows that \be
\left(\frac{ \cV}{c_2}\right)^{c_3}= \left(\frac{ \cU}{c_4}\right)^{c_1}\ee
   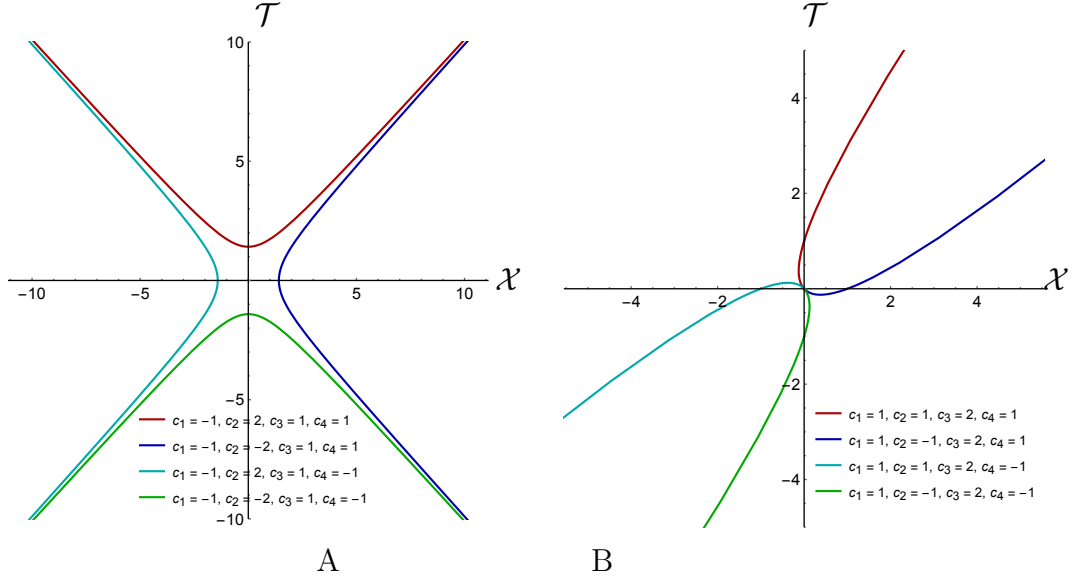
\begin{figure}[h!]
  \centering
\includegraphics[scale=0.25]{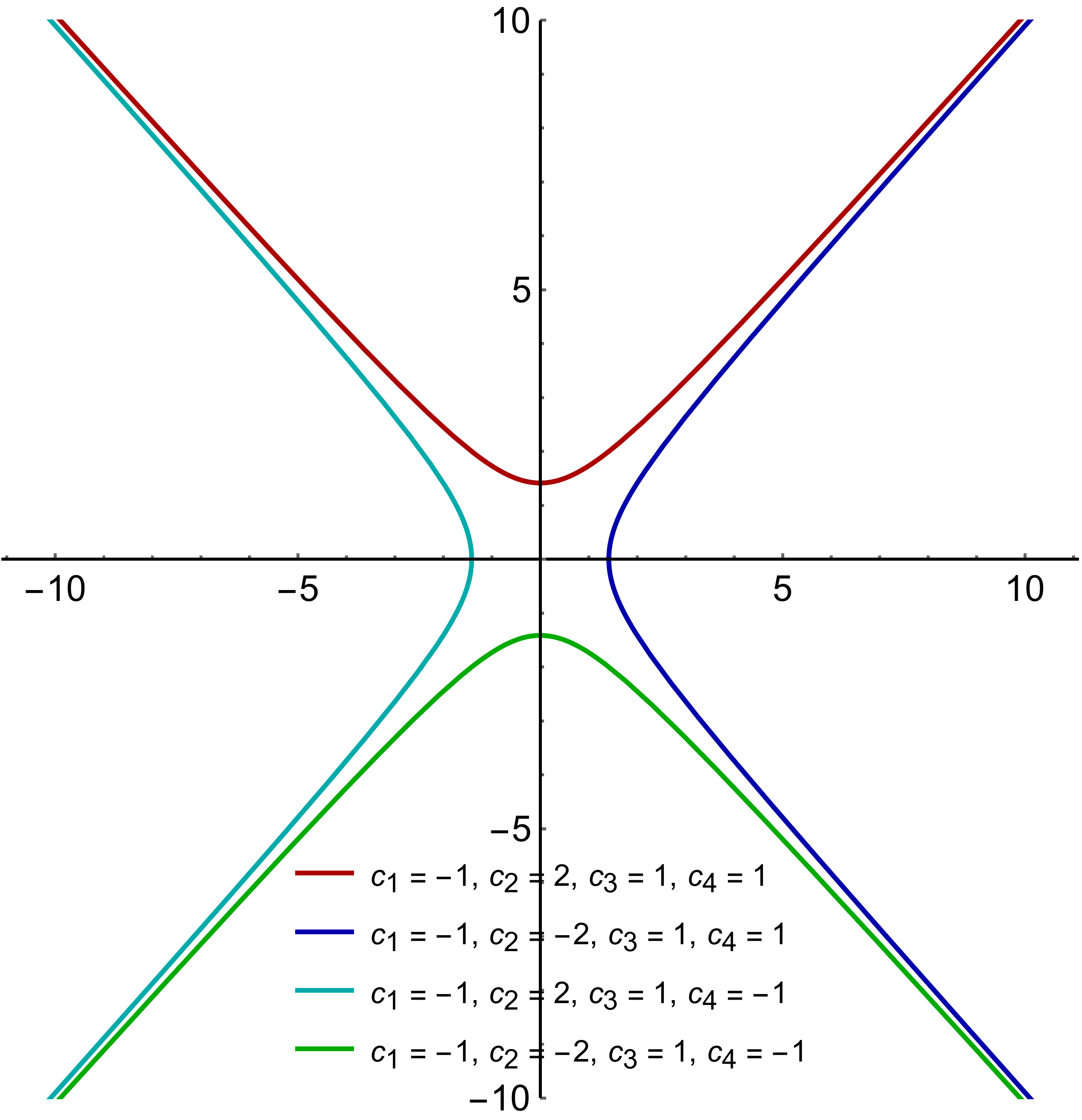} \qquad     \includegraphics[scale=0.25]{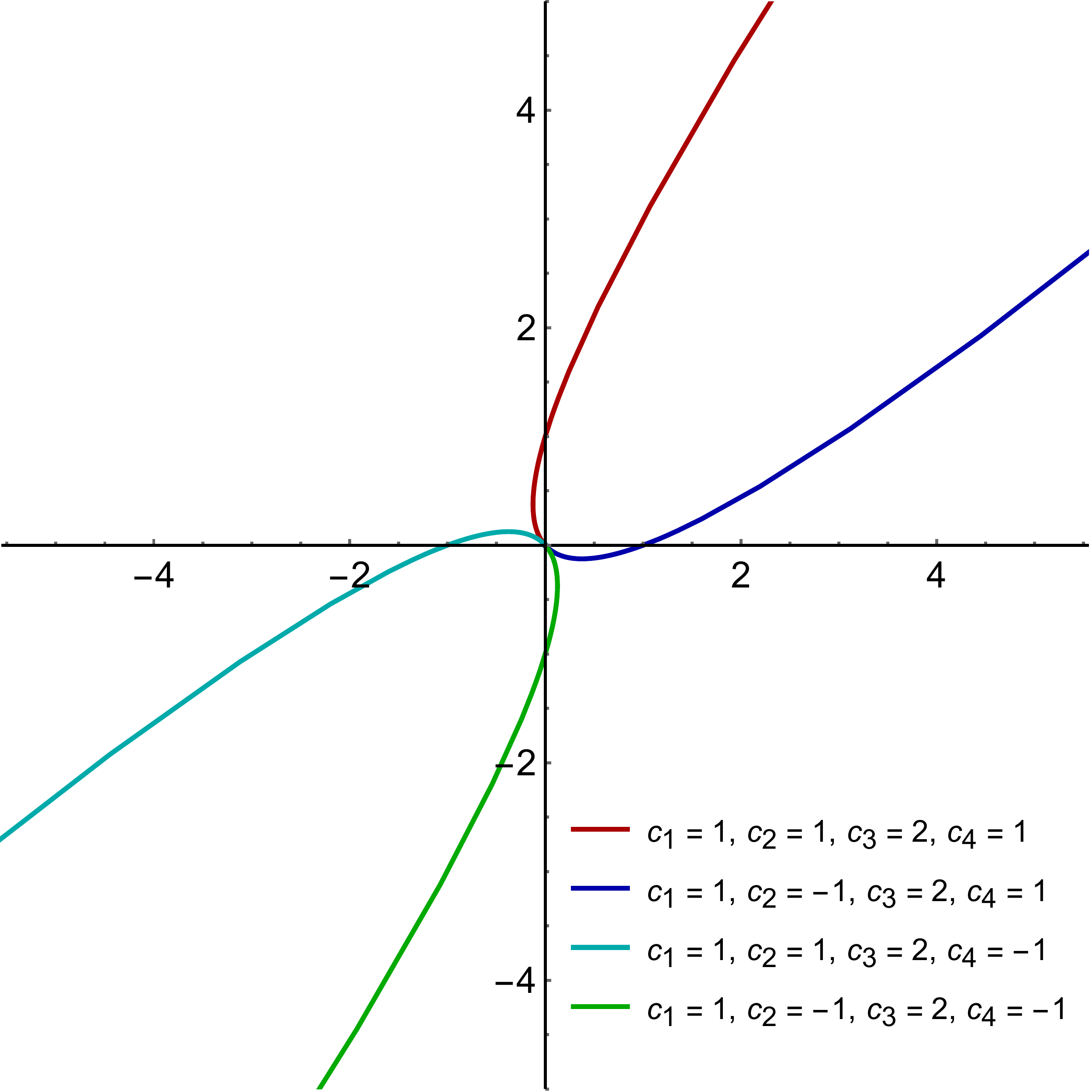}
   \begin{picture}(-50,170)\put(-95,190){$ \cT $}  
     \put(-300,190){$ \cT$}
     \put(-5,90){$ \cX $}  
     \put(-210,90){$ \cX$}
    \end{picture} \\
A\qquad\qquad\qquad\qquad B
      \caption{Geodesics for different values of $c_i, i=1,2,3,4$. 
  }
  \label{Fig:geod-XX-TT}
\end{figure}

 \begin{figure}[h!]
  \centering

     \includegraphics[scale=0.25]{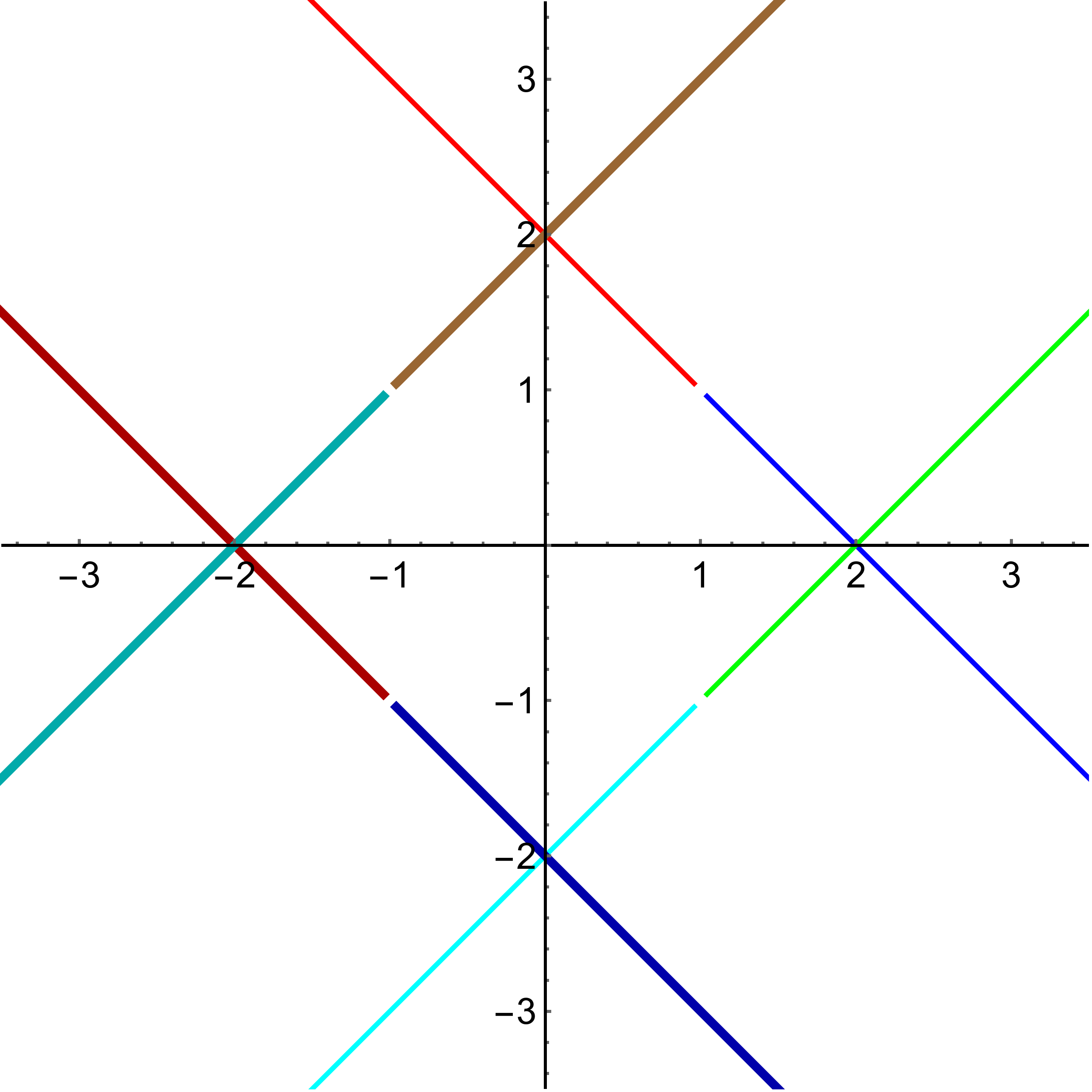}  \qquad   
     \includegraphics[scale=0.25]{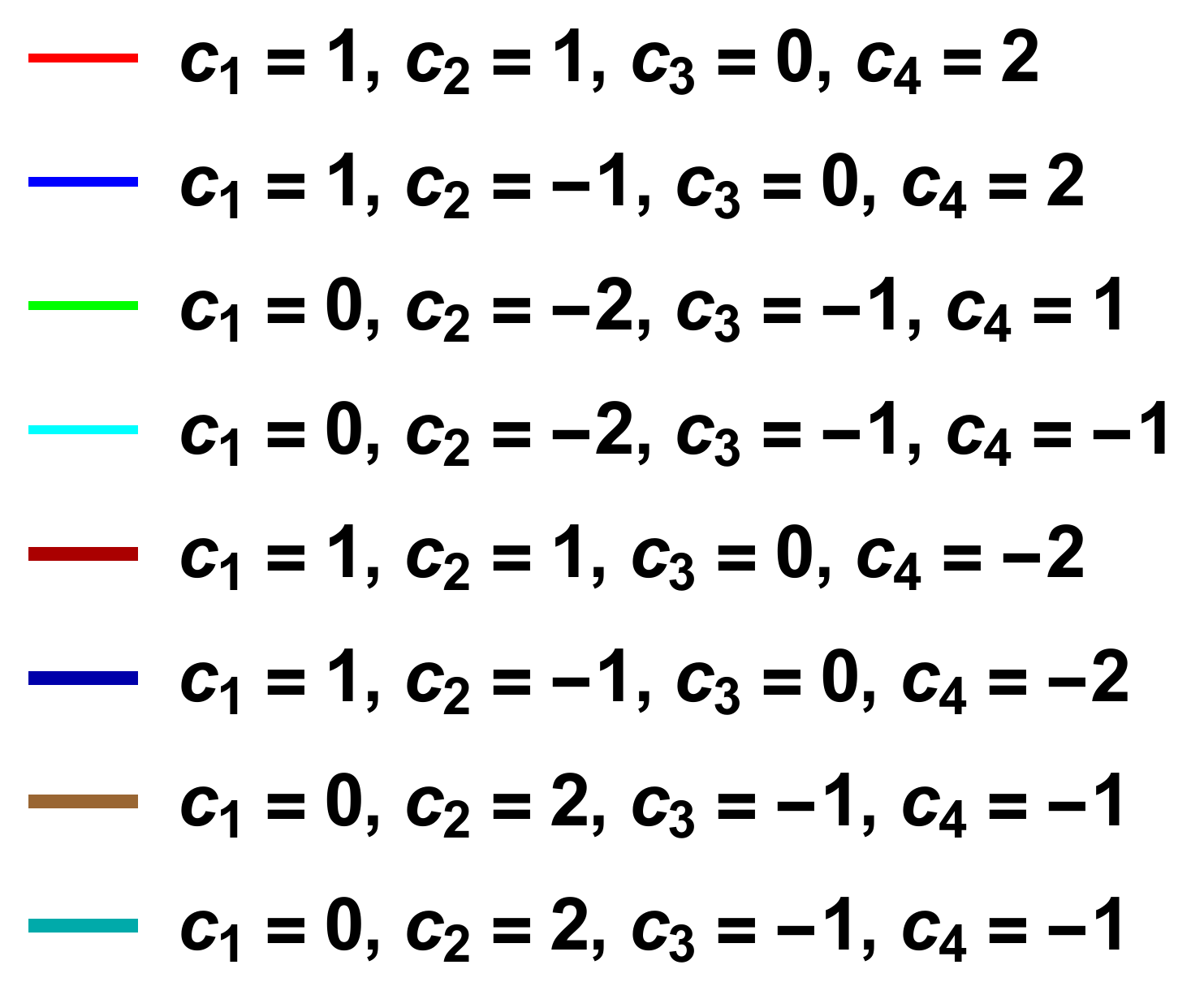}
  
  \caption{ Geodesics for the case when $c_1$ or $c_3$ is equal to 0.    
  }
  \label{Fig:geod-XX-TT}
\end{figure}

\newpage
\newpage

\subsection{Acceleration in E-coordinates}\label{Acc-QK}
 Let us consider  the time-like trajectory with located at  $ \cX= \cX_0$. The proper 
   time
   $\tau$  
   \bea
 \tau= b \arctan \left(\frac{ \cT}{\sqrt{ \cX_0^2- \cT^2}}\right)+\tau_0\label{sfT}
   \eea
From  \eqref{sfT}
one has
\be
 \cT=\pm \cX_0\sin \left(\frac{\tau-\tau_0}{ b}\right)\label{sfTm}\ee 
  
 We see that it takes the finite time $\cT$, or $\tau$, to reach the characteristic $\cX=\cT$.
    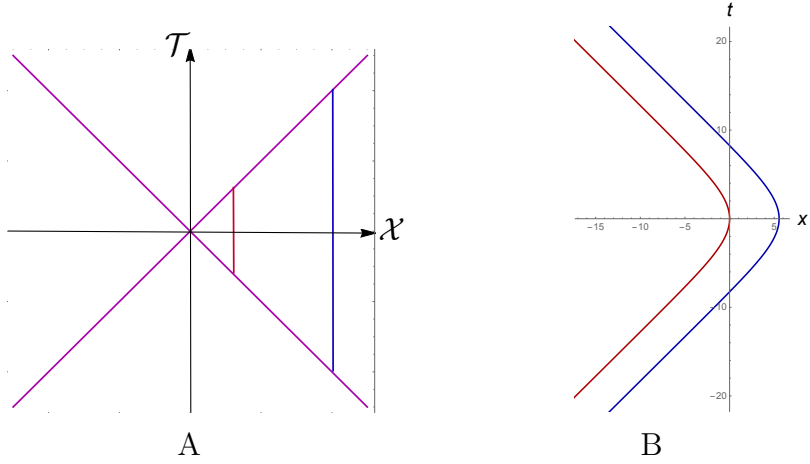
\begin{figure}[h!]
  \centering
\includegraphics[scale=0.2]{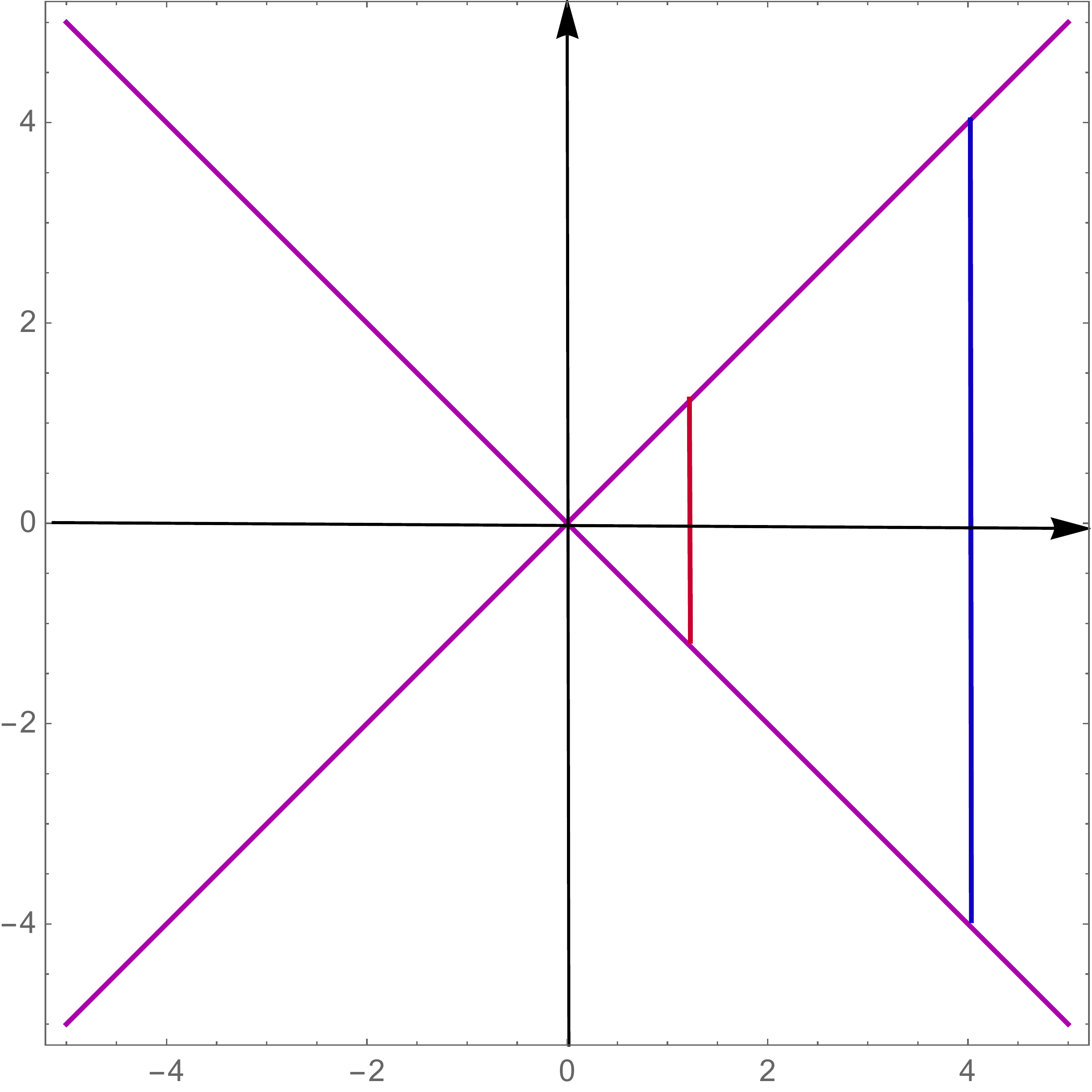}  \qquad \qquad \qquad\includegraphics[scale=0.18]{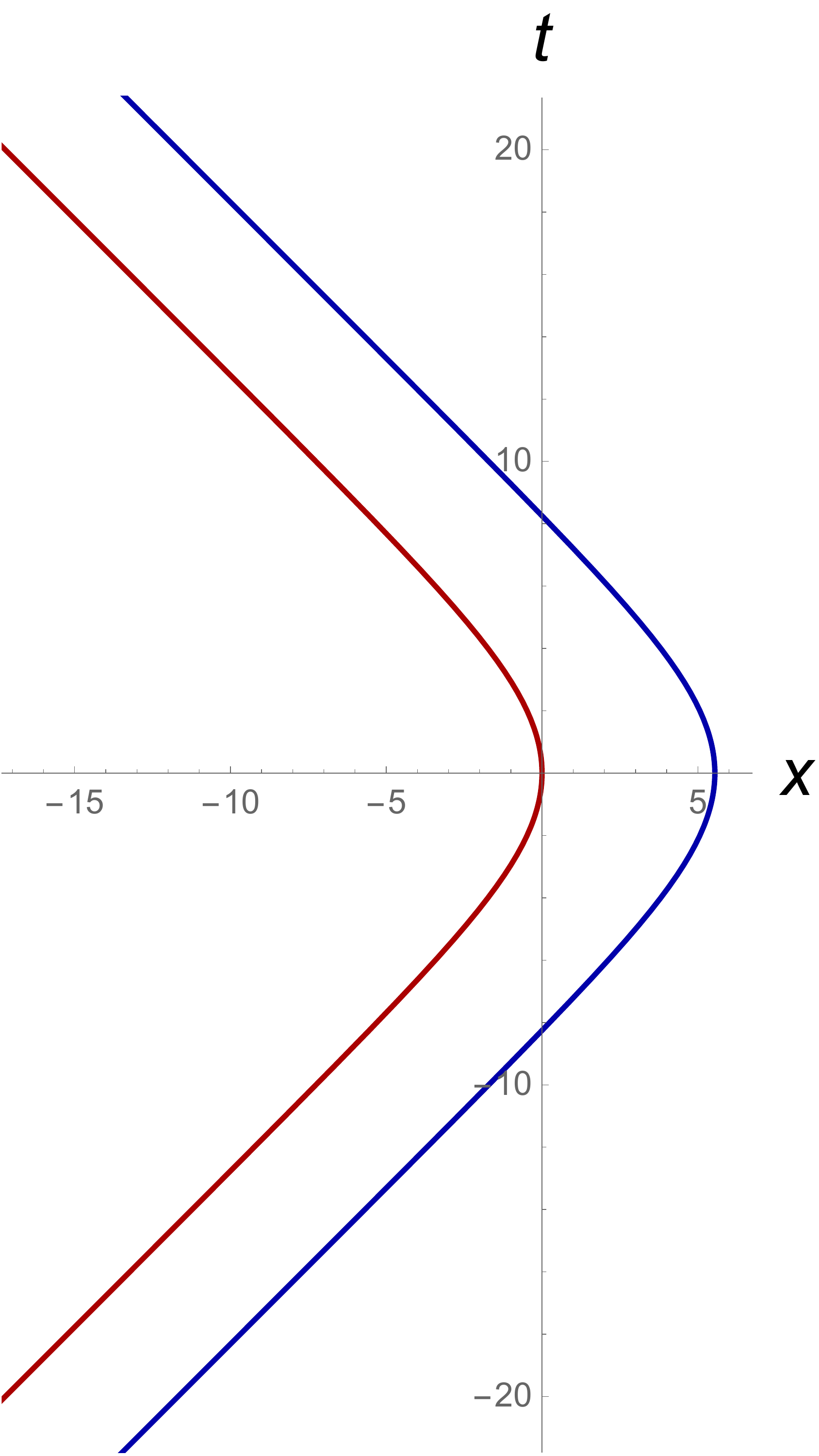}  
       \begin{picture}(5,150)\put(-245,135){$ \cT$}\put(-165,65){$ \cX$}
      \end{picture}
     \\A\qquad\qquad\qquad\qquad \qquad \qquad \qquad B
  \caption{Trajectories   with $ \cX_0=1$ (red) and   $ \cX_0=4$ (blue) in $( \cX, \cT)$ plane (A)  and $(x,t)$ plane (B). We can take $\tau_0=0$.
}
  \label{Fig:sT}
\end{figure}

In Fig.\ref{Fig:sT}.A and Fig.\ref{Fig:sT}.B. we plot trajectories  with $ \cX_0=1$ (red) and   $ \cX_0=4$ (blue) in $( \cX, \cT)$  and   $(t,x)$ planes, respectively. 
We can parameterize these  trajectories in the Minkowski  by  $\tau$:
 \bea
 t&=& b\,\mbox{arctanh}\left(\sin \left(\frac{\tau-\tau_0}{b}\right)\right)\label{t-Ttau-m}\\
 r&=&  \frac{b}2\,\log\left( \cX_0^2\cos^2 \left(\frac{\tau-\tau_0}{b}\right)\right)\label{r-Ttau-m}
 \eea
We get
\bea
V^0&=&\frac{dt}{d\tau}=\frac{ \cX_0}{\sqrt{ \cX_0^2- \cT^2}},\qquad \qquad \, V^1=\frac{dx}{d\tau}=\frac{ \cT}{\sqrt{ \cX_0^2- \cT^2}},\label{V0}\\
W^0&=&\frac{dV^0}{d\tau}=\frac{ \cX_0  \cT}{b \left( \cX_0^2- \cT^2\right)},  \qquad   W^1=\frac{dV^1}{d\tau}=\frac{ \cX_0^2}{ b \left( \cX_0^2- \cT^2\right)}.\label{W0}\eea 
We have
\be
W^2\equiv-(W^0)^2+(W^1)^2=
\frac{ \cX^2_0 }{ b^2 \left( \cX_0^2- \cT^2\right)}.\label{W}\ee
One can also rewrite \eqref{W} in term of $t$ 
\bea
W^2&=&\frac{1}{ b^2 }\cosh^2\frac{t}{b},\label{Wm}\eea
or  in term of $\tau$
\bea
W^2&=&\frac{1}{ b^2\cos^2 \left(\frac{\tau-\tau_0}{b}\right)} .\label{QK-Sch}
\eea
These calculations show that the acceleration $W$ of the E-trajectory with  "$\cX=\cX_0$" in the inertial coordinate system
($t,x$)  increases with increasing of the inertial time $t>0$ as $W \sim \frac1{b}e^{t/b}$. If $b\to 0$
this acceleration increases to infinity, and decreases to 0 when $b\to \infty$.

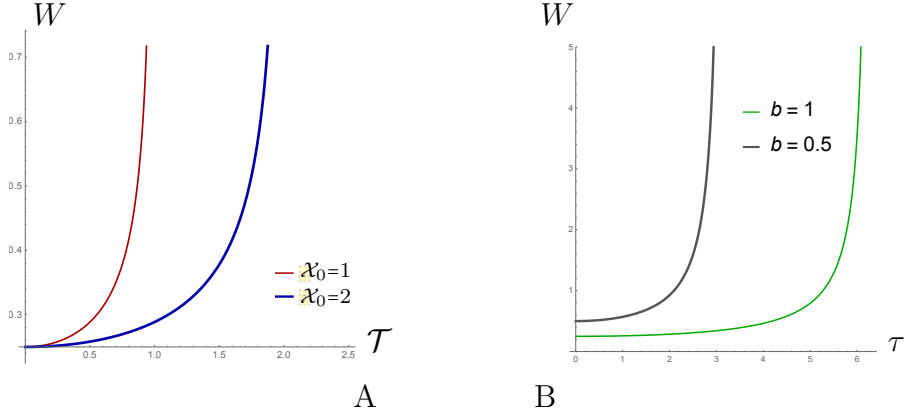
\begin{figure}[h!]
  \centering
\includegraphics[scale=0.18]{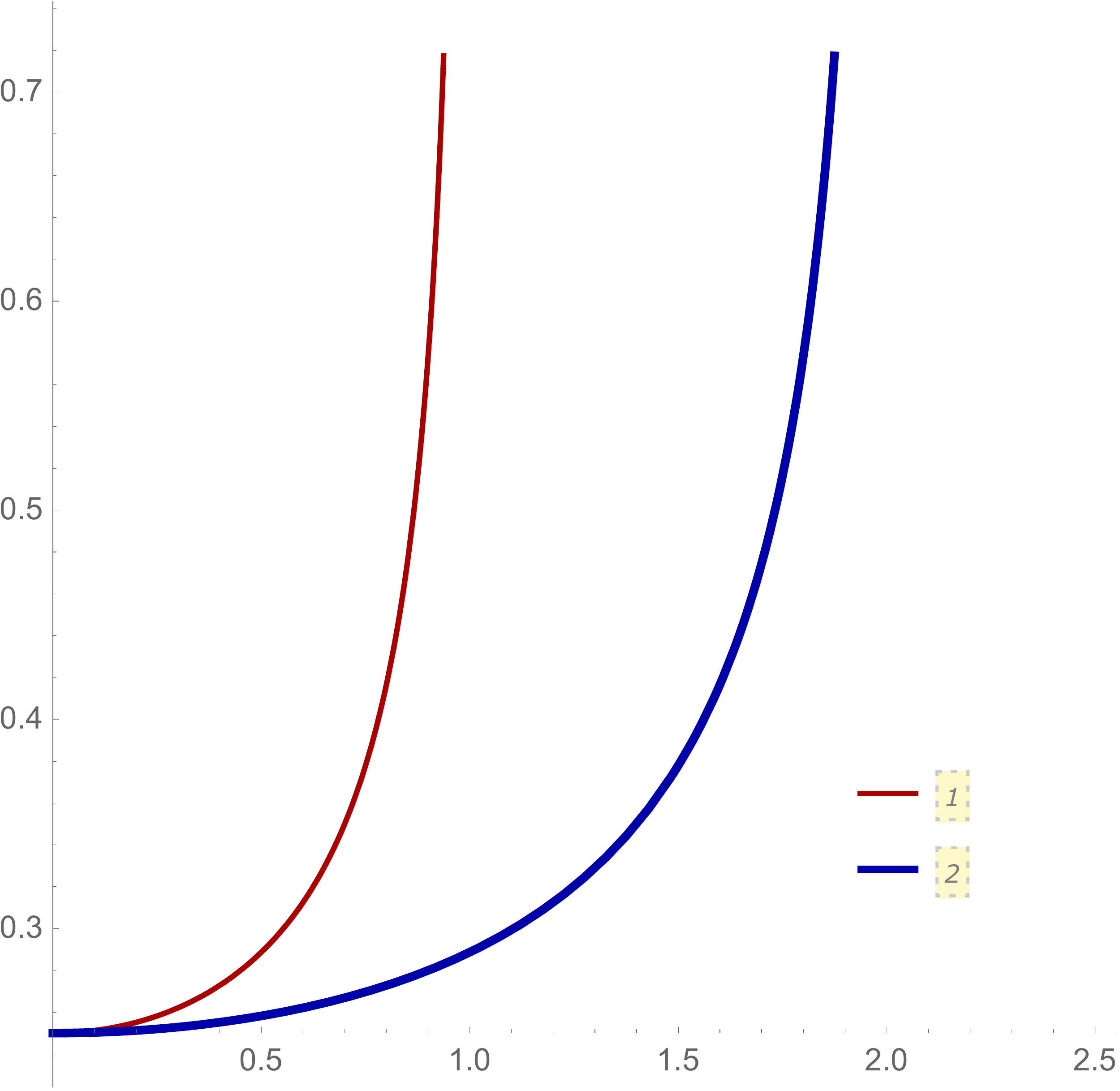}  
 \begin{picture}(5,150)\put(-125,128){$ W$}\put(-0,5){$ \cT$}\put(-25,35){$ _{\cX_0=1}$}
       \put(-25,25){$ _{\cX_0=2}$}
      \end{picture}\qquad\qquad\qquad
      \includegraphics[scale=0.16]{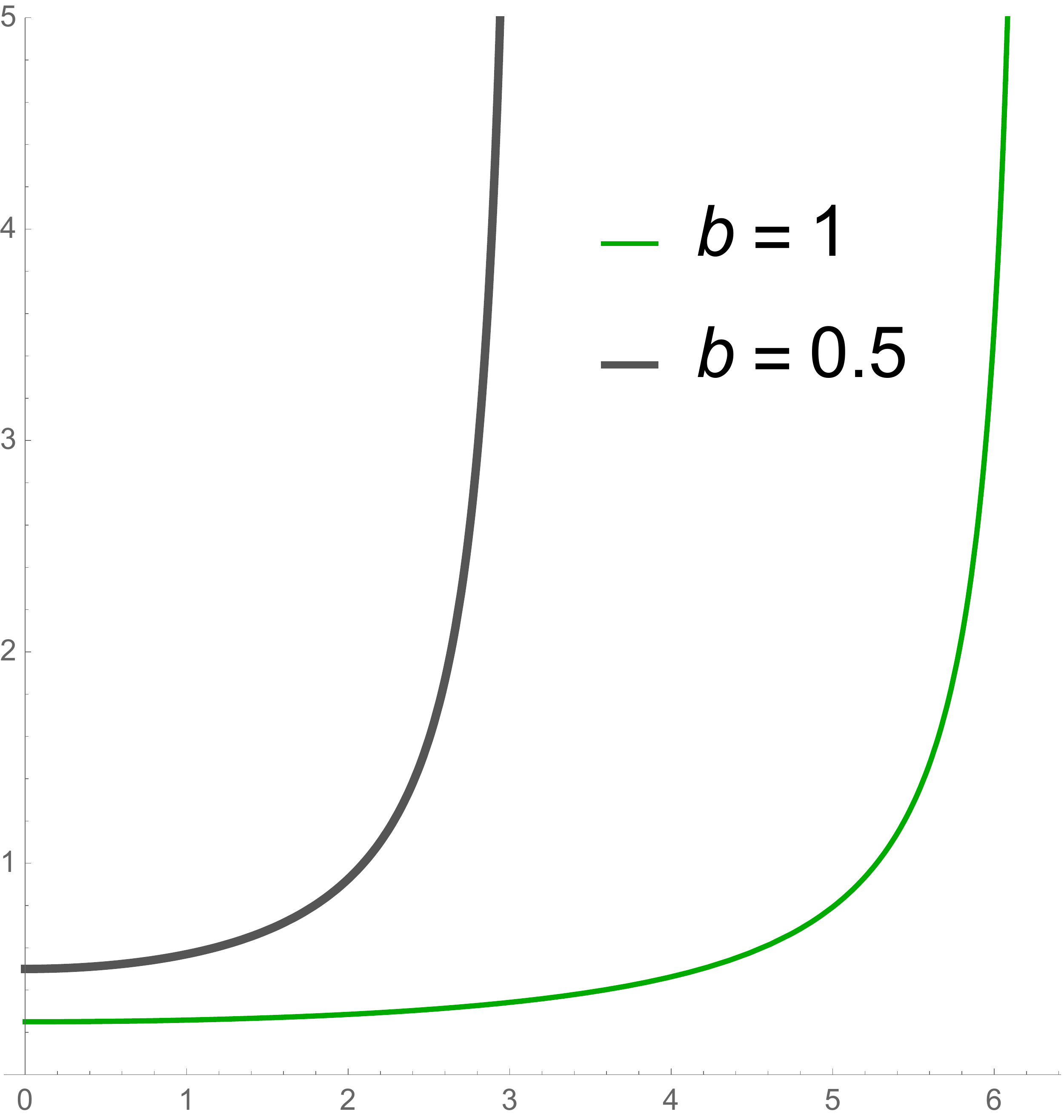} 
       \begin{picture}(5,150)\put(-130,128){$ W$}\put(-0,5){$ \tau$}      \end{picture}\\
       A\quad\quad\quad\quad\quad B
  \caption{Accelerations of  E-observers  located at  $ \cX_0=1$ (red) and   $ \cX_0=2$ (blue) agains  their E-time $\cT$, $b=1$ (A) and proper time $\tau$, $b=0.5$ (gray) and $b=1$ (green) (B). 
  }
  \label{Fig:WT}
\end{figure}
Acceleration of  the E-observer  located at  $ \cX_0$ depends on its E-time $\cT$ (see Fig.\ref{Fig:WT}) as well as it own proper time. The  proper time does not depend on the location. In all cases, the acceleration is inversely proportional to the $ b $ parameter.

\subsection{Comparison of Exponential and Rindler coordinates}\label{qK-Rindler}
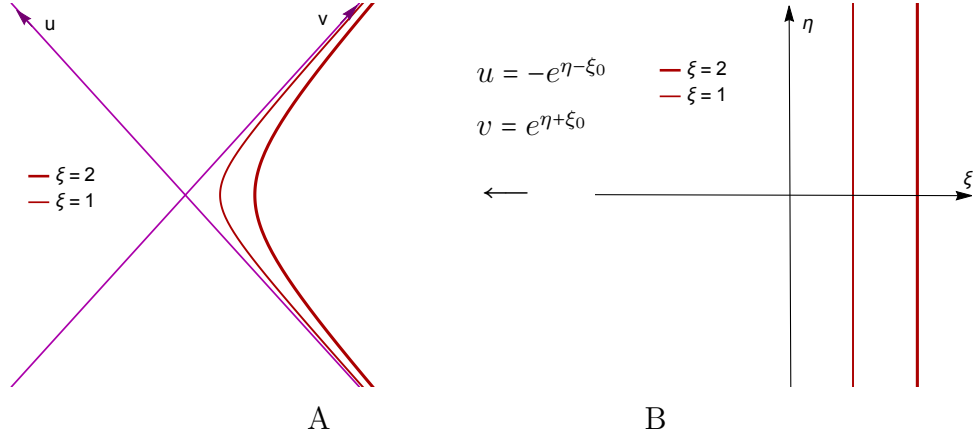
\begin{figure}[h!]
  \centering
\includegraphics[scale=0.2]{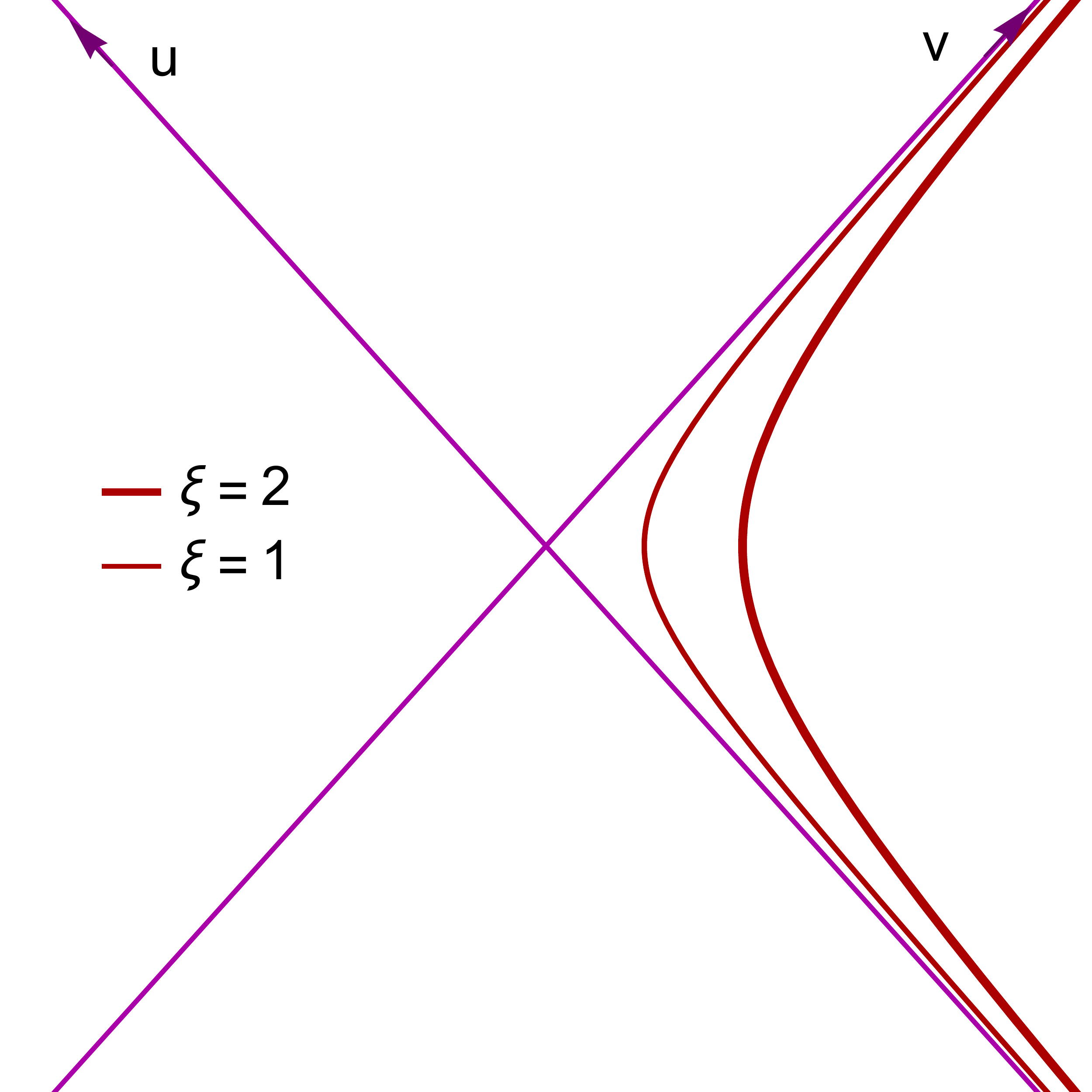}  \begin{picture}(50,150)\put(35,70){$ \longleftarrow $
\put(-20,25){$  v=e^{\eta +\xi_0}$}
\put(-20,45){$u =-e^{\eta-\xi_0}$}
} \end{picture} \qquad \includegraphics[scale=0.2]{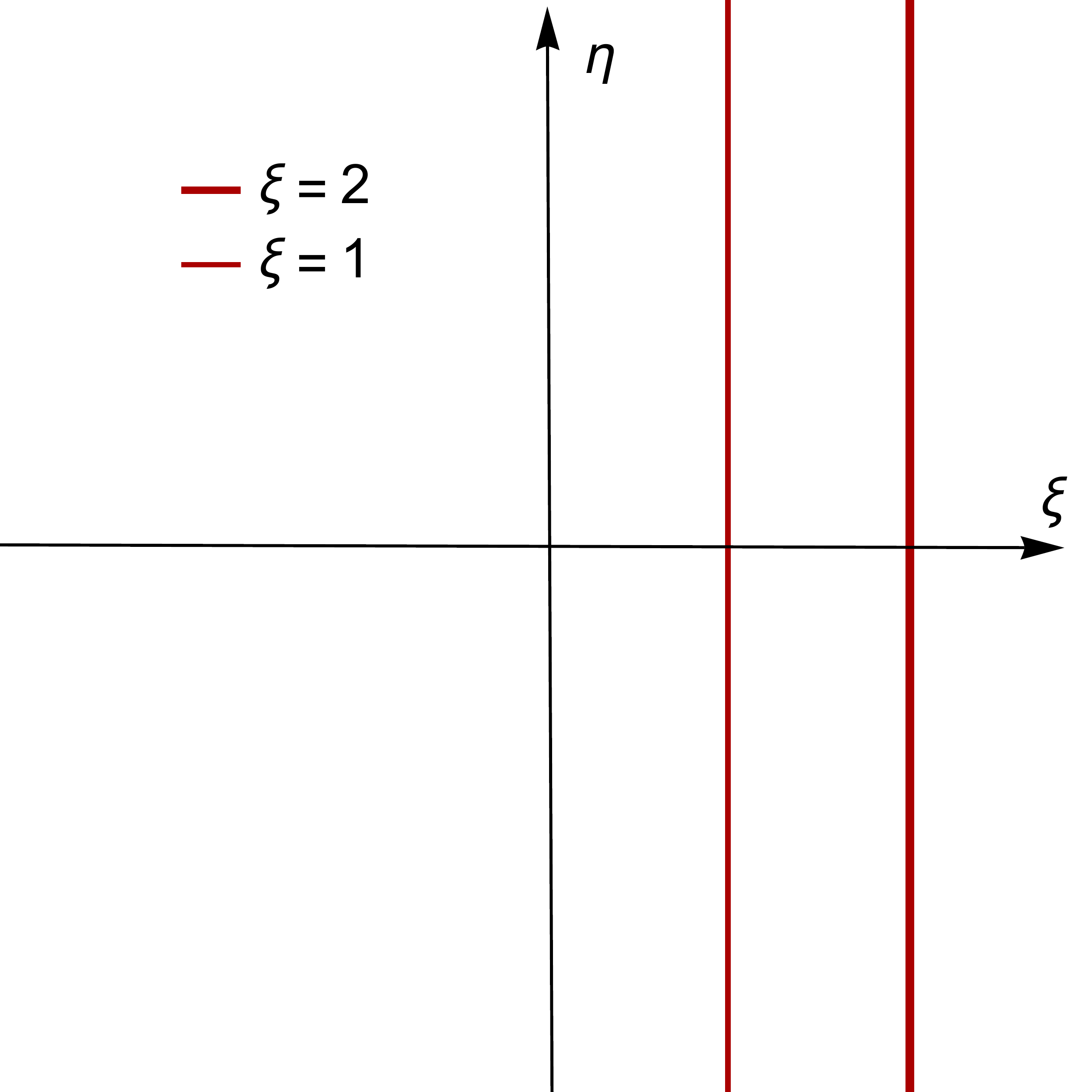}   \\
      A\qquad \qquad \qquad \qquad \qquad  B
    \caption{Rindler observers "in the rest" (located at $\xi=1,2$) move along hyperbolae in the inertial coordinates with constant acceleration.   
    }
  \label{Fig:RR}
\end{figure}

The accelerated observer  traveling in Minkowski with constant acceleration $a$  is described by 
 the Rindler coordinates 
 $(\nu,\vartheta)$ related with the inertial coordinates $(u,v)$ \cite{Rindler:1966,BD}
 \be
u=-\frac{1}{a}e^{-a\nu},\qquad v=\frac{1}{a}e^{a\vartheta},\,\,a>0. \label{uv-nuvartheta}
\qquad
[[-\frac{\log( - au)}{a}=\nu]]\ee
These transformations define the Rindler observer as an observer that is "at rest" in Rindler coordinates, i.e., maintaining constant $\xi$ and only varying $\eta$; $a$  represents the proper acceleration (along the hyperbola $\xi=0$) of the Rindler observer,  whose proper time is defined to be equal to Rindler coordinate time). The Rindler observer in the rest in $(\eta,\xi)$ Rindler coordinate travels along the hyperbola 
\be
uv=e^{\xi_0}\ee
in the inertial coordinates $(t,x)$, see Fig.\ref{Fig:RR}.

\begin{figure}[h!]
$$\,$$

 $$\,$$ 
  \centering
    \includegraphics[scale=0.25]{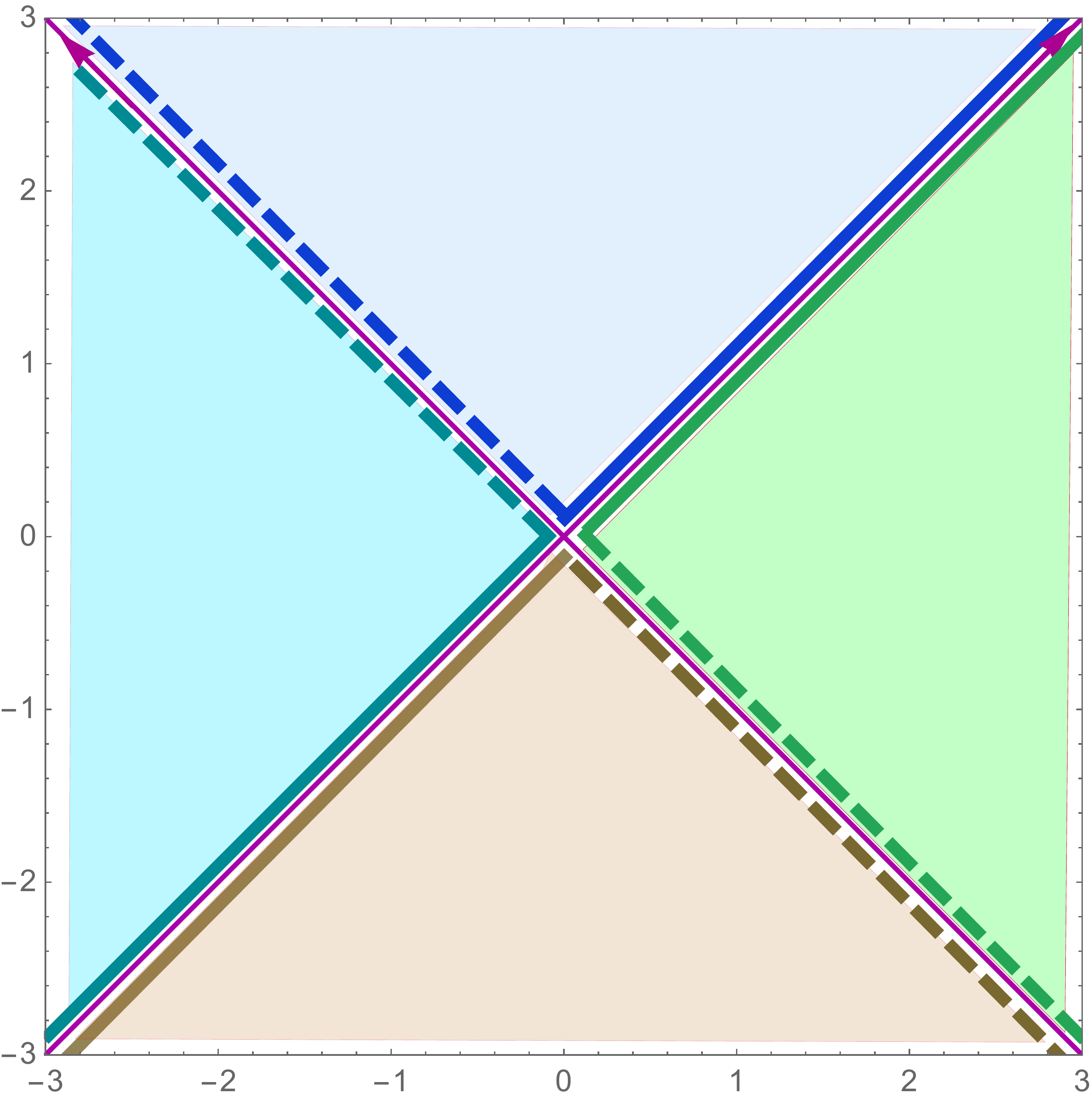}  
     \begin{picture}(50,150)\put(-35,165){$ v $}  
     \put(-155,165){$u$}
        \put(-45,235){$\small{\xi}$}   \put(-85,285){$\small{\eta}$}
      \put(-140,195)  { \includegraphics[scale=0.13]{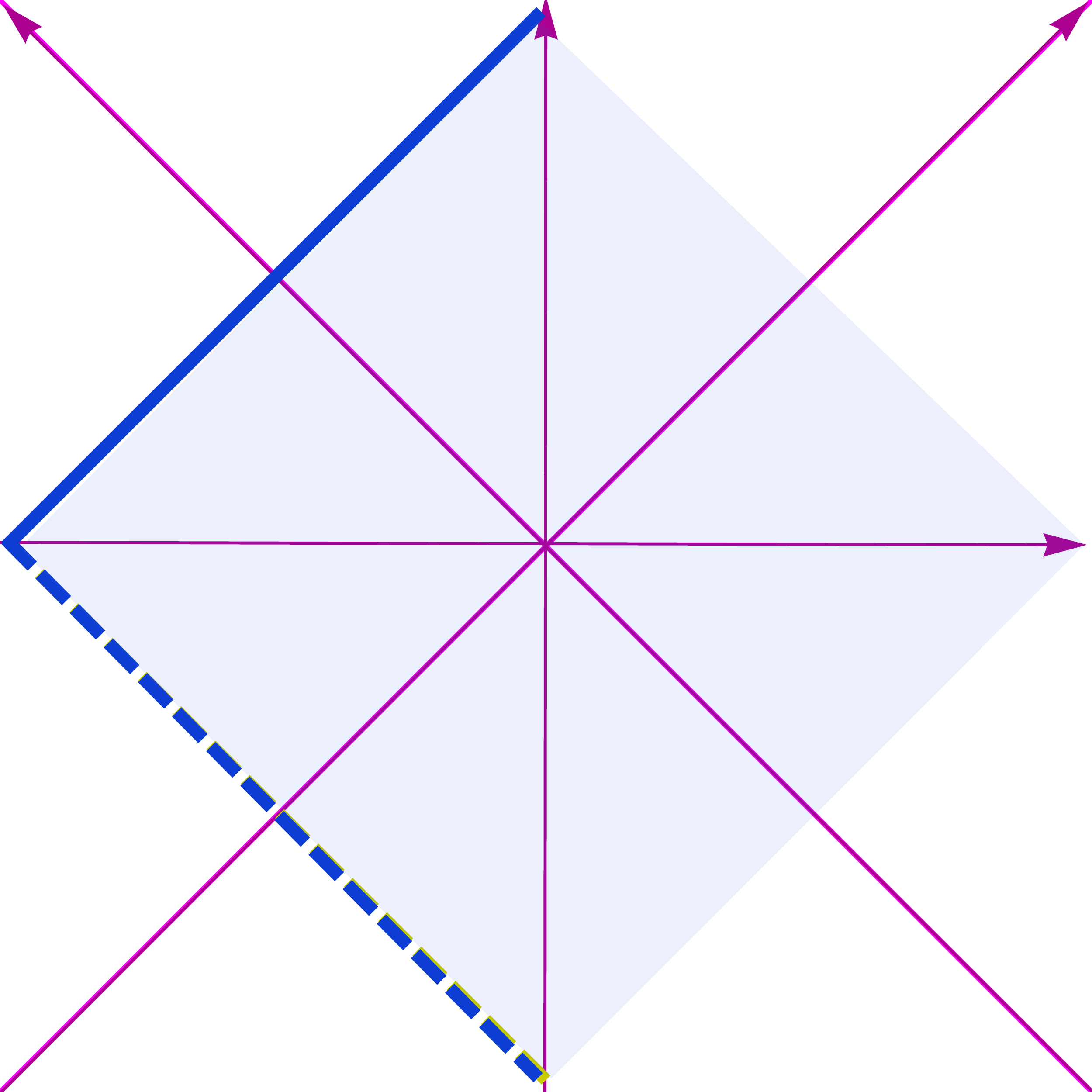} }
           \put(-140,-120)  { \includegraphics[scale=0.13]{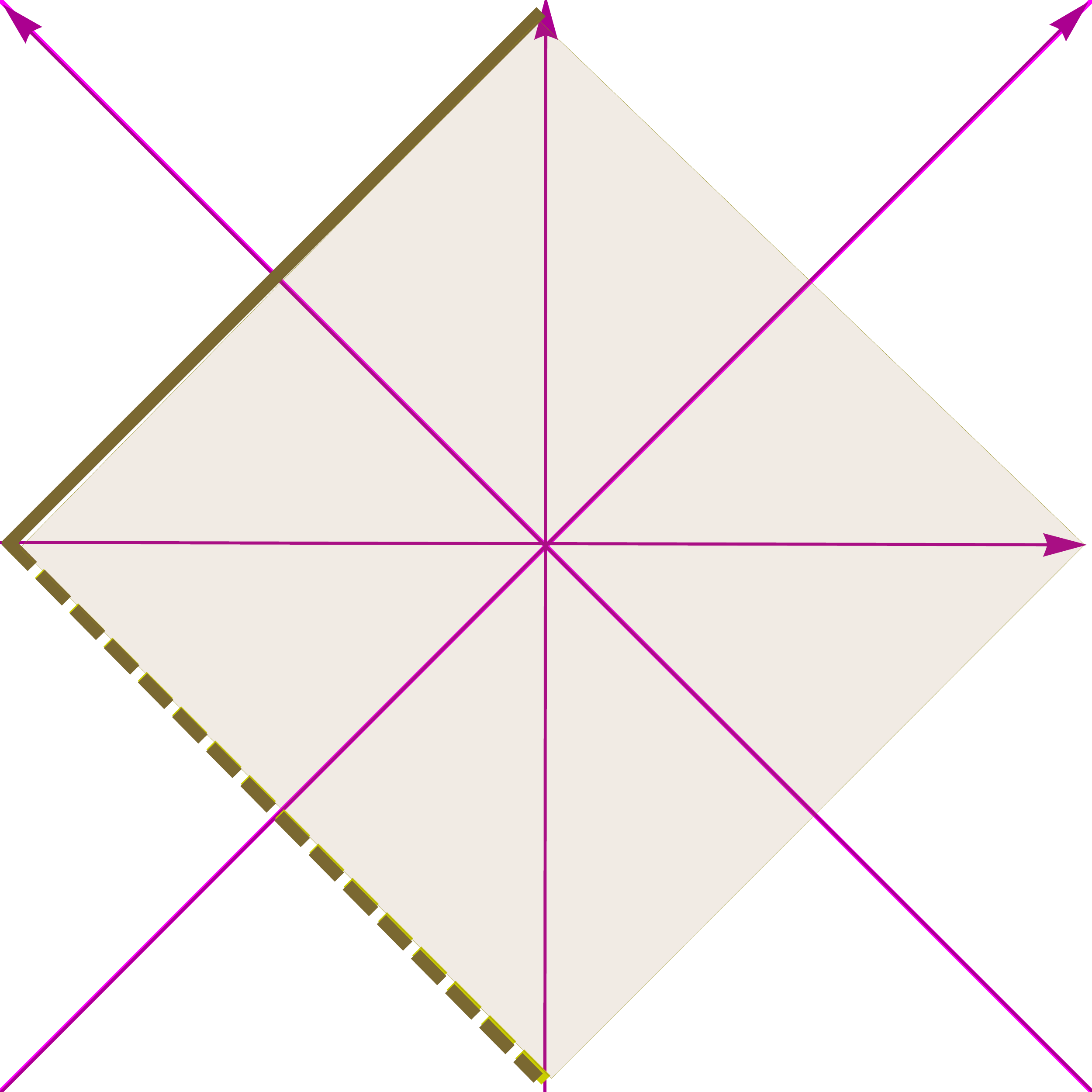} }
            \put(190,105){$\small{\xi}$}   \put(140,175){$\small{\eta}$}\put(185,175){$\small{\vartheta}$}
            \put(75,175){$\small{\nu}$}   \put(60,135){$\small{\nu=\infty}$}
             \put(90,50){$\small{\vartheta=-\infty}$}

   \put(-110,185){$\downarrow\qquad u =e^{-\nu}  \quad v=e^{ \vartheta}$}     
   \put(-120,-15){$ \uparrow\qquad
   u =-e^{-\nu},\quad v=-e^{  \vartheta}$}	
        \put(-310,55)  { \includegraphics[scale=0.15]{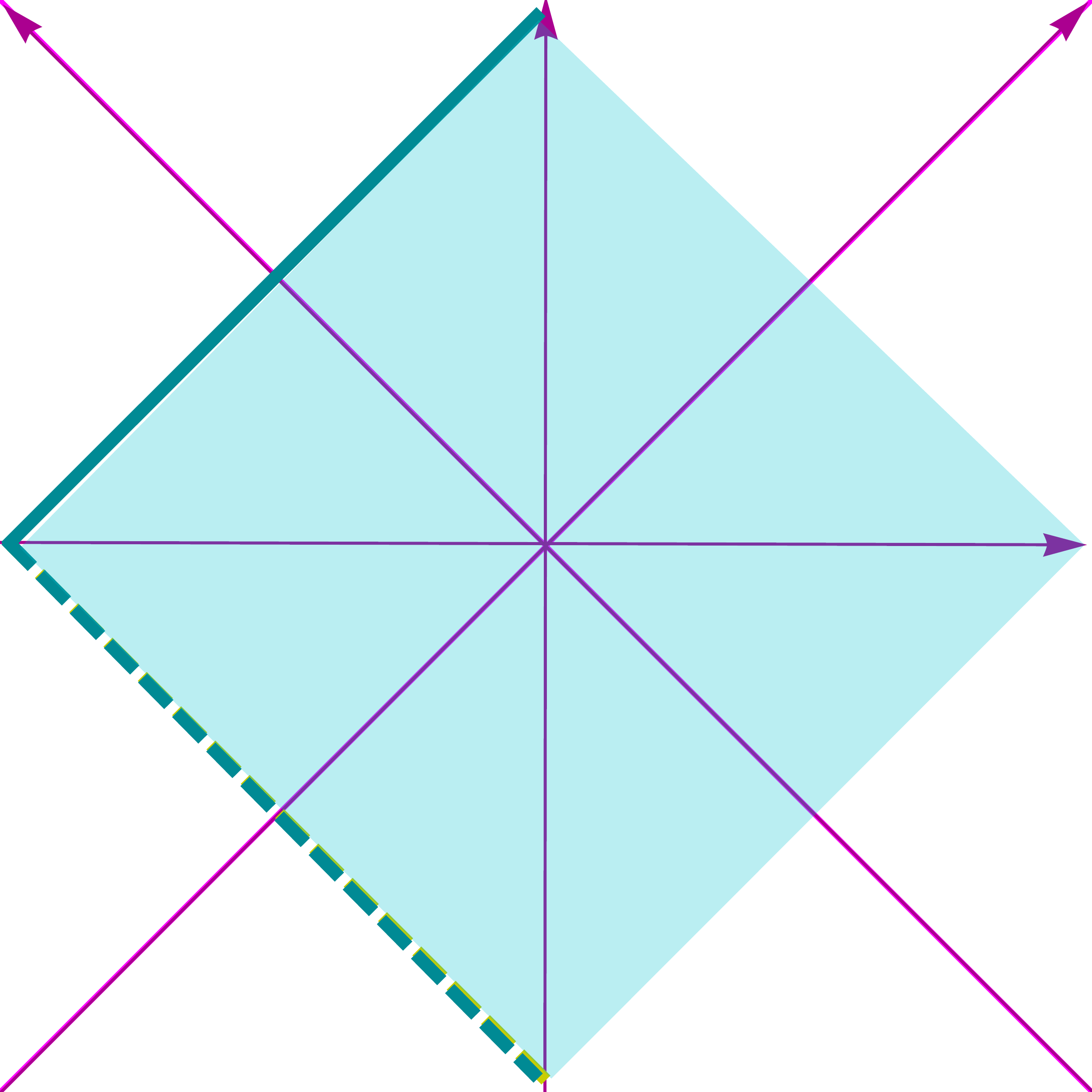}} 
     \put(65,50)  { \includegraphics[scale=0.18]{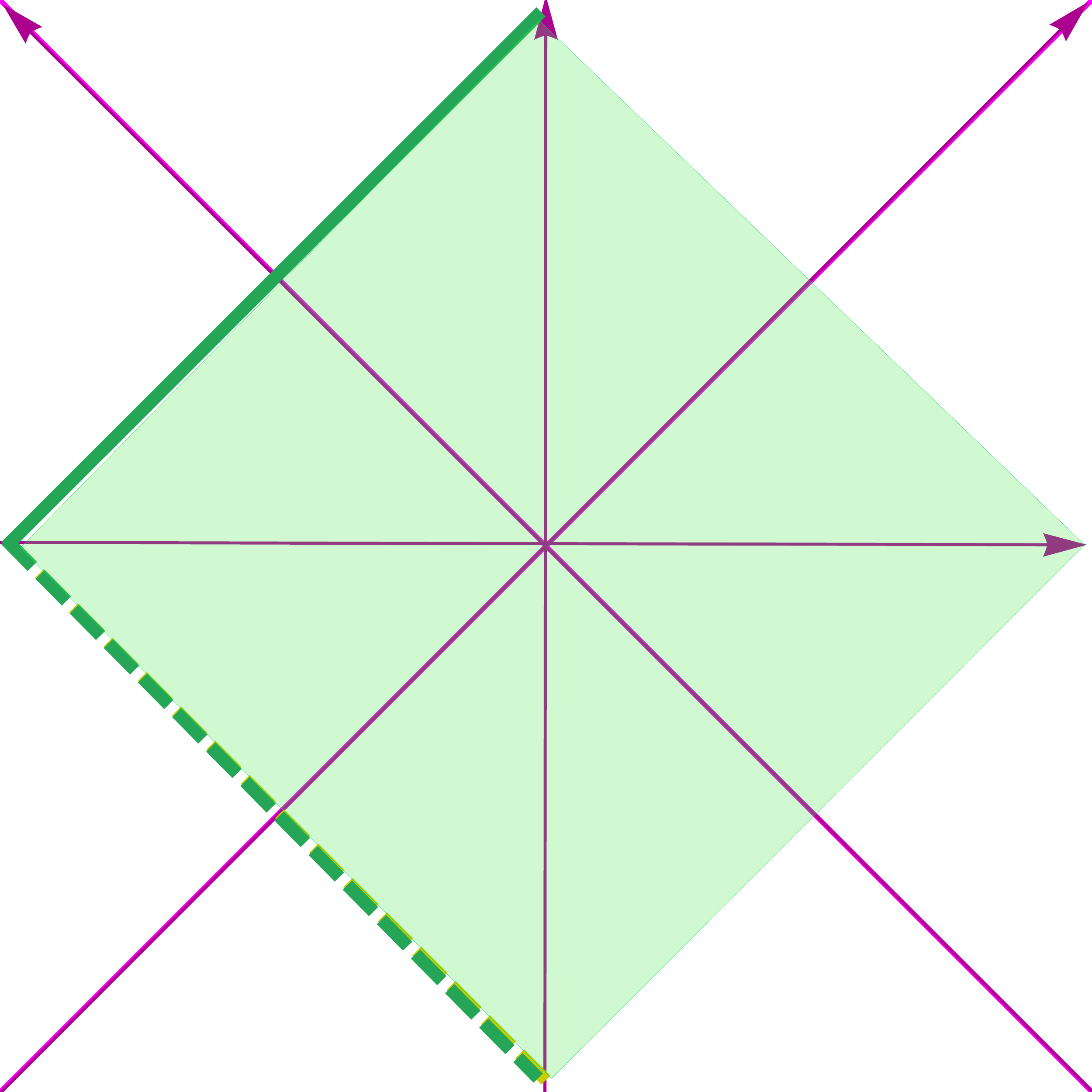} }
\put(-200,115){$\rightarrow$}
\put(-235,95){{\bf L}$:u =e^{-\nu}$}
\put(-218,75){$  v=-e^{ \vartheta}$}
\put(20,115){$ \leftarrow$}
\put(5,95){{\bf R}$:u =-e^{-\nu}$}
\put(24,75){$  v=e^{  \vartheta}$}
\end{picture}
 $$\,$$
  $$\,$$
   $$\,$$
    $$\,$$
  \caption{Map of 4 copies of ${\mathbb M}^{1,1}$  to one  $\mathbb {M}^{1,1}$. Here $a=1$}
 \label{Fig:Rindler}
\end{figure}
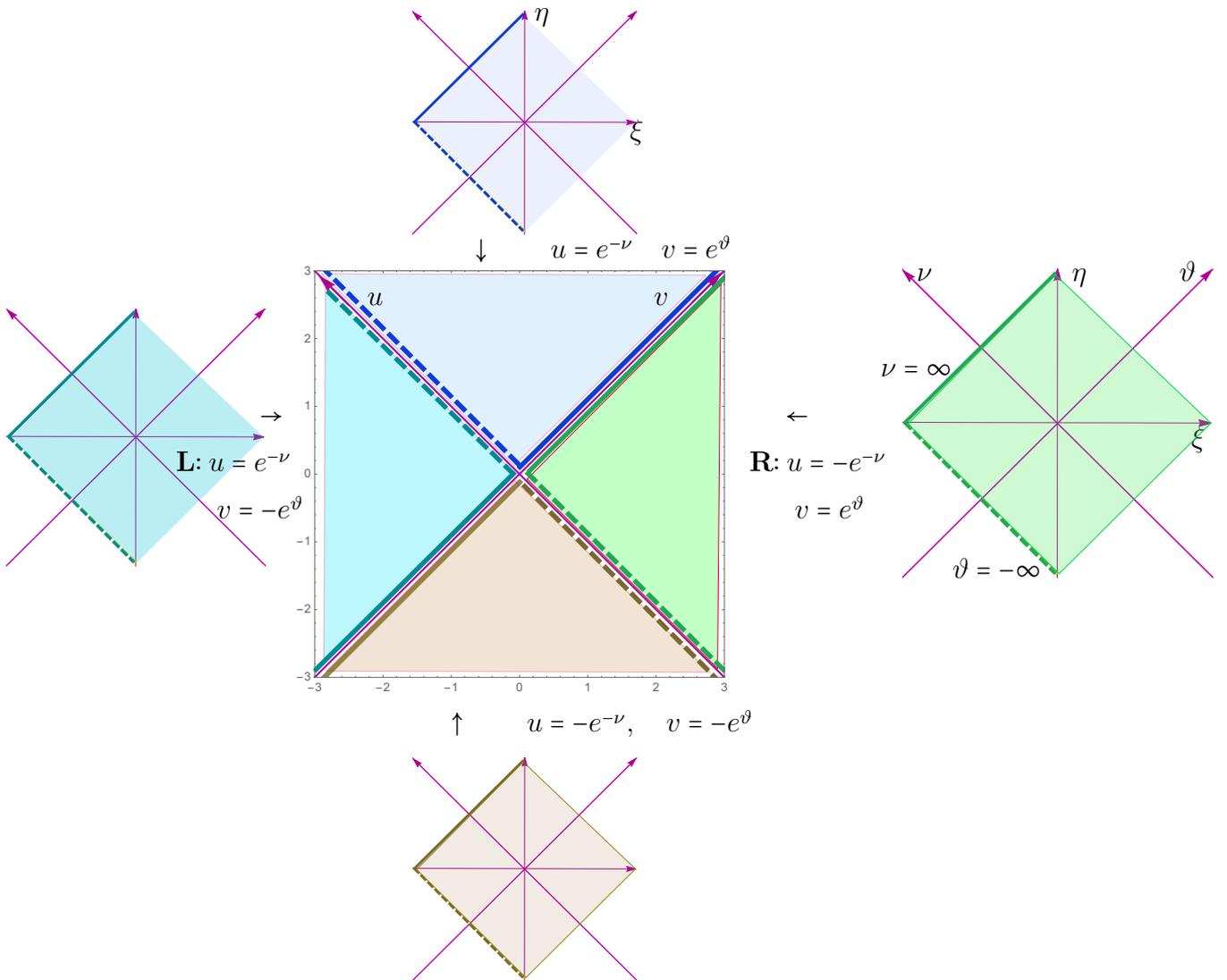

The transformation formulas between the inertial coordinates $(t,x)$ and the  Rindler coordinates $(\xi,\eta)$ (for simplicity we consider 2-dimensional case)
are different in four parts of the inertial coordinates plane $\MM^{1,1}$. Four different maps of $\MM^{1,1}$ to four different parts  of inertial plane  \cite{Rindler:1966,BD}. Below we present them in light-cone coordinates $u,v$ and $\nu, \vartheta$:
 \bea
u&=&t-x,\qquad v=t+x,\label{uv}
\\
 \nu &=&\eta-\xi,\qquad \vartheta=\eta+\xi,\label{omegaupsilon}\eea
 and
  \bea
 {\mbox {\bf R}}:\qquad u&=&-\frac {1}{ a }e^{- a \nu},\qquad \qquad v=\frac {1}{ a }e^{ a \vartheta},\label{UupsilonR}
\\ 
  {\mbox {\bf F}}:\qquad u &=&\frac {1}{ a }e^{-  a \nu },\qquad \qquad v =\frac {1}{ a }e^{  a \vartheta },\label{UupsilonF}\\
 {\mbox {\bf L}}:\qquad u &=&\frac {1}{ a }e^{- a \nu },\qquad \qquad v =-\frac {1}{ a }e^{  a \vartheta },
 \label{upsilonUL}\\
 {\mbox {\bf P}}:\qquad u &=&-\frac {1}{ a }e^{-  a \nu },\qquad \qquad  v =-\frac {1}{ a }e^{ a \vartheta }.\label{UupsilonP}
 \eea
 In all cases $ \vartheta\in(-\infty,+\infty),\quad\nu\in (-\infty,+\infty)$, see  Fig.\ref{Fig:Rindler}.


 As it is well known,  the velocity and acceleration  along the trajectory $\xi=\xi_0$ in the inertial coordinates 
\bea
 t&=&\frac{e^{a\xi_0}}{a} \sinh (a\eta),\qquad
 x=\frac{e^{a\xi_0}}{a}( \cosh (a\eta)-1)\eea 
 are 
\bea
u^0&=&\frac{dt}{d{ s}}=
\cosh(a\eta),\qquad 
u^1=\frac{dx}{d{ s}}=
 \sinh(a\tau), \quad \mbox{since} \quad ds=e^{a\xi_0}\,d\eta\\
 w^0&=&\frac{du^0}{d{s}}=
 a\sinh(a\tau)e^{-a\xi_0},\quad
  w^1=\frac{du^1}{d s}= a\cosh(a\tau)e^{-a\xi_0}\nn\eea
and  we get, 
that the velocity  squared  is equal to 1 and  the acceleration squared is equal to
$a^2e^{-2a\xi_0}$.

   We see that the formulas for maps \eqref{UupsilonR}-\eqref{UupsilonP} are the same as in the E-case,  but there is the essential difference  in the forms of the corresponding metrics.
 In Rindler case the metrics in all 4 spaces with $(\xi,\eta)$ coordinates  are non trivial:
 \bea
ds_{Rindler}^{2}&=&\pm \alpha^2 e^{a(\vartheta-\nu)} d\vartheta\, d\nu\eea 
 with $"-"$ for $(u,v)\in$ {\bf R} or {\bf L}, and $"+"$ for   $(u,v)\in $ {\bf F} or {\bf P}.
 
  In the case of E-thermal, the metrics in all 4 spaces with coordinates $(u,v)$  are trivial, as the first formula in \eqref{dsM-QK}, but the metric in 2-dimensional space with coordinates $( \cT, \cX)$
 is nontrivial and given by \eqref{ds4}, or in details by \eqref{bUV} and \eqref{FUV}.
Therefore, the maps between the pairs of coordinates $( \cU, \cV)\leftrightarrow(u,v)$ and 
 $(\Um, \Vm)\leftrightarrow( \nu, \vartheta)$  are given by the same formulas, but the metrics are different.

 $$\,$$ 
  $$\,$$
$$\,$$

 $$\,$$ 
 
\newpage
\section{General E-coordinates}\label{Gen-QK}
\subsection{E-coordinates}\label{ThermalQK}
 We define here an extension of E-coordinates to the arbitrary static metric of the form 
 \be
ds^2=-f(r)dt^2 +f(r)^{-1}dr^2+r^2d\Omega^2\label{dsf}
\ee
Define also
\be
r_*=r_*(r)=\int\frac{dr}{f(r)} \label{rsG}
\ee
and general Eddington-Finkilstein coordinates 
\be
 u=t-r_*,\,\, v=t+r_*.\label{GEF}
\ee
Then the 2-dim part of the metric becomes
\be
ds^2_2=-f(r)d u d v. 
\label{dsvn}
\ee
E-coordinates $\Um ,\Vm $ are defined by the relations
\be
\Um =-e^{- u/B},\,\,\Vm =e^{ v/B},\,\,B>0.\label{UmVm}
\ee
Now  the the 2-dim part of the metric \eqref{dsvn} reads
\be
ds^2_2=-f(r)d u d v=f(r)B^2\frac{d\Um d\Vm }{\Um \Vm },\label{Krnk}
\ee
where $r$ is implicitly defined by the relation
\be
e^{2r_*/B}=-\Um  \Vm. \label{exprs}
\ee

By introducing the coordinates $\Tm =(\Vm +\Um )/2$ and $\Xm =(\Vm -\Um )/2$ the metric can be written in the  form
\be
ds^2_2=f(r)B^2\frac{-d\Tm ^2+d\Xm ^2}{-\Tm ^2+\Xm ^2},\label{dsXn}
\ee
where $\Xm ^2-\Tm ^2>0.$

Note that if $r_*$ and  $r$ are constants, then from \eqref{exprs}
rewritten as
\be
\Xm ^2-\Tm ^2=e^{2r_*/B}=const,\label{hyp}
\ee
It follows that we deal with  the motion on a hyperbola \eqref{hyp}.

In the same way as in Sect.\ref{quasiK-Sch-T},  one can show that in the  E-coordinates introduced above 
for a rather  general function $f(r)$ the E-observer will see the Planck distribution of quanta with  temperature 
\be
T=\frac{1}{2\pi B}.\label{T-B}
\ee
It is interesting that one can get the temperature distribution for the function $f(r)$ that has no zeros, i.e for a metric \eqref{ds-f} that does not describe a black hole.
For instance, take 
\be
f(r)=e^{-\alpha r}, \quad r>0,\quad \alpha>0\ee
in this case 
\be
r_*=\frac1\alpha (e^{-\alpha r}-1), \,\,\,u=t-r_*,\quad  v=t+r_*.\ee
In E-coordinates \eqref{UmVm} one gets  the temperature \eqref{T-B}.

Note that if there is a black hole with horizon at $r_h$, $f(r_h)=0$, $f'(r_h)>0$, the metric for arbitrary $B$ has a coordinate singularity. 
One can  fix $B=B_0$ to avoid this singularity,
\be
B_0=\frac{2}{f'(r_h)}, \qquad \kappa=\frac12 f'(r_h)\Rightarrow B_0=\frac1\kappa,\label{B-f}\ee
where $\kappa$ is the surface gravity for metric \eqref{dsf}.  Indeed, taking into account, that from \eqref{rsG} it follows that as $r \to r_h$
\be
r_*=\frac{\ln (r-r_h)}{f'(r_h)}+...=\frac{\ln (r-r_h)}{2\kappa}+...,\ee
we get that
\be
\Um \Vm \sim e^{2r_*/B}\sim e^{\frac{\ln (r-r_h)}{B\kappa}}\sim (r-r_h)^{\frac{1}{B\kappa}}\ee
To compensate the first order zero coming from $f(r)$ near $r_h$ we take $B$ as in \eqref{B-f}.
We denote
\be
\Um \Big|_{B=B_0}=U,\qquad \Vm \Big|_{B=B_0}=V.\label{UmVm-kappa}\ee
    Note that these formula reproduce usual formula, in particular, for the Schwarzschild metric \eqref{Scw}.

The coordinates with 
\be B=B_0+b, \quad B_0=\frac{1}{\kappa},
\label{GQK}\ee
we call the  general  E-coordinates.

\subsection{Acceleration of the E-observer in   black holes}\label{Sect:QK-accel}

Now we can fix $\Xm=\Xm_0$.  This trajectory can be parametrized by $\Tm $,
i.e. in the Schwarzschild coordinates 
\bea
t=t(\Tm),\qquad r=r(\Tm),\eea 
\bea
t=t(\Tm)&=&\frac{B}{2}\log (\frac{\Tm + \Xm_0}{\Xm_0-\Tm}),\label{t-T}\\
r_*=r_*(\Tm)&=&\frac{B}2 \log(\Xm_0 ^2-\Tm ^2).\label{rs-T}\eea
Due to \eqref{dsXn}
 along the trajectory we have
\bea
\frac{d\Tm }{ds}=\frac{\sqrt{\Xm_0 ^2-\Tm ^2}}{B\sqrt{f(r)}}\label{dsXn-m}
\eea
Therefore    the velocity components for the observer moving along this trajectory are 
\bea
 u^0&=&\frac{dt}{ds}=\frac{\Xm_0}{\sqrt{f(r)} \sqrt{\Xm_0^2-\Tm^2}},\label{v0}\\
u^1&=&\frac{dr}{ds}=-\frac{\sqrt{f(r)} \Tm}{\sqrt{\Xm_0^2-\Tm^2}}.\label{v1}\eea
We have
\bea
-f u^0 u^0+f^{-1}u^1 u^1=-1\eea
 The components of the  moving   observer acceleration  are defined as
 \bea
 &w^0&=\frac{du^0}{ds}+\Gamma^0_{\mu\nu}u^\mu u^\nu
 =\frac{du^0}{dr}\frac{dr}{ds}+\frac{du^0}{d\Tm}\frac{d\Tm}{ds}+\frac{f'}{f}u^0 u^1
  \\
&=&
\Big(  -\frac12 f' + \frac{1}{B}\Big)
 \frac{1}{f} \frac{\Tm\Xm_0}{\Xm_0^2-\Tm^2}
\eea
\bea
  w^1&=&\frac{du^1}{ds}+\Gamma^1_{\mu\nu}u^\mu u^\nu=\frac{du^1}{dr}\frac{dr}{ds}+\frac{du^1}{d\Tm}\frac{d\Tm}{ds}+\frac{f'}{2f}(f^2 u^0 u^0-u^1 u^1)\\
&=&\left(\frac{f'}{2}-\frac1B\right)\frac {\Xm_0^2}{(\Xm_0^2-\Tm^2)}
\eea
The square of the acceleration is
\bea
 w^2&=&-f(w^0)^2+f^{-1}(w^1)^2=\frac1f\left(\frac{f'}{2}-\frac1B\right)^2\cdot\frac {\Xm^2_0}{\Xm_0^2-\Tm^2}\label{QK-acc}
 \eea
 One can check that 
$
-fu^0\cdot w^0 +f^{-1}u^1\cdot w^1=0$.

\subsection{Examples. Black holes in E-coordinates}
\subsubsection{E-observer in the Schwarzschild black hole } \label{SchQK}
\begin{figure} [h!] \centering
   \includegraphics[scale=0.28]{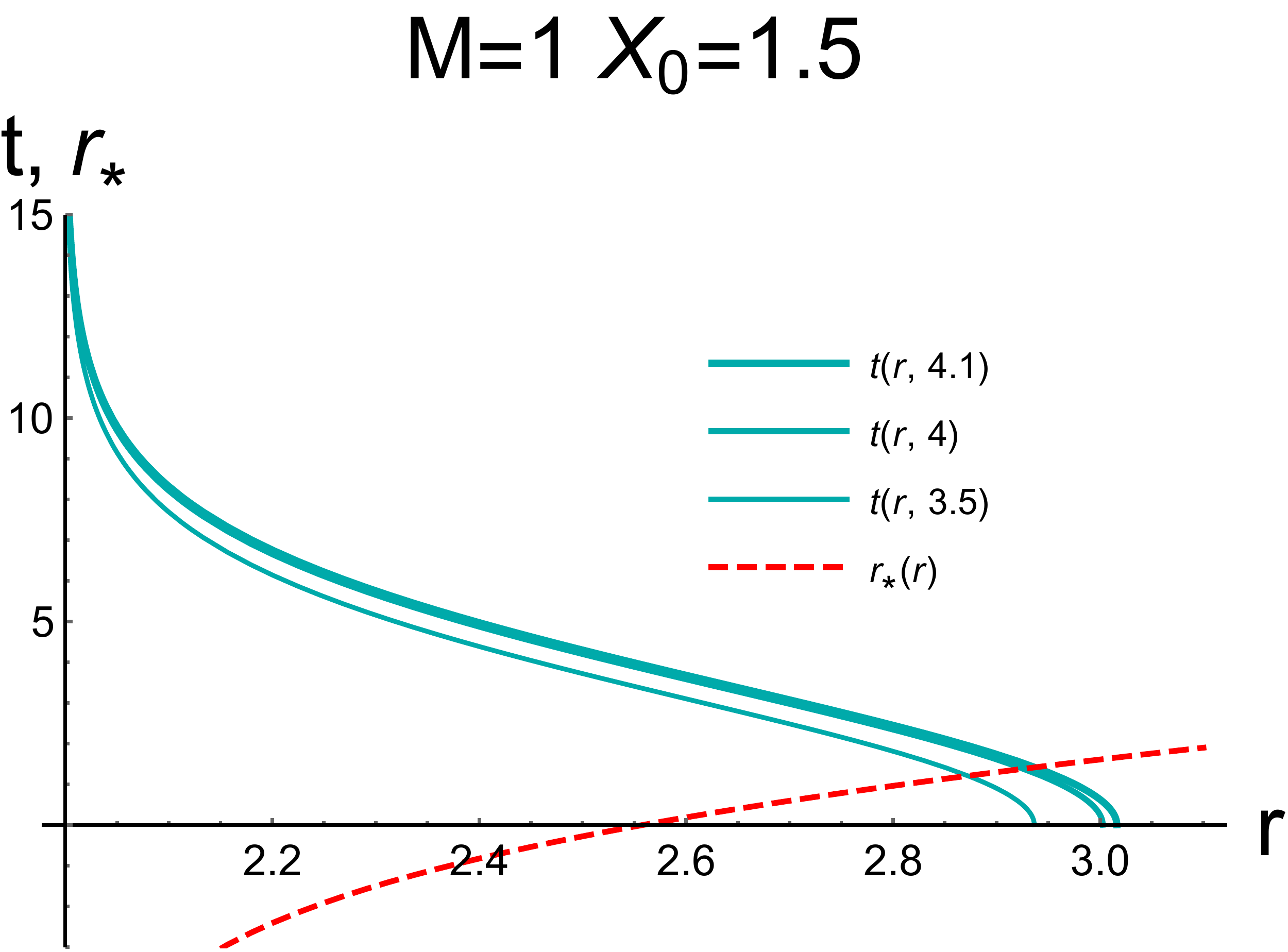}
  \includegraphics[scale=0.18]{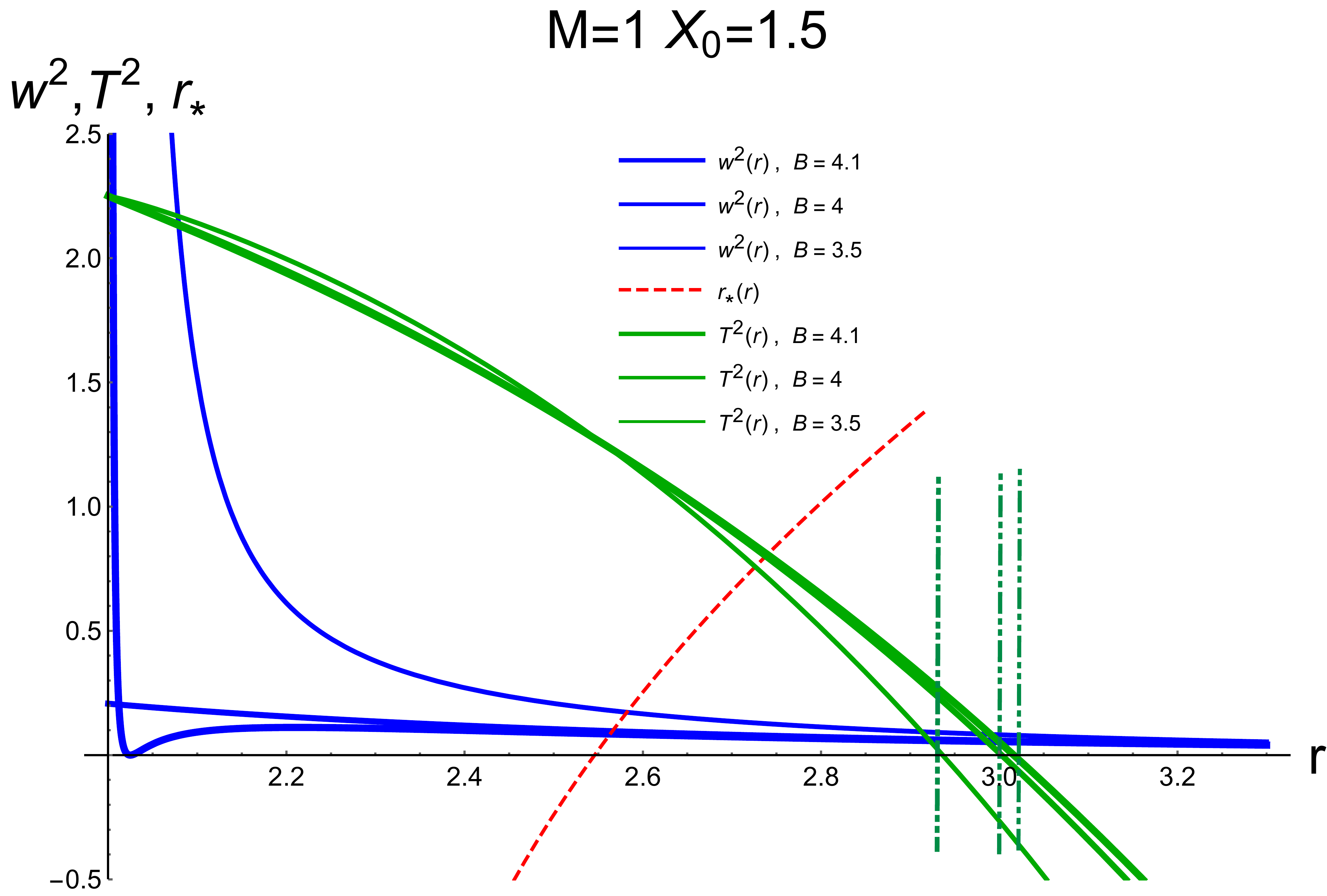}\\A\qquad\qquad\qquad\qquad\qquad B
  \caption{ A) The trajectories of stationary observer at $(\Tm,\Xm)$ coordinates with $\Xm=\Xm_0$ in the the Schwarzschild coordinates for different $B$ and $M=1$, $X_0=1.5$.
  B) The acceleration $ w ^ 2 $ vs $ r $ (blue lines) and $T^2$ vs $r$ (green lines)  for trajectories in shown on A).  
The red dashed line shows $r_*=r_*(r)$ for the same $M$ and $X_0$. The dashed-dotted lines show the restrictions on $r$ following from the requirement $T^2>0$. }
  \label{Fig:SchQK}
\end{figure}

\begin{figure} [h!] \centering
  \includegraphics[scale=0.4]{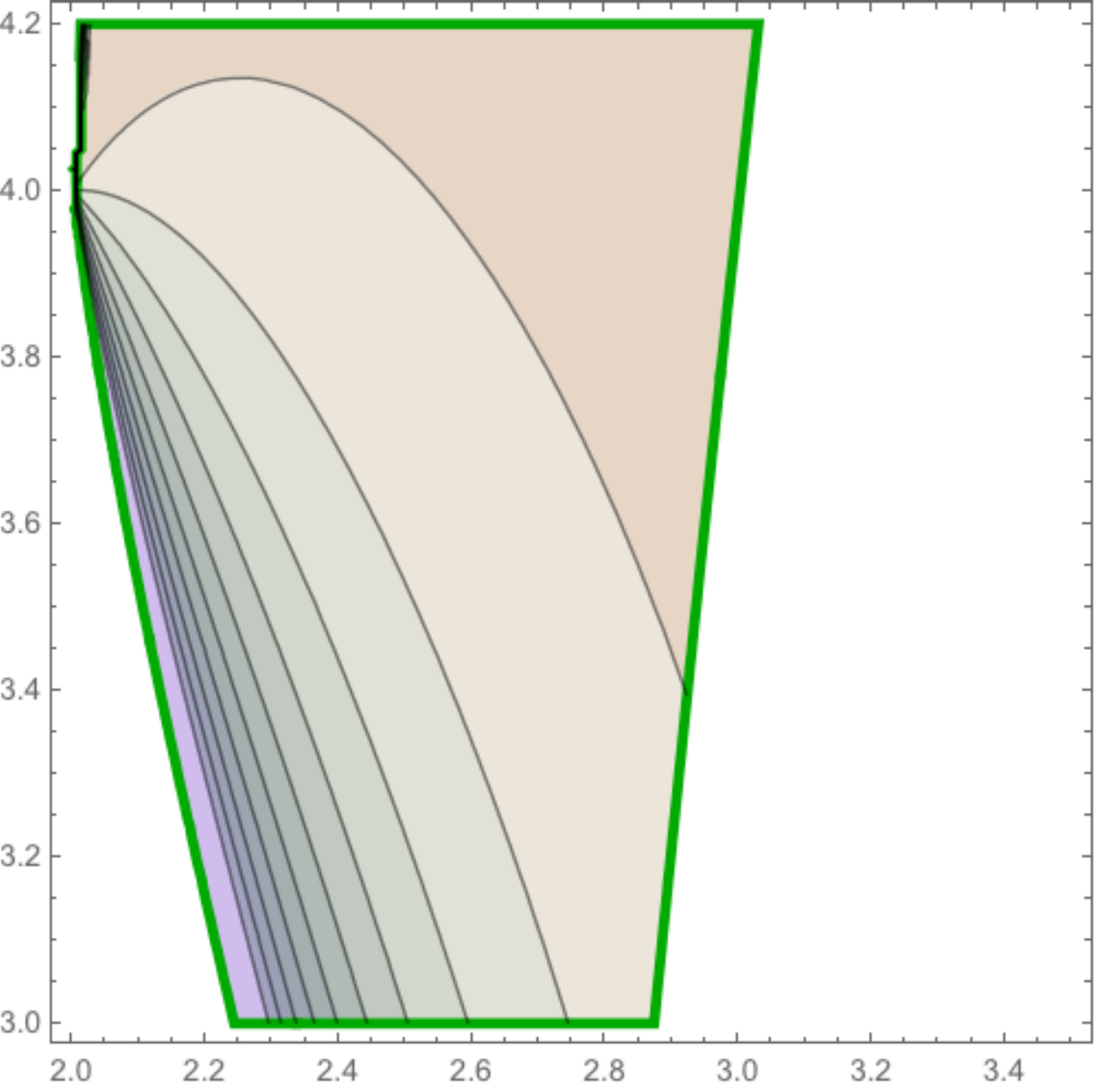}
    \includegraphics[scale=0.2]{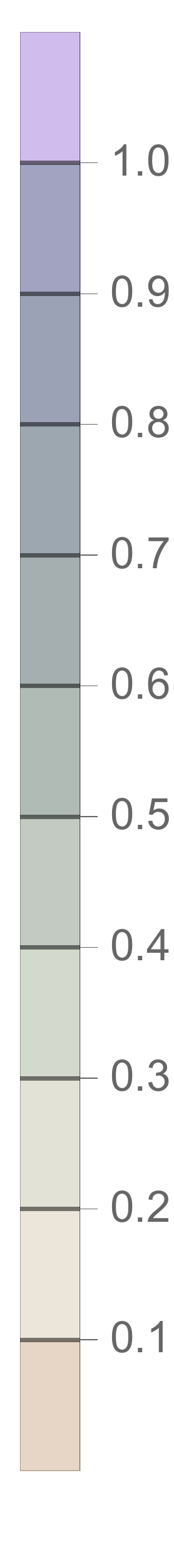}\qquad\qquad
     \includegraphics[scale=0.4]{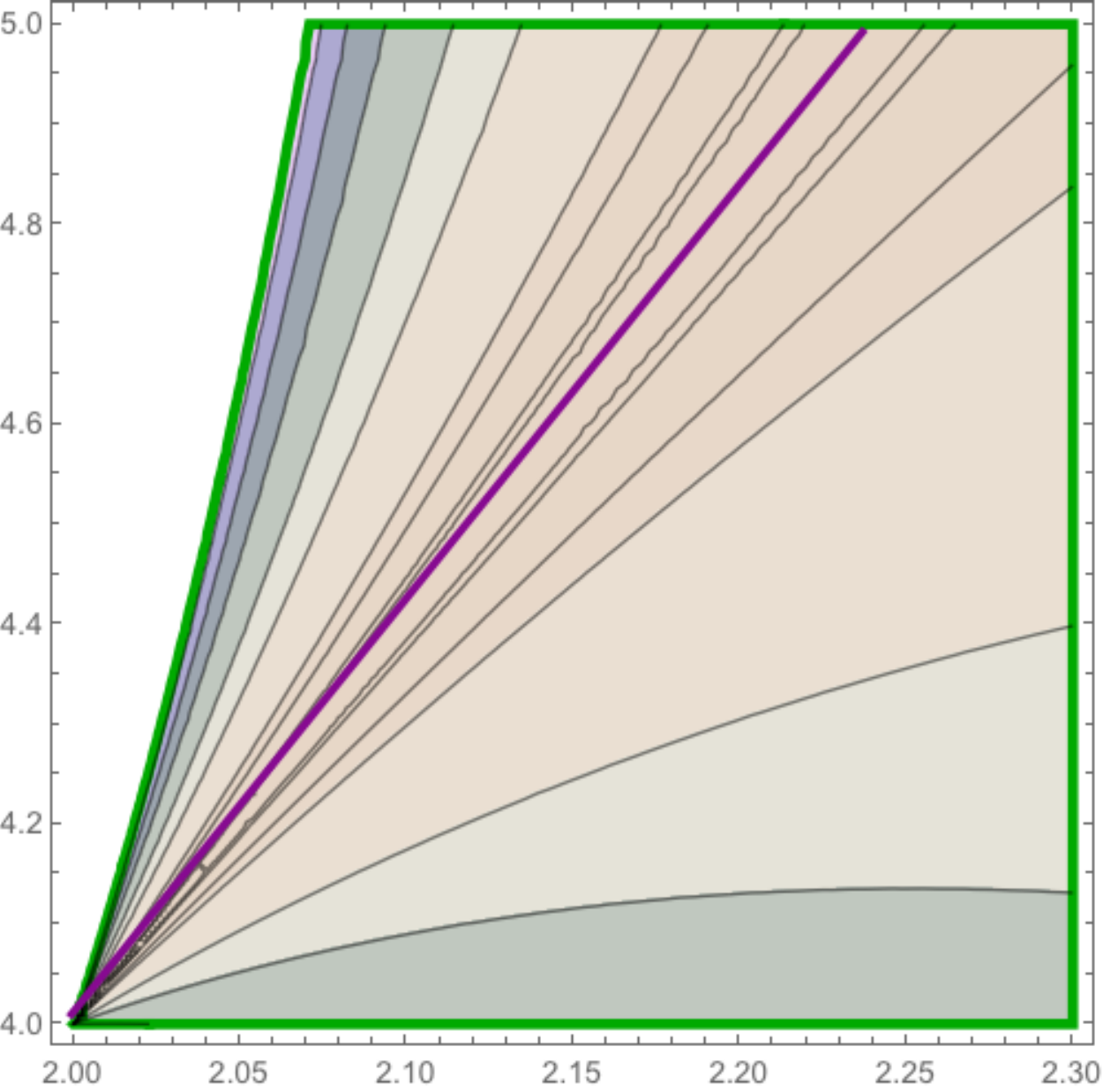}\includegraphics[scale=0.2]{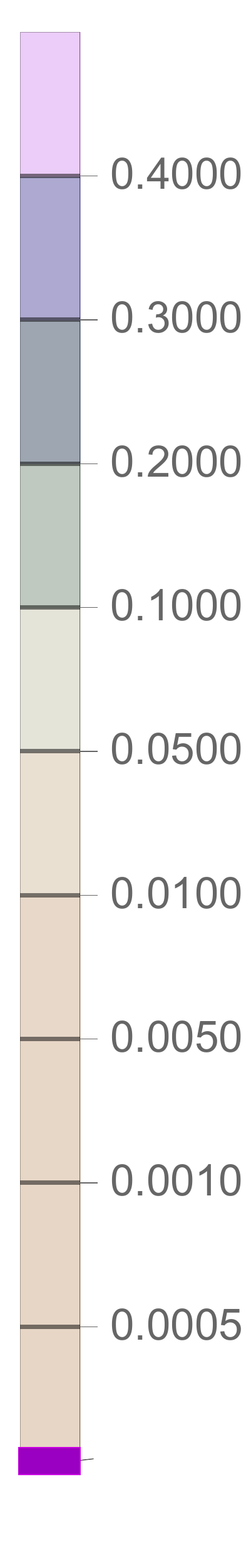}
     \begin{picture}(50,15)\put(35,80){$ B $}  \put(140,10){$ r $}  
     \put(-15,160){$w^2$}
     \put(-175,80){$ B $} 
      \put(-100,10){$ r $}  
     \put(195,160){$w^2$}
     \end{picture} \\
    A\qquad\qquad\qquad \qquad\qquad\qquad B
    \caption{
The contourplot versions 
  of the plot presented in Fig.\ref{Fig:SchQK}.  Here on horizontal axes we show $r$ and on vertical $B$, $M=1$, $X_0=1.5$. The acceptable region with $T^2>0$ is bounded by solid green line.  Behaviour of $w^2$ mainly  for $B<4M$ (A)
  and  for $B>4M$ only (B). Here
  the magenta line shows locations of points with $w^2=0$. These points exist only for $B>4M$.  
  }
  \label{Fig:SchQK-AB}
\end{figure}
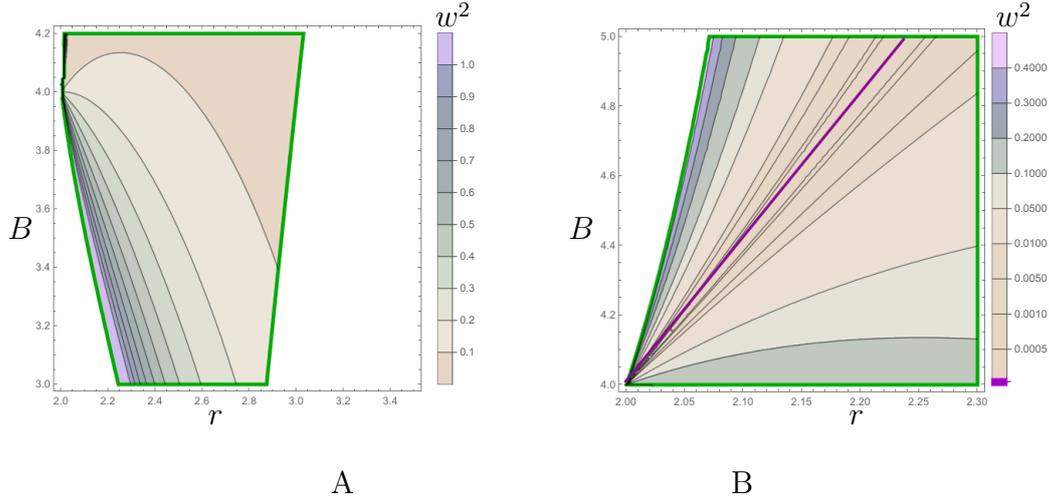

For the Schwarzschild  solution 
\be
f(r)=1-\frac{2M}{r}\label{Sch}
\ee
and 
\be
r_*=r+2 M \log \left(\frac{r}{2 M}-1\right),\ee
and due to \eqref{rs-T}
\bea
r_*=r_*(\Tm)&=&\frac{B}2 \log(\Xm_0 ^2-\Tm ^2)\label{rs-T-m}
\eea
and
\bea
\Tm^2&=&\Xm_0^2-e^{2r/B}
   \left(\frac{r}{2 M}-1\right)^{4M/B}\eea
  For the acceleration we have
   \bea
   w^2=\frac{\Xm_0^2
   \left(\frac{1}{B}-\frac{M}{r^2}\right)^2
   }{1-\frac{2 M}{r}}e^{-2r/B}
   \left(\frac{r}{2 M}-1\right)^{-4M/B}\label{accel-Sch}
\eea
The dependence of acceleration \eqref{accel-Sch} on $r$ for fixed $M=1$ and along the trajectory \eqref{t-T}, \eqref{rs-T}
with $\Xm_0=1.5$ is presented in Fig.\ref{Fig:SchQK}. We see that for $B<4M$ the acceleration is infinite at the horizon
$z_h=2M$. For $B=4M$ the acceleration at the horizon   is related with  the surface gravity $\kappa=1/4M$,
\be
w^2|_{r=r_h}=\frac{4\Xm_0^2}{\e}\kappa^2 \ee where $\kappa=1/4$ and $w^2=0.20693$. For $B>4M$ the  acceleration near horizon is infinite and tends to  zero 
at $r=r_0=\sqrt{BM}$; for $B=3.5$ $r_0=\sqrt{4.1}=2.02485$. The locations of these zeros are shown at the contourplot  in Fig.\ref{Fig:SchQK-AB}.B by the magenta line.

$$\,$$
\newpage
\subsubsection{E-observer in the Reissner-Nordstrom black hole} \label{RNQK}
For the RN solution one has
\be
f(r)=1-\frac{2M}{r}+\frac{Q^2}{r^2},\label{RN}
\ee
\be
r_*=r+\frac{r_+}{r_+-r_-}\ln(r-r_+)-\frac{r_-}{r_+-r_-}\ln(r-r_-),
\ee
where $r_\pm=M\pm\sqrt{M^2-Q^2}$, $M\geq Q$.  $u,v$ are defined as  $u=t-r_*,\,v=t+r_*$, \cite{HE}.

\begin{figure} [h!] \centering
 \includegraphics[scale=0.25]{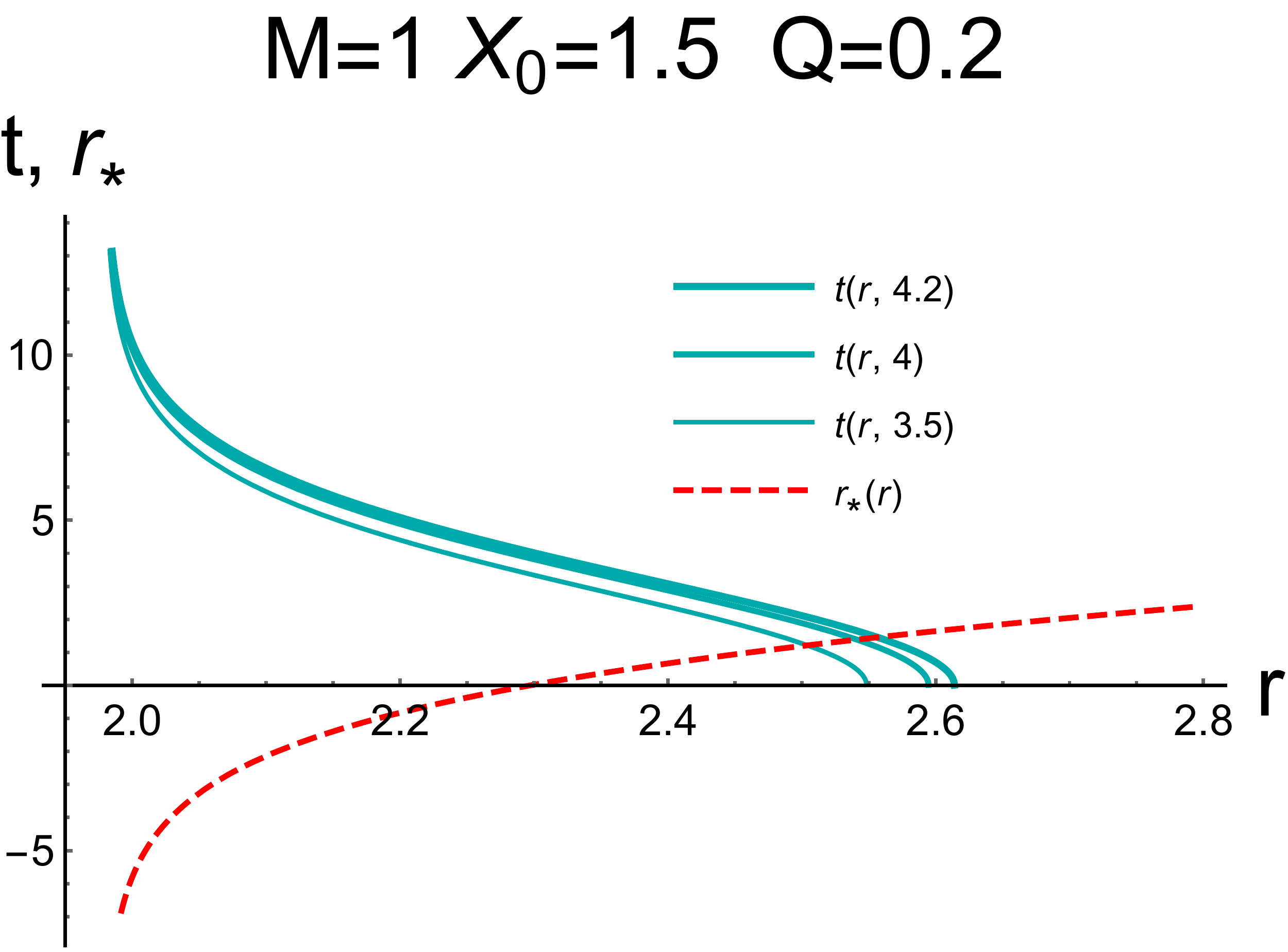} 
  \includegraphics[scale=0.18]{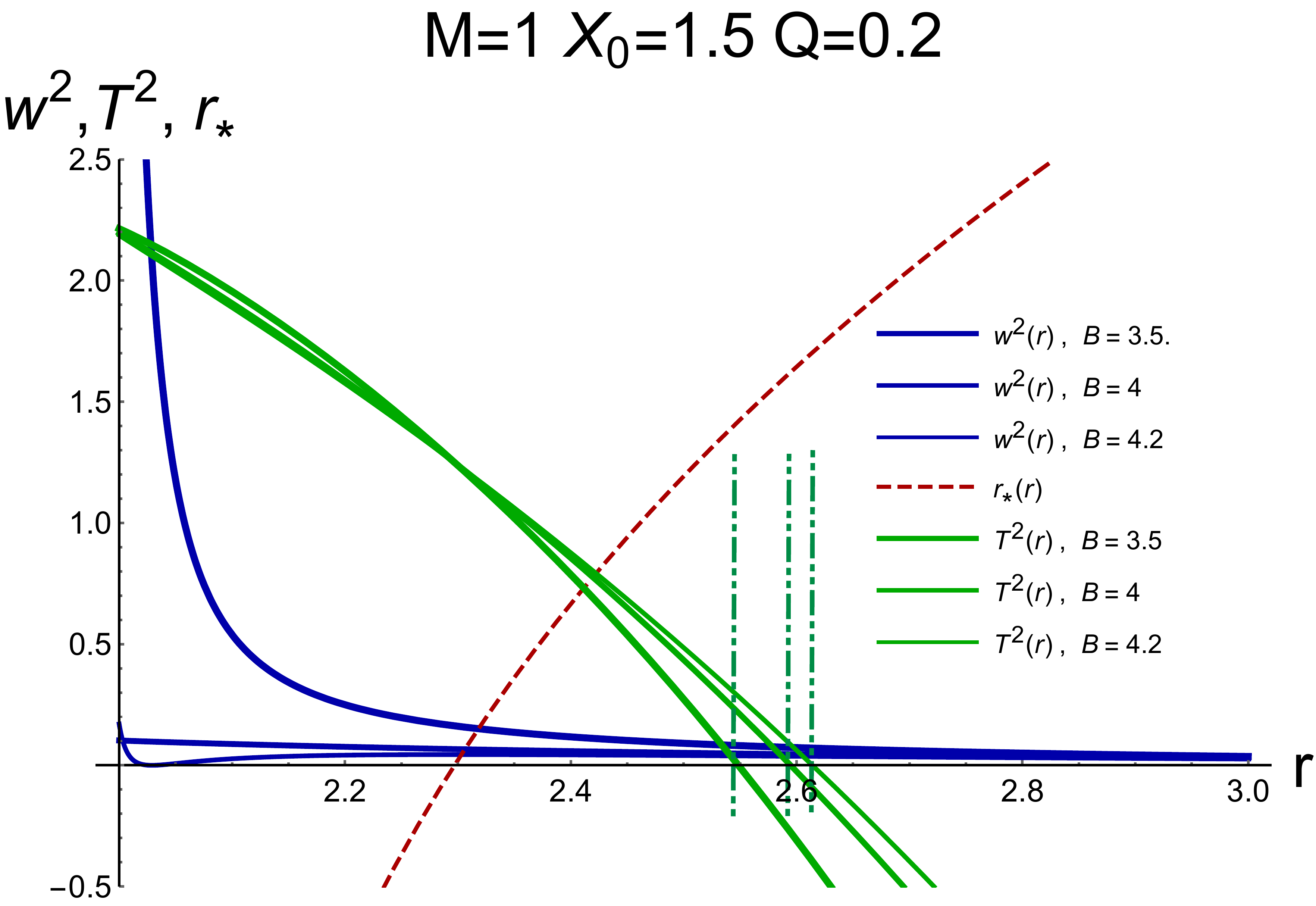} \\A\qquad\qquad\qquad \qquad\qquad\qquad\qquad
  B\\\caption{A)   
    The trajectories  with $\Xm=\Xm_0$ in the   RN spacetime in the Schwarzschild coordinates for different $B$ and $M=1$, $X_0=1.5, Q^2=0.2$.  B)
$w^2$ vs $r$ (blue) and $T^2$ vs $r$ (green)  for trajectories shown in A).
The red dashed line shows $r_*=r_*(r)$ for the same $M,X_0$. The dashed-dotted lines show the restrictions on $r$ following from the requirement $T^2>0$.   
  }
  \label{Fig:RNQK}
\end{figure}

Using the general formula \eqref{QK-acc} we find the dependence of the acceleration on $r$ along the trajectory $\Xm=\Xm_0$ and the results are presented in Fig.\ref{Fig:RNQK}. We see that the qualitative behaviour of acceleration dependence  on $r$ is similar to the  previous case considered in Sect.\ref{SchQK}.
For $B<1/\kappa(M,Q)$,  here $1/\kappa(1,0.2)=4.00042$, the  acceleration near horizon is infinite and 
monotonically decreases to  zero  for  $r\to \infty$.
For $B=1/\kappa(M,Q)$ the acceleration at the horizon $r=r_+$ is finite, here $r_+=1.9798$ and $w^2|_{r_+}=  0.10448$,
 and for $r\to \infty$ decreases to $0$. For $B>1/\kappa(M,Q)$ the acceleration becomes   infinite at the horizon and decreases up to zero at $r=r_0(M,Q)$. 

\newpage

\subsubsection{E-observer in Schwarzschild-AdS}
For the Schwarzschild-AdS solution one has
\be
f(r)=1-\frac{2M}{r}+k^2r^2\label{Sch-AdS}
\ee
and
\bea
r^*(r)=\int_0^r\frac{dr^\prime}{f(r^\prime)}&=&\frac{r_h}{3 k^2 r^2_h+1}\Big[ \log
   \left|1-\frac{r}{r_h}\right|-\frac{1}{2}
    \log \left(1+\frac{k^2r
   \left(r_h+r\right)}{k^2r_h^2+1}\right)\Big]\nn\\
   &+&\frac{\left(3 k^2r_h^2+2\right) 
   }{k\sqrt{3 k^2r_h^2+4}}\arctan\left(\frac{k r \sqrt{3k^2
   r^2_h+4}}{2
   \left(k^2r^2_h+1\right)+k^2r
   r_h}\right)\label{Sch-AdS-rs}
\eea
where
\bea
r_h=\sqrt[3]{\frac{M}{k^2}}\left(\sqrt[3]{1-\sqrt{\frac{1}{27 k^2
   M^2}+1}}+\sqrt[3]{\sqrt{\frac{1}{27 k^2
   M^2}+1}+1}\right) 
   \eea



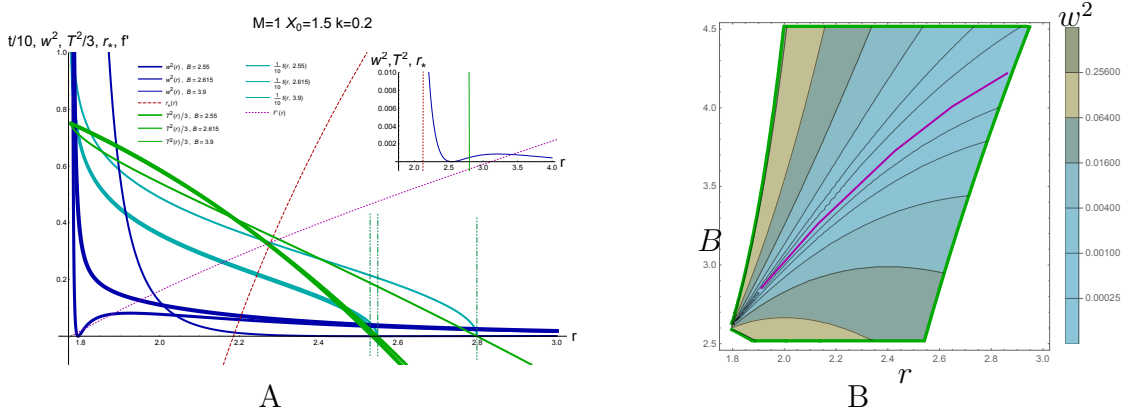
\begin{figure} [h!] \centering
  \includegraphics[scale=0.115]{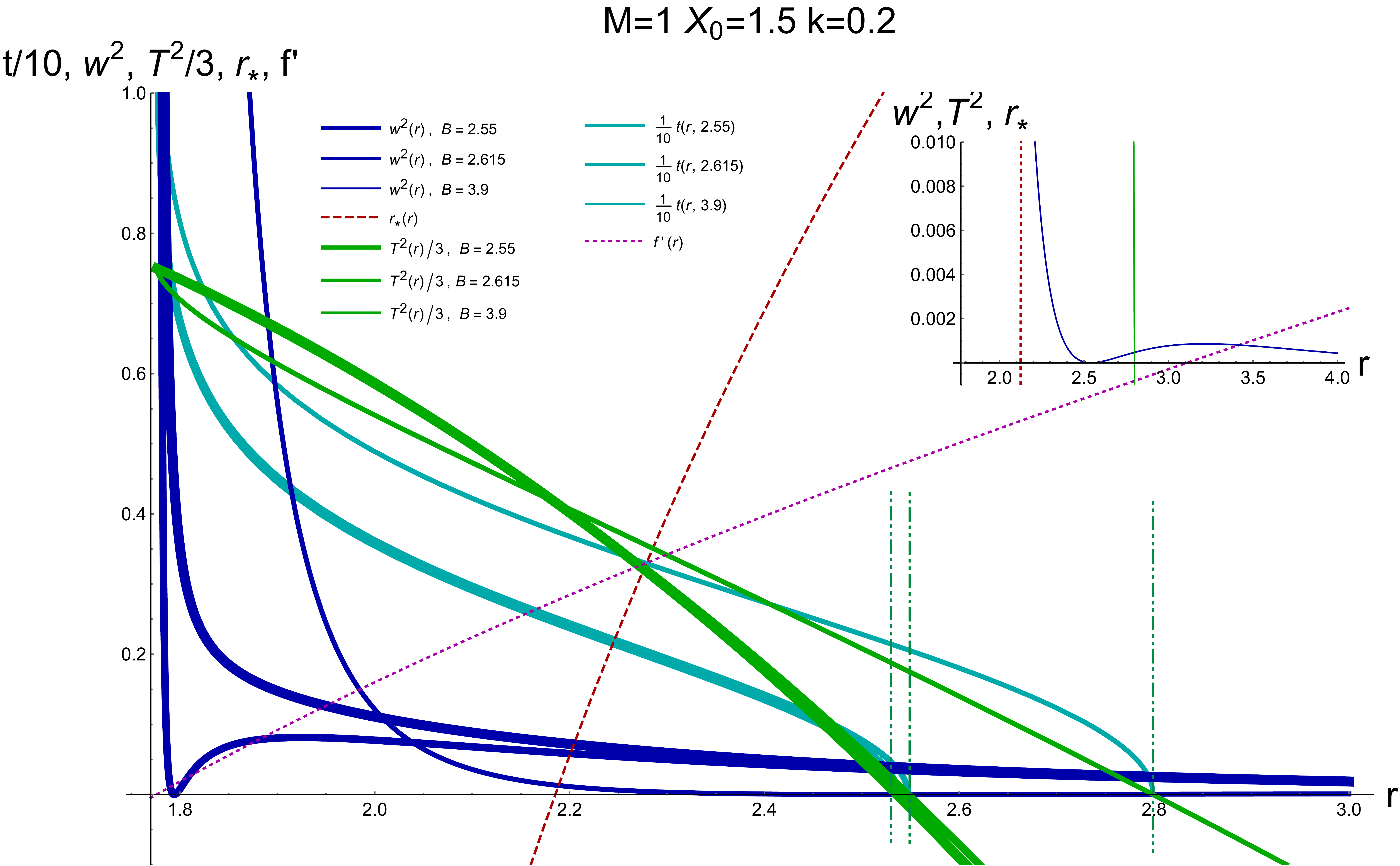}\qquad\qquad
  \includegraphics[scale=0.18]{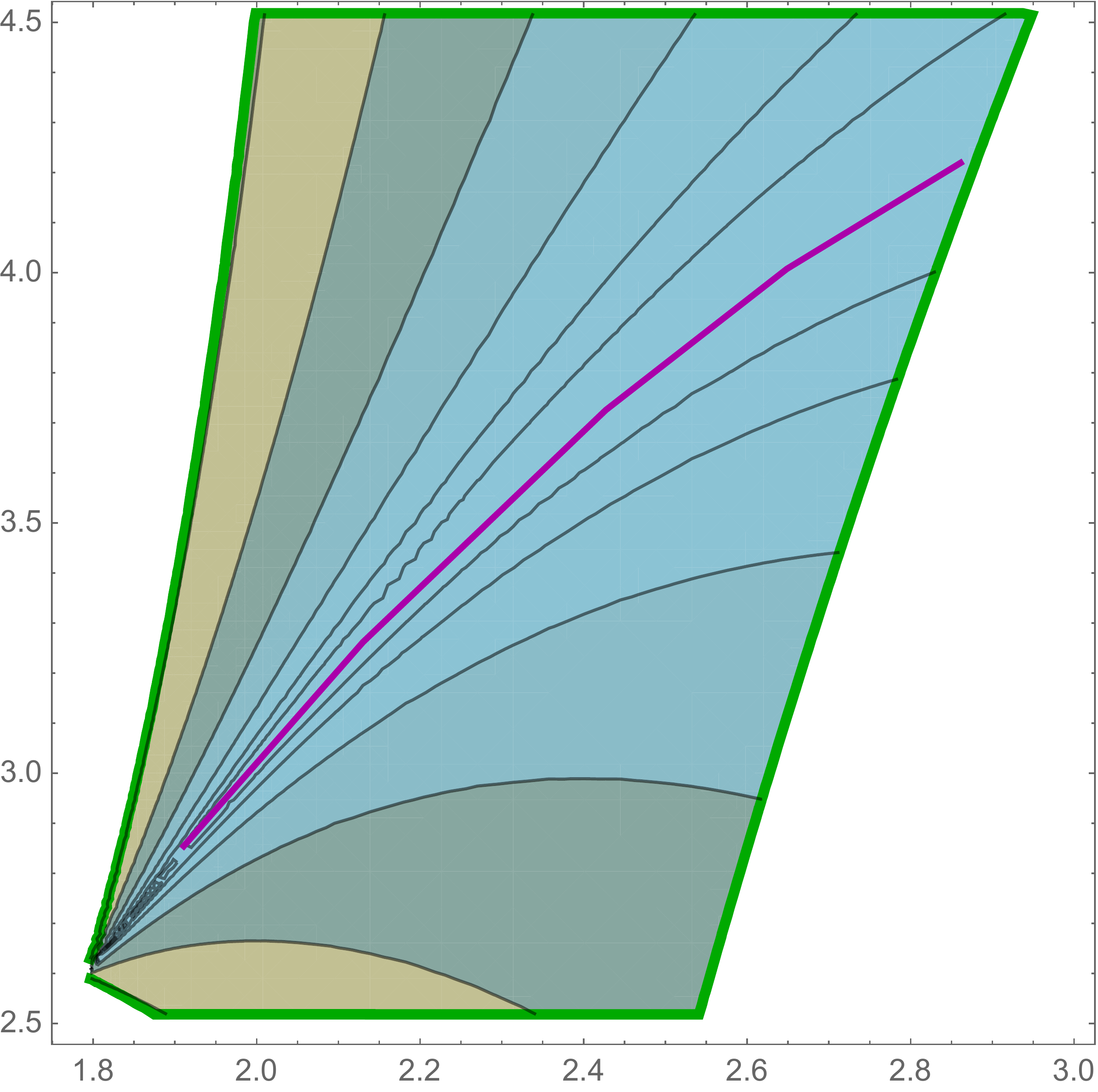}
  \includegraphics[scale=0.18]{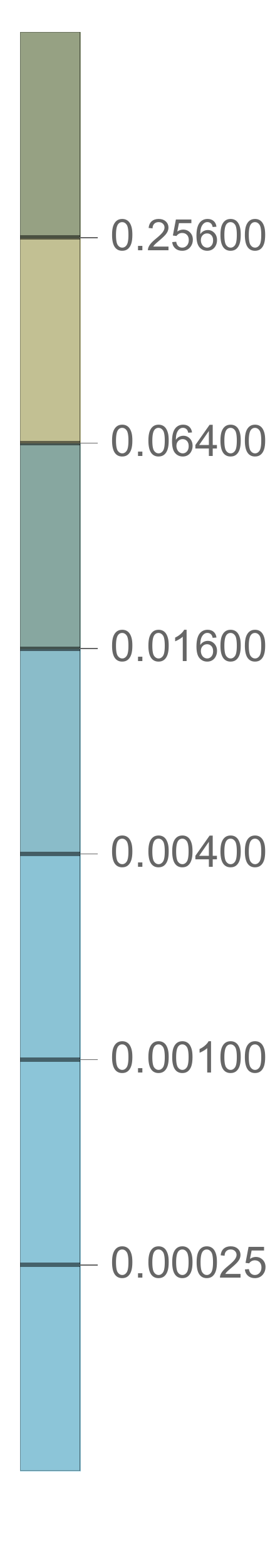}
  \\A\qquad\qquad\qquad\qquad\qquad\qquad\qquad \qquad\qquad B\\          
  \begin{picture}(50,15)\put(75,75){$B$} \put(210,160){$w^2$}
 \put(150,25){$ r $}        \end{picture}
\caption{A) The trajectories  with $X_0=1.5$ in the AdS-Schwarzschild  spacetime in the Schwarzschild coordinates 
  for different $B$ and $M=1$ are shown by darker cyan lines. 
The acceleration $w^2$ along these trajectories are shown by blue lines. $T^2$ for these trajectories   for the same   $M=1$ and $X_0$ are shown by green lines.
The red dashed line shows $r_*=r_*(r)$ for the same $M,X_0$. The dashed-dotted lines show the restrictions on $r$ following from the requirement $T^2>0$. The in-plot  shows the zoom of the original plot for $B=2.615$. B) The contour plot version of A) with $r$ in the horizontal direction and $B$ on the vertical one. The darker magenta line shows the acceleration $w^2$  zeroes locations.  
}
  \label{Fig:QK-AdS}
\end{figure}

Fig.\ref{Fig:QK-AdS} shows the dependence of the acceleration along trajectories with $\Xm_0=1.5$ and different $B$ in the AdS-Sch metric 
with $M=1$  and varying $B$. We see that the acceleration becomes infinite near the horizon,  decreases monotonically for $B<B_c$
(the thick line in Fig.A) and has minimum equal to zero for $B>B_c$ (the line of moderate thickness in Fig.A). For $B>B_r$ this minimal value is reached out of the acceptable region (the thin line in Fig.A). 

\newpage
\subsubsection{E-observer in  Schwarzschild-dS}
For the Schwarzschild-dS solution one has
\be
f(r)=1-\frac{2M}{r}-k^2r^2\label{Sch-dS}
\ee
and for $0<27k^2 M^2<1$ 
there exist
two positive roots  $r_1$ and $r_{2}$ of $f(r)$
such that $ 0<2M<r_1<3M<r_{2}$, 
\bea
r_1=\frac2{k\sqrt3 }\cos(\alpha/3+4\pi/3),\quad r_{2}=\frac2{k\sqrt3 }\cos(\alpha/3)\quad
\mbox{with} \quad \cos\alpha=-3Mk\sqrt3
\eea 
There also is  a negative root 
\bea
r_3=\frac2{k\sqrt3 }\cos(\alpha/3+2\pi/3).
\eea
Here $r_1$ and $r_2$ describe the black-hole event
horizon and the cosmological event horizon, respectively. Now we can write  $r_*$ in the form
\bea
r_*=\int \frac{dr}{f(r)}=
-\frac{1}{k^2}\left(A \ln(r_1-r )+B\ln(r-r_{2})+C\ln(r-r_{3})\right)+D,\label{Sch-dS-rs}\eea 
where \bea
A=\frac{r_1}{(r_1-r_{-})(r_1-r_{2})},\quad
B=\frac{-r_{2}}{(r_{2}-r_{-})(r_{1}-r_{2})},\quad
C=\frac{r_{-}}{(r_{2}-r_{-})(r_{1}-r_{-})}\eea
We adjust $D$ to remove the imaginary part from expression for $r_*$.

\begin{figure} [h!] \centering
\includegraphics[scale=0.25]{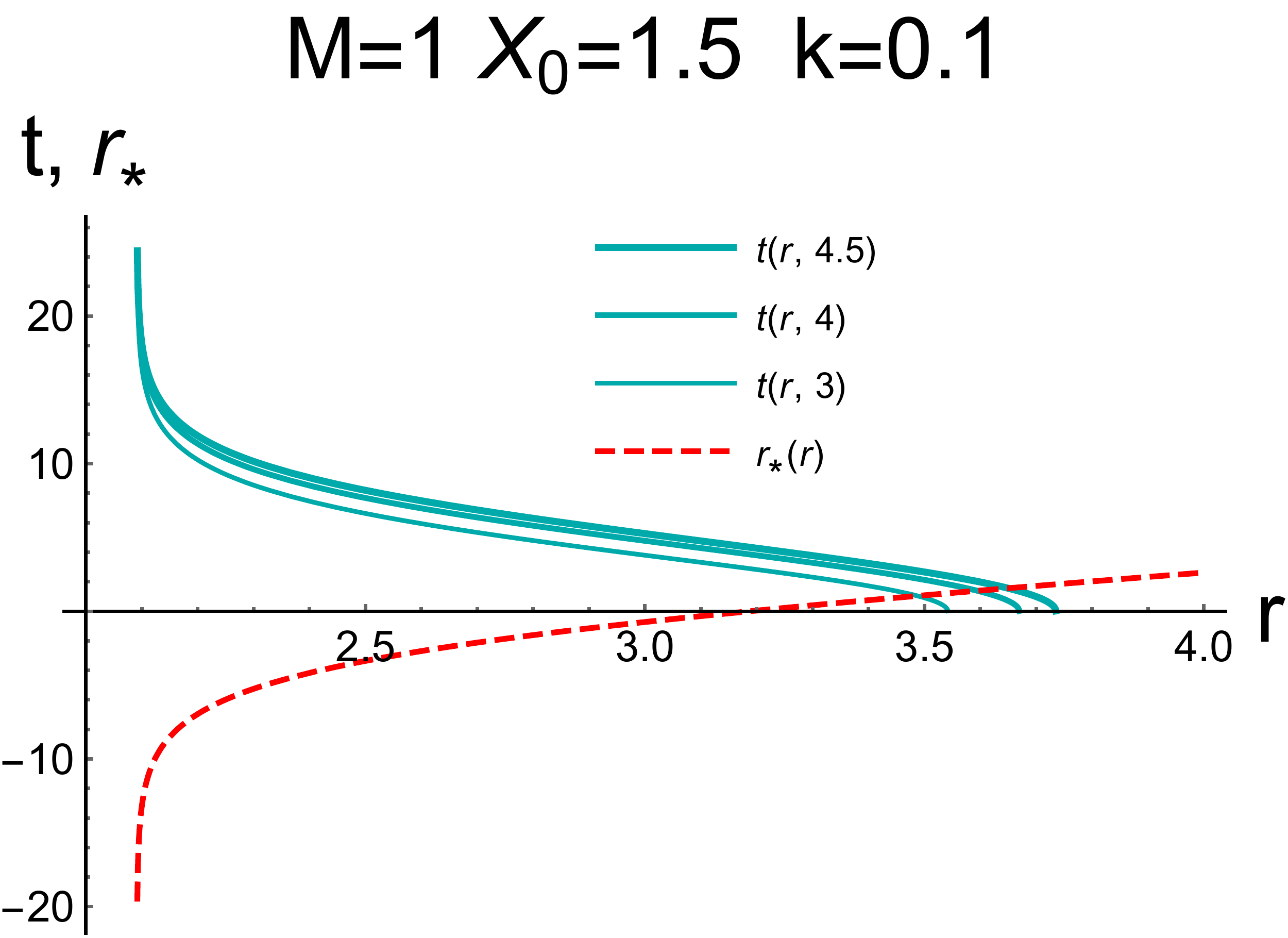}
\includegraphics[scale=0.2]{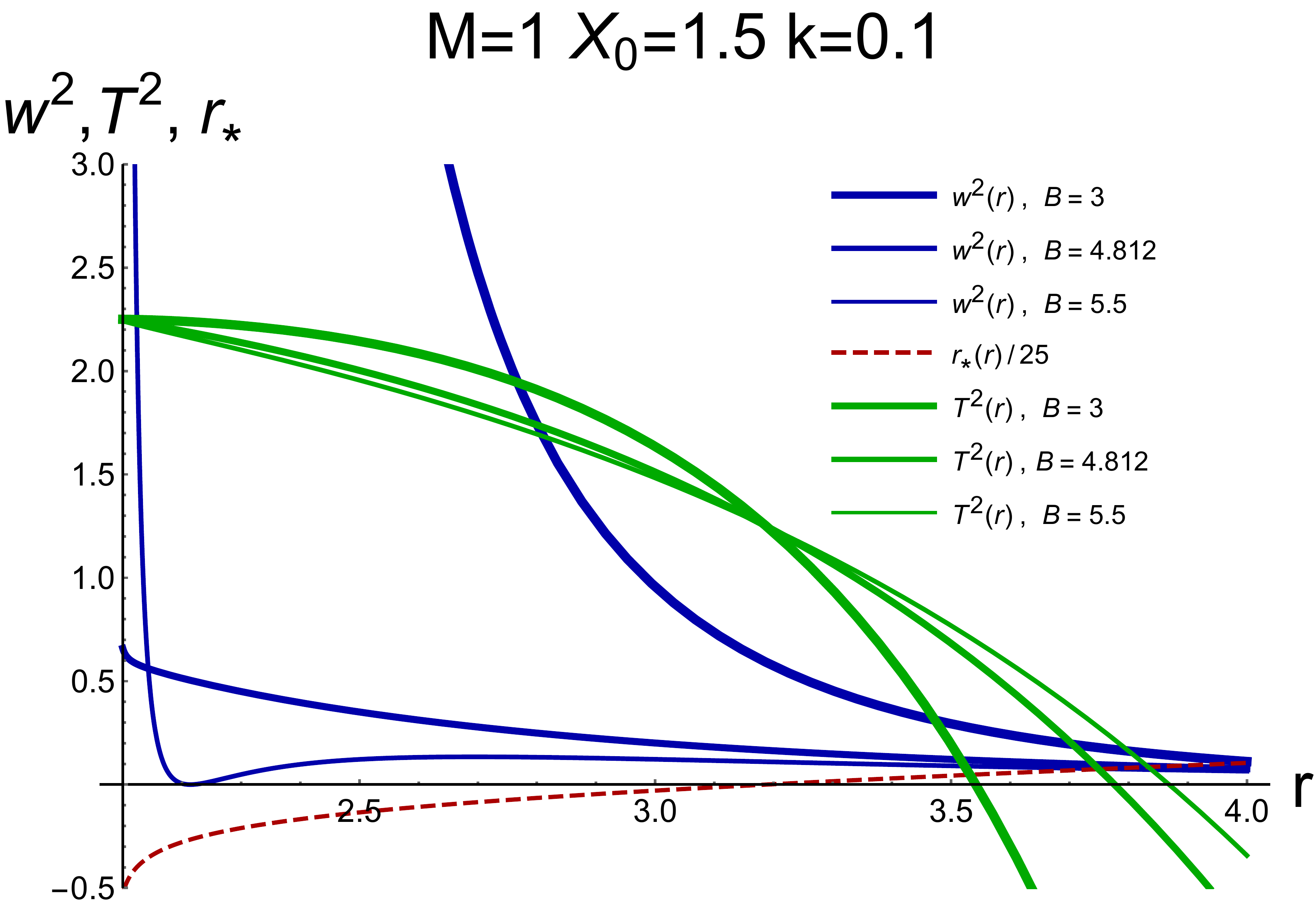}
 \\ A\qquad\qquad\qquad\qquad\qquad\qquad B\caption{ A) The trajectories with $X_0=1.5$ in the Schwarzschild coordinates  for dS-Schwarzschild metric. B)
$w^2$ vs $r$ (blue) and $T^2$ vs $r$(green)  for different $B$ and $M=1$.
The red dashed line shows $r_*=r_*(r)/25$ for the same $M,X_0$. 
}
  \label{Fig:dSQK}
\end{figure}

Fig.\ref{Fig:dSQK} shows the dependence of the acceleration along trajectories with $\Xm_0=1.5$ and different $B$ in the dS-Sch metric 
with $M=1$  and varying $k$. We see that the acceleration becomes infinite near the horizon and decreases monotonically for $B<B_0$
(the thick line in Fig.A). For  $B=B_0$ the acceleration is finite at the horizon and for $B>B_0$ it is infinite at the horizon and has minimum equal to zero. For $B$ large enough  this minimal value is reached out of the acceptable region (the thin line in Fig.A). 

\newpage
\section{General L-coordinates }\label{GenQR}
\subsection{L-coordinates}\label{TermalGenQR}
Having in mind formulas \eqref{ds-f}, \eqref{rsG}, \eqref{GEF} and \eqref{dsvn} we define  L-coordinates $\nu,\vartheta$  by the relations\footnote{To have a possibility to send $a\to 0$ one can use a modified definition
\bea  u&=&-\frac{1}{a}(e^{-a\nu} \IA{-1}),\qquad v=\frac{1}{a}(e^{a\vartheta} \IA{-1}),\,\,a>0,
\label{unn}
\eea}
\bea
 \nu &=&-\frac1a\log (-a u), \qquad \vartheta=\frac1a\log(av).\,\,a>0\label{un}
\eea

In term of  coordinates  $(\vartheta,\nu)$ the metric \eqref{dsvn} reads
\be
ds_2^2=-f(r)d u d v=-f(r)e^{a( \vartheta- \nu)}d \nu d \vartheta,\qquad  \nu,  \vartheta \in (-\infty,\infty)
\label{nuvar}\ee
where $r=r(\vartheta, \nu)$ is  defined in two steps: first $r$ is defined as $r=r(r_*)$ by the relation \eqref{rsG} and then  $r_*$ as a
 function of $\vartheta, \nu $ using \footnote{In the case of \eqref{unn} 
$
r_*=\frac{1}{2a}(e^{a \vartheta}+e^{-a \nu}-2).
$}
\be
r_*=\frac{1}{2a}(e^{a \vartheta}+e^{-a \nu}).
\label{mmR1}\ee

By introducing the coordinates 
\bea 
\eta&=&( \vartheta+ \nu)/2, \qquad \xi=( \vartheta- \nu)/2\label{eta-xi}\eea
 the metric \eqref{nuvar} can by rewritten as
\bea\label{mR}
ds^2=-f(r)e^{a(\vartheta-\nu)}d\nu d\vartheta=f(r)e^{2a\xi}(-d\eta^2+d\xi^2),\label{QRindler-ds}\eea
that is up to the conformal factor $f(r)$  the Rindler metric  on the $(\eta, \xi)$-plane,
\be
ds_{Rindler}^2=e^{2a\xi}(-d\eta^2+d\xi^2).
\ee

From \eqref{mmR1} and \eqref{eta-xi} follow relations

\be
r_*=\frac{e^{a\xi}\cosh (a\eta)}{a},\,\,\,t=\frac{e^{a\xi}\sinh (a\eta)}{a}.
\label{mRR3}\ee
Hence in the case of \eqref{mmR1} one gets
\be
r_*^2-t^2=\frac{e^{2a\xi}}{a^2},
\ee
in the case of \eqref{unn} one gets
\be
(r_*+\frac{1}{a})^2-t^2=\frac{e^{2a\xi}}{a^2}
\ee
and in both cases if $\xi$ is a constant one has a motion along hyperbola. In particular, if one takes $\xi=0$
then the parameter $1/a$ is a semi-axis of the hyperbola. So, for the two-dimensional part of general spherically symmetric metric \eqref{QRindler-ds} the parameter $1/a$  is a semi-axis of the hyperbola.

It will be shown below, in Sect.\ref{Sect: q-Rindler-T}, that for a rather general metric in the form \eqref{QRindler-ds}
the temperature is 
\be
T=\frac {a}{2\pi}.\label{T-a}\ee
The temperature \eqref{T-a} does not depend of the form of the function $f(r)$ in the metric \eqref{ds-f}. But the trajectory of the L-observer 
does depend on $f(r)$ through the form of $r_*$.

\subsection{Acceleration along trajectories $\xi=\xi_0$ in black hole backgrounds}

Let us  consider an observer moving along this hyperbola, i.e. along the  trajectory \eqref{mRR3}
with $\xi=\xi_0$ in $(t,r_*)$-plane. One can parametrize this trajectory as 
\bea
r_*=r_*(\eta)&=&\frac{e^{a\xi_0}\cosh (a\eta)}{a},\label{rs-xhi0}\\ t=t(\eta)&=&\frac{e^{a\xi_0}\sinh (a\eta)}{a}.
\label{mR3-tr}\eea 
This parametrization means that $r_*>0,$ or, more precisely, that  $r_*>\frac{e^{a\xi_0}}{a}$.

One can find  velocity and acceleration   along this trajectory. 
Indeed, along this trajectory the interval is 
\bea ds=\sqrt{f(r)}e^{a\xi_0}d\eta, \quad\mbox{here}\quad  ds=|ds|,\eea
and  components of the velocity along this trajectory are
\bea
u^0&=&\frac{dt}{ds}=\frac{1}{\sqrt{f(r)}}\cosh(a\eta),\nn\\
u^1&=&\frac{dr}{ds}=\sqrt{f(r)}\sinh(a\eta).
\eea
We see
that the square of velocity is equal to -1, $
-f(u^0)^2+f^{-1}(u^1)^2=-1$.

The components of acceleration are 
\bea
 w^0&=&\frac{du^0}{ds}+\Gamma^0_{\mu\nu}u^\mu u^\nu =\frac{\sinh(a\eta)}{f}\left(a e^{-a\xi_0}+\frac{f'}{2}\cosh a\eta\right),\eea
and
\bea
  w^1&=&\frac{du^1}{ds}+\Gamma^1_{\mu\nu}u^\mu u^\nu =\cosh(a\eta)\left(\frac{f'}{2} \cosh (a\eta)+ae^{-a\xi_0}\right),\\
 w^\vartheta&=&\Gamma^\vartheta_{\mu\nu}u^\mu u^\nu=0,\\
 w^\varphi&=&\Gamma^\varphi_{\mu\nu}u^\mu u^\nu=0,\eea
 and
 \bea
 w^2&\equiv&-f (w^{0})^2+\frac{1}{f}(w^{1})^2=
 \frac{a^2e^{-2a\xi_0}}{f}\left(\frac{f'}{2} r_*(r)+1\right)^2.\label{GenRW2}
 \eea
 One can also check the orthogonality condition 
 $-f u^0 w^0+f^{-1} u^1 w^1=0$.

\subsection{Examples.}
 \subsubsection{Schwarzschild in  L-coordinates}
 According \eqref{GenRW2} the acceleration
along the trajectory \eqref{rs-xhi0},\eqref{mR3-tr} is given by
\bea
w^2&=&
\frac{a^2e^{-2a\xi_0}}{1-\frac{2M}{r}}\left(1+\frac{M}{r}\left(1+\frac{2 M}{r} \log (\frac{r}{2 M}-1)\right)\right)^2.\label{acceleration-sch}
\eea
Acceleration along these trajectories  as function of $r$ is presented in 
Fig.\ref{Fig:a-Sch}.A and as function of $\eta$ in Fig.\ref{Fig:a-Sch}.B.

   \begin{figure} [h!] \centering
\includegraphics[scale=0.25]{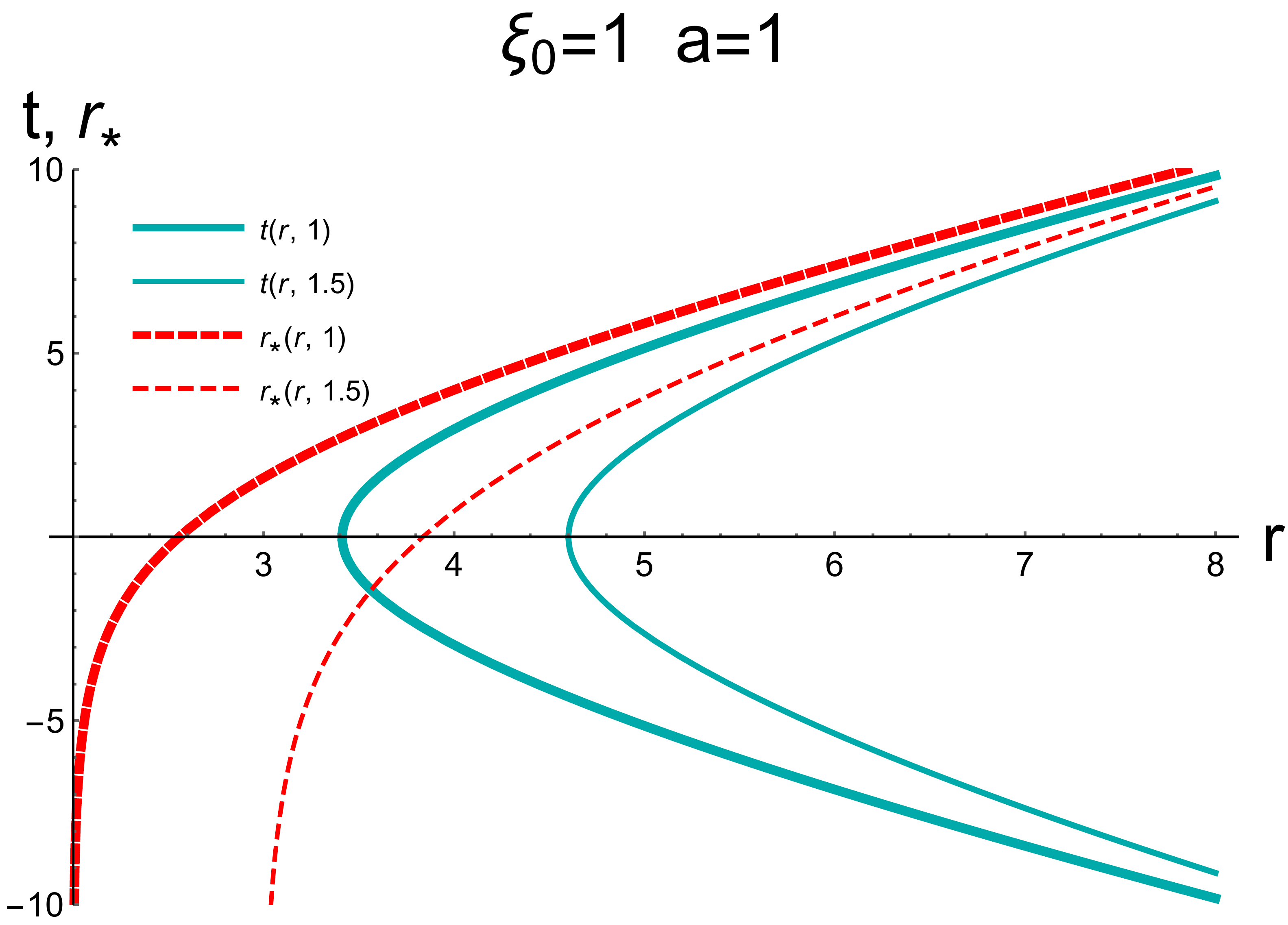}\
  \caption{Trajectories with $\xi=\xi_0$ in the Schwarzschild spacetime are shown by darker cyan lines. The red lines show $r_*=r_*(r)$.  
  }
  \label{Fig:a-traj-Sch}
\end{figure}
  \begin{figure} [h!] \centering
\includegraphics[scale=0.25]{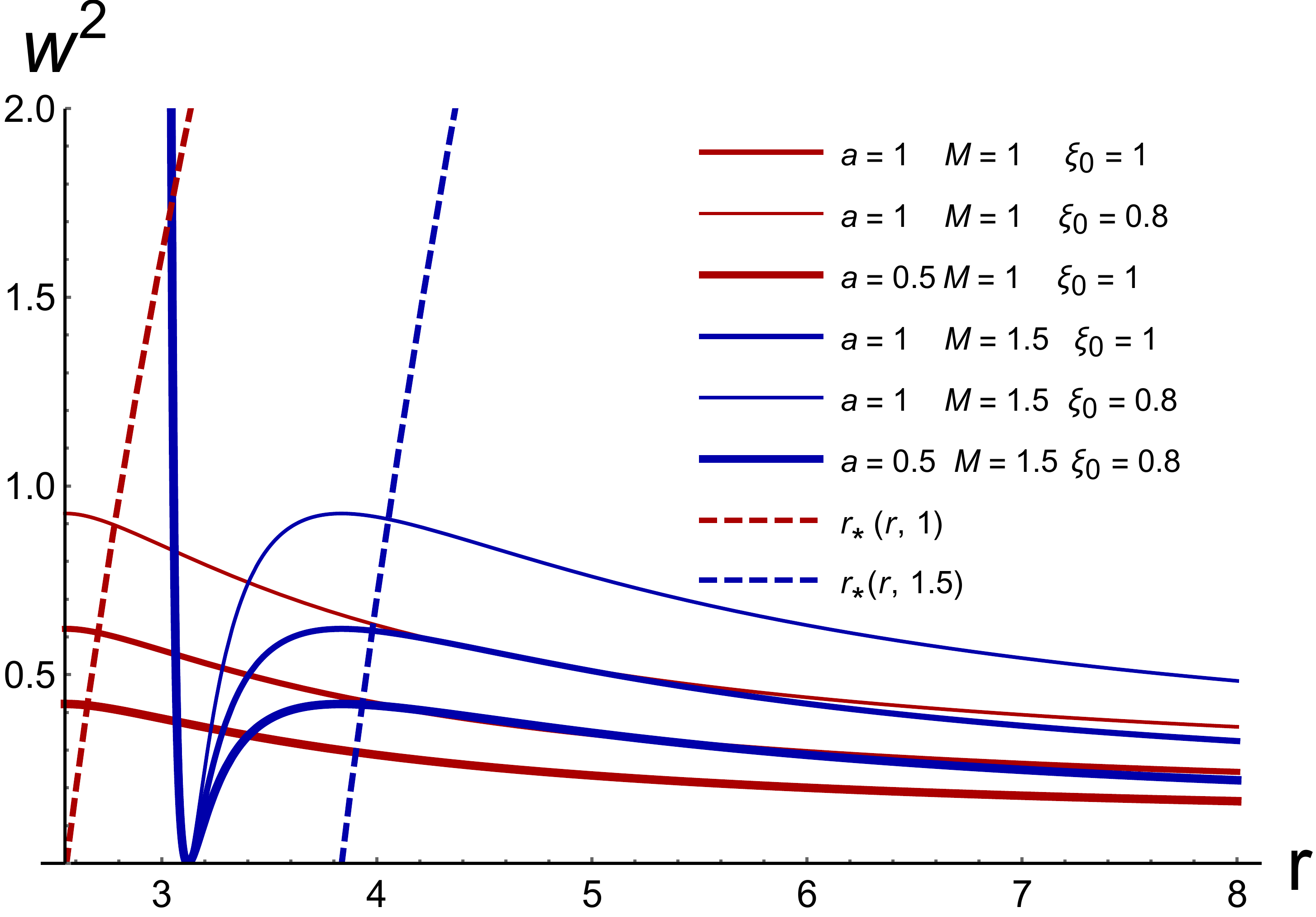}\qquad
\includegraphics[scale=0.2]{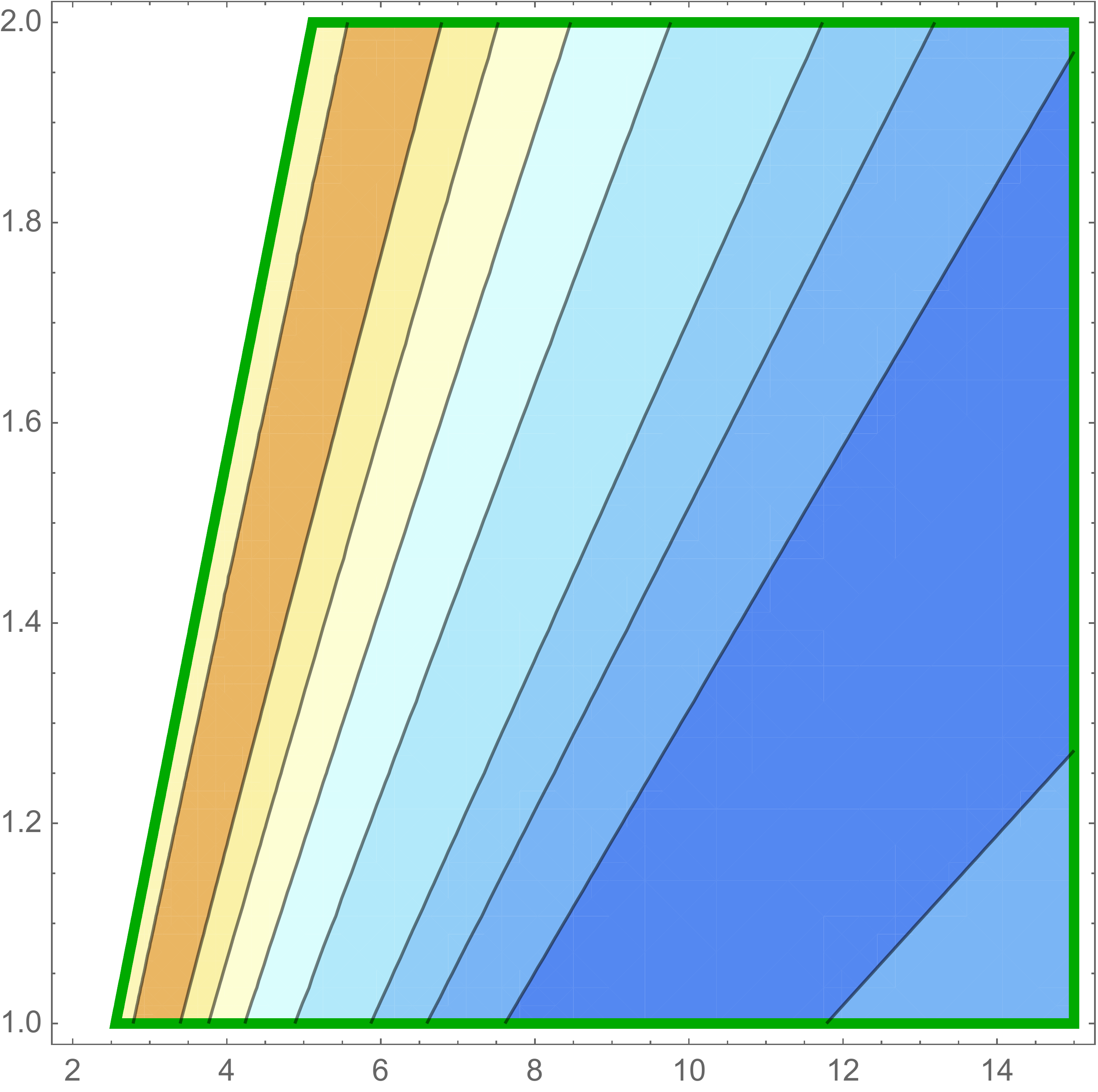}
\includegraphics[scale=0.2]{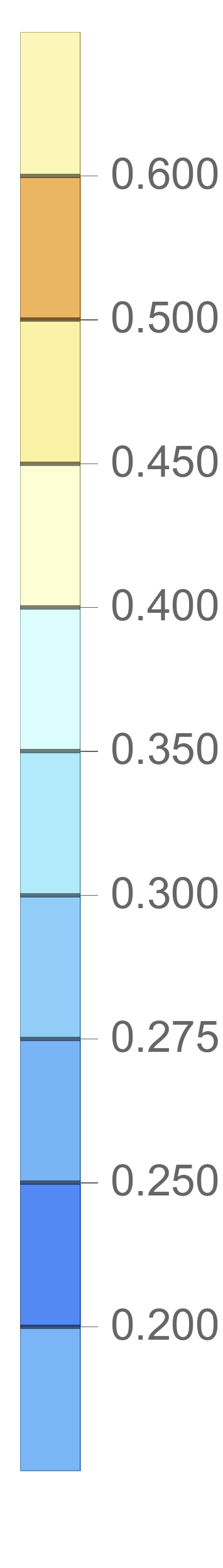} 
\begin{picture}(50,15) 
   \put(205,160){$w^2$}
     \put(45,80){$ M $} 
      \put(129,10){$ r $}  
     \end{picture}\\A
\qquad\qquad\qquad\qquad\qquad B\\
\caption{The acceleration along the trajectories shown in Fig.\ref{Fig:a-traj-Sch}  as function of $r$ B) Contourplot for varying $r$ (horizontal) and $M$ (vertical). We see that in the admissible area 
  the acceleration $w^2$ decreases when $r$ increases. 
  }
  \label{Fig:a-Sch}
\end{figure}
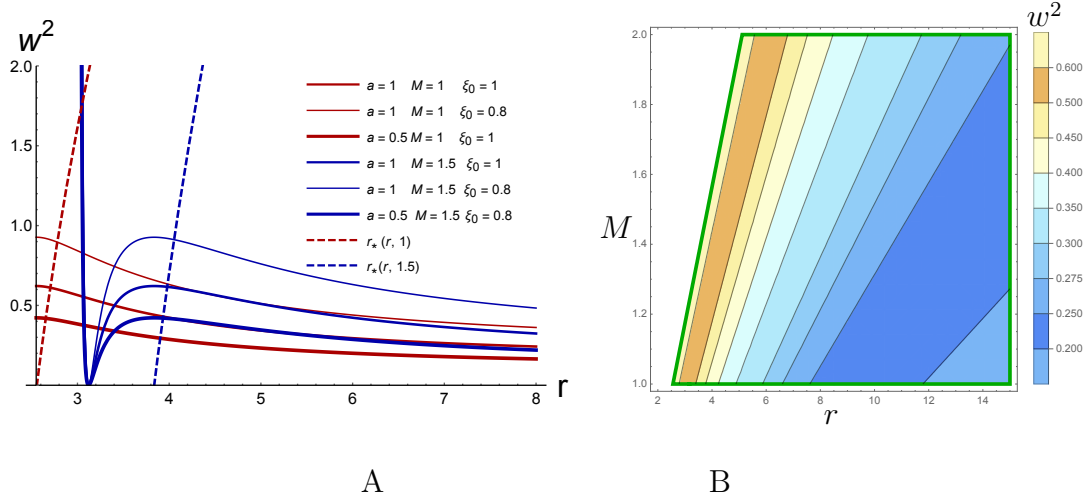

\subsubsection{Reissner-Nordstrom in  L-coordinates}
\begin{figure}[h!]
  \centering
     \includegraphics[scale=0.25]{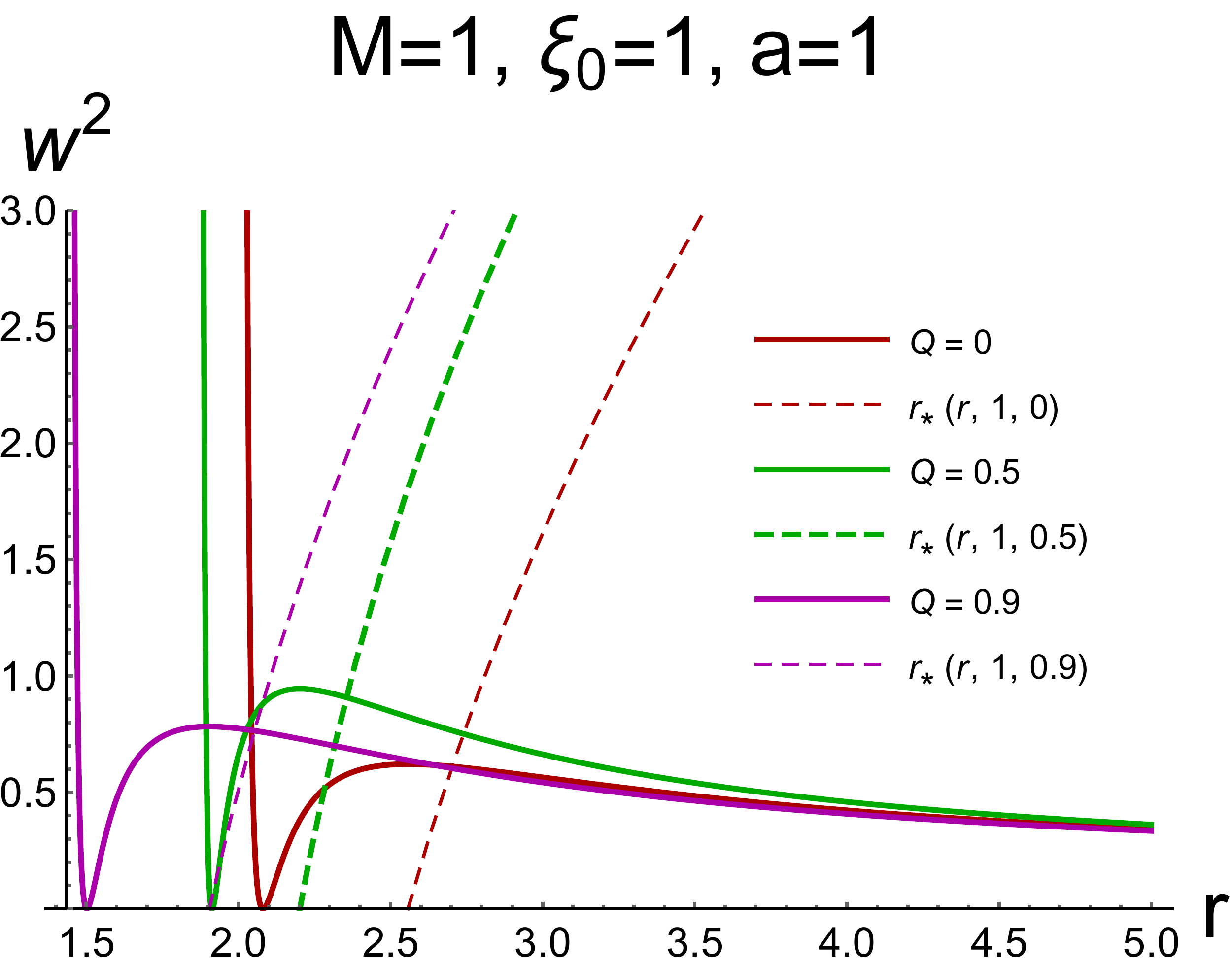} \qquad\includegraphics[scale=0.2]{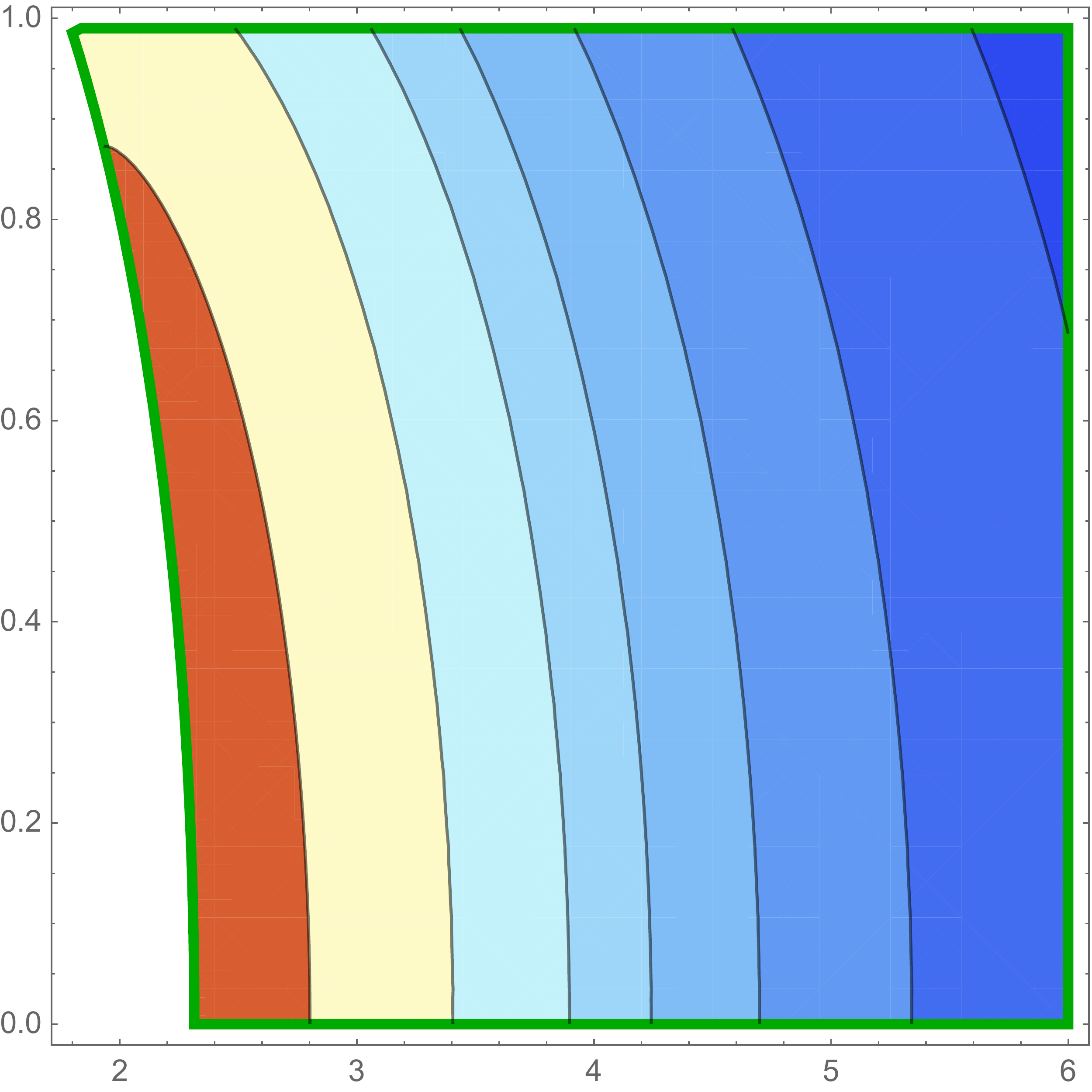} \includegraphics[scale=0.2]{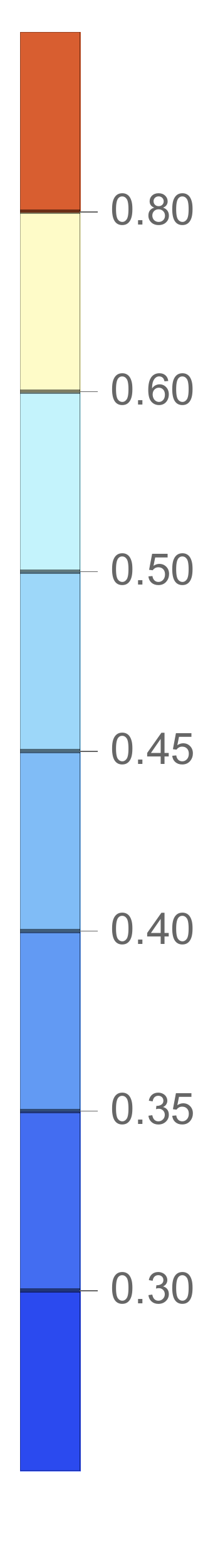} 
     \begin{picture}(50,15) 
   \put(-25,150){$w^2$}
     \put(-185,80){$ Q $} 
      \put(-45,-5){$ r $}  
     \end{picture} \\ A\qquad \qquad \qquad \qquad \qquad \qquad B
\\
\caption{A) $w^2$ as function on $r$ along the trajectories with $\xi=\xi_0$ for  different $Q$. The physical acceptable regions are on the right of dashed lines 
shown $r_*$, here $M=1$ . B) Countour plot of A) for varying $r$ and $Q$, $M=1$. The physiacally acceptable domain is bounded by the green line.
We see that the character  of the acceleration dependence on $r$ is the same as for the Schwarzschild   case shown in Fig.\ref{Fig:a-Sch}.
    }
  \label{Fig:RN-acc}
\end{figure}
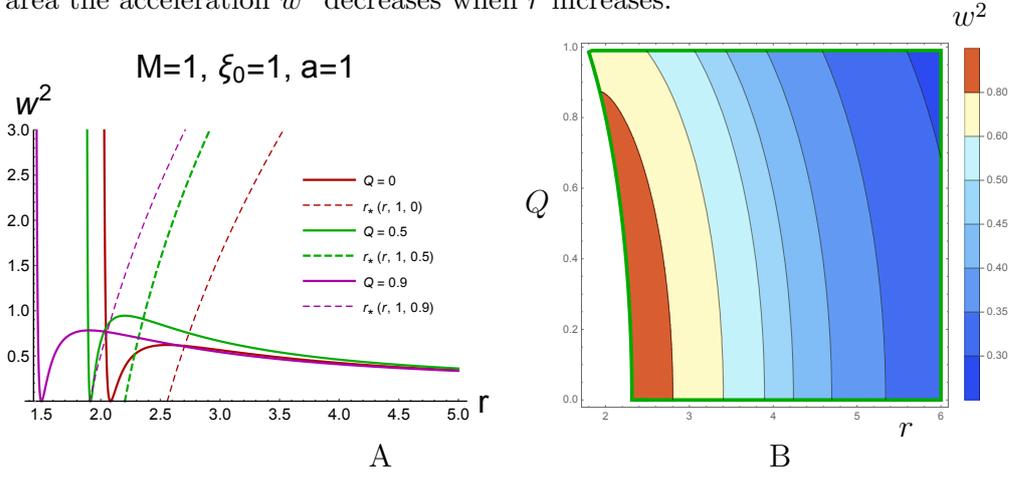

For the RN solution with $f(r)$ given by \eqref{RN} the acceleration dependence on $r$ can be calculated using general formula 
\eqref{GenRW2}. The results are  presented in Fig.\ref{Fig:RN-acc}.A. 

\newpage
\subsubsection{Schwarzschild-AdS in  L-coordinates}
Here we consider  the Schwarzschild-AdS metric in  L-coordinates.
For the Schwarzschild-AdS solution $f(r)$ and $r_*=r_*(r)$ are given by \eqref{Sch-AdS} and \eqref{Sch-AdS-rs}, respectively.
Using general formula \eqref{GenRW2} we calculate the acceleration. The result is presented in Fig.\ref{Fig:RindlerAdS-accel}.

  \begin{figure} [h!] \centering
\includegraphics[scale=0.25]{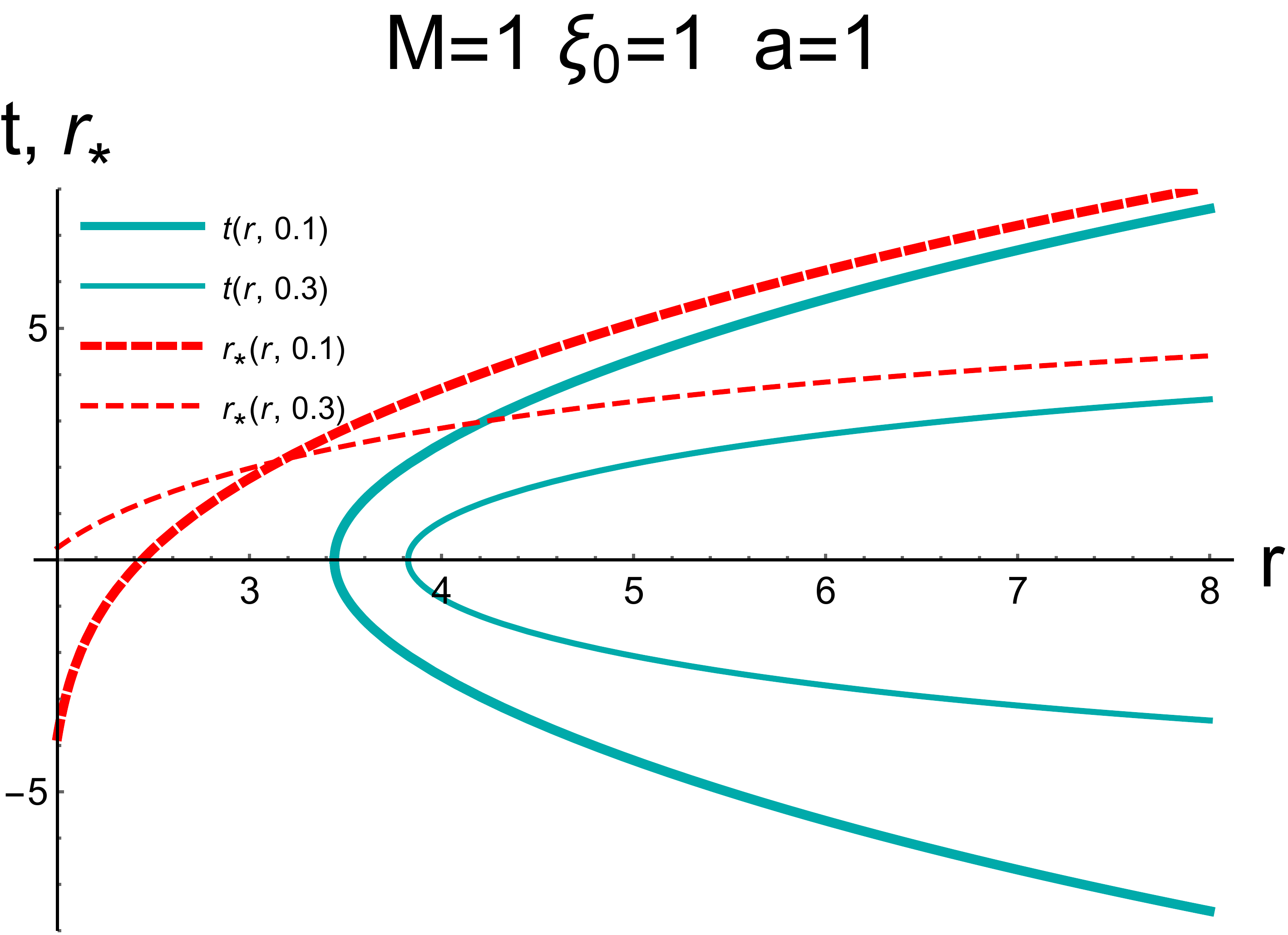}\\A\\
 \includegraphics[scale=0.23]{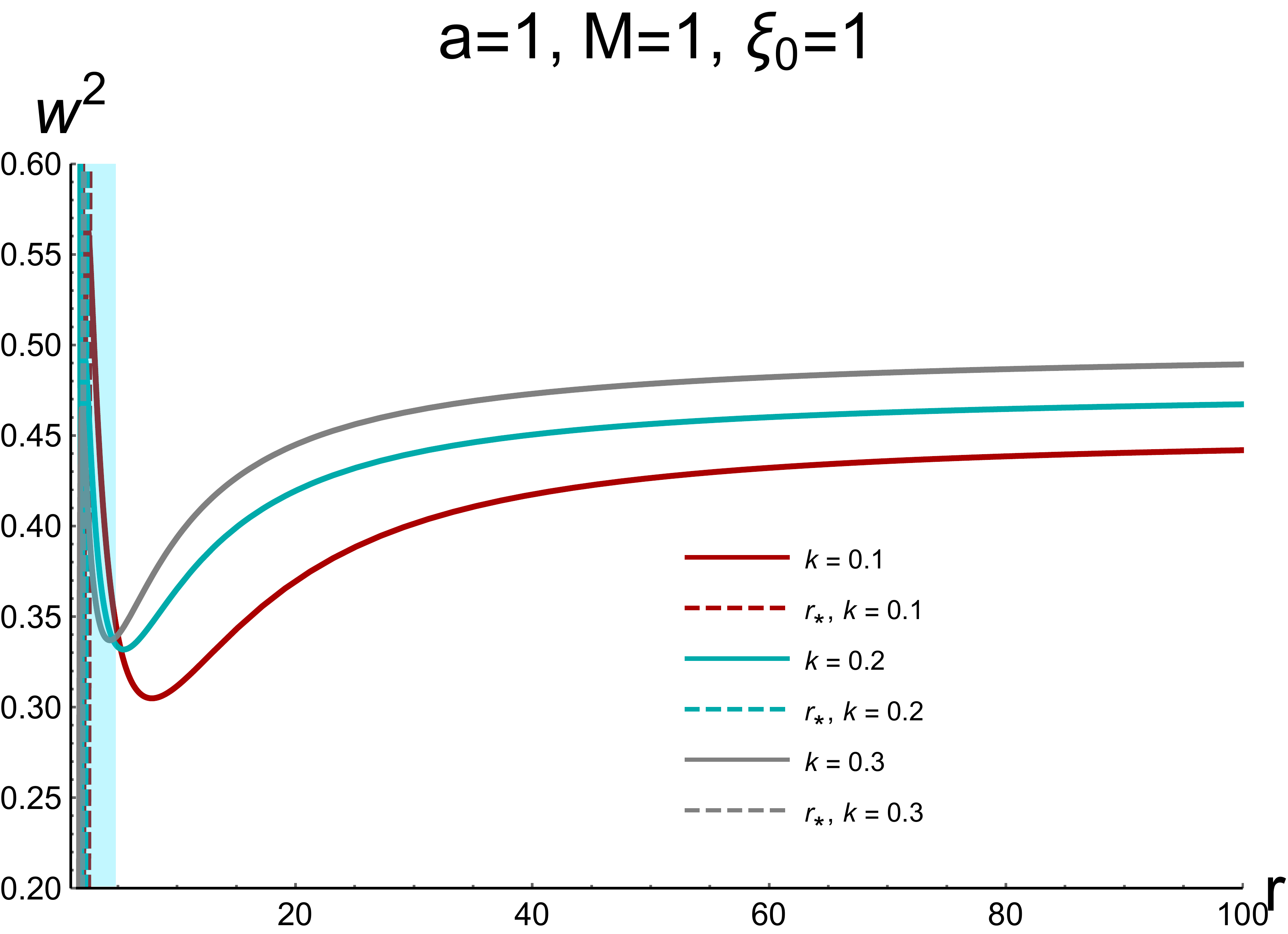} \qquad  
      \includegraphics[scale=0.15]{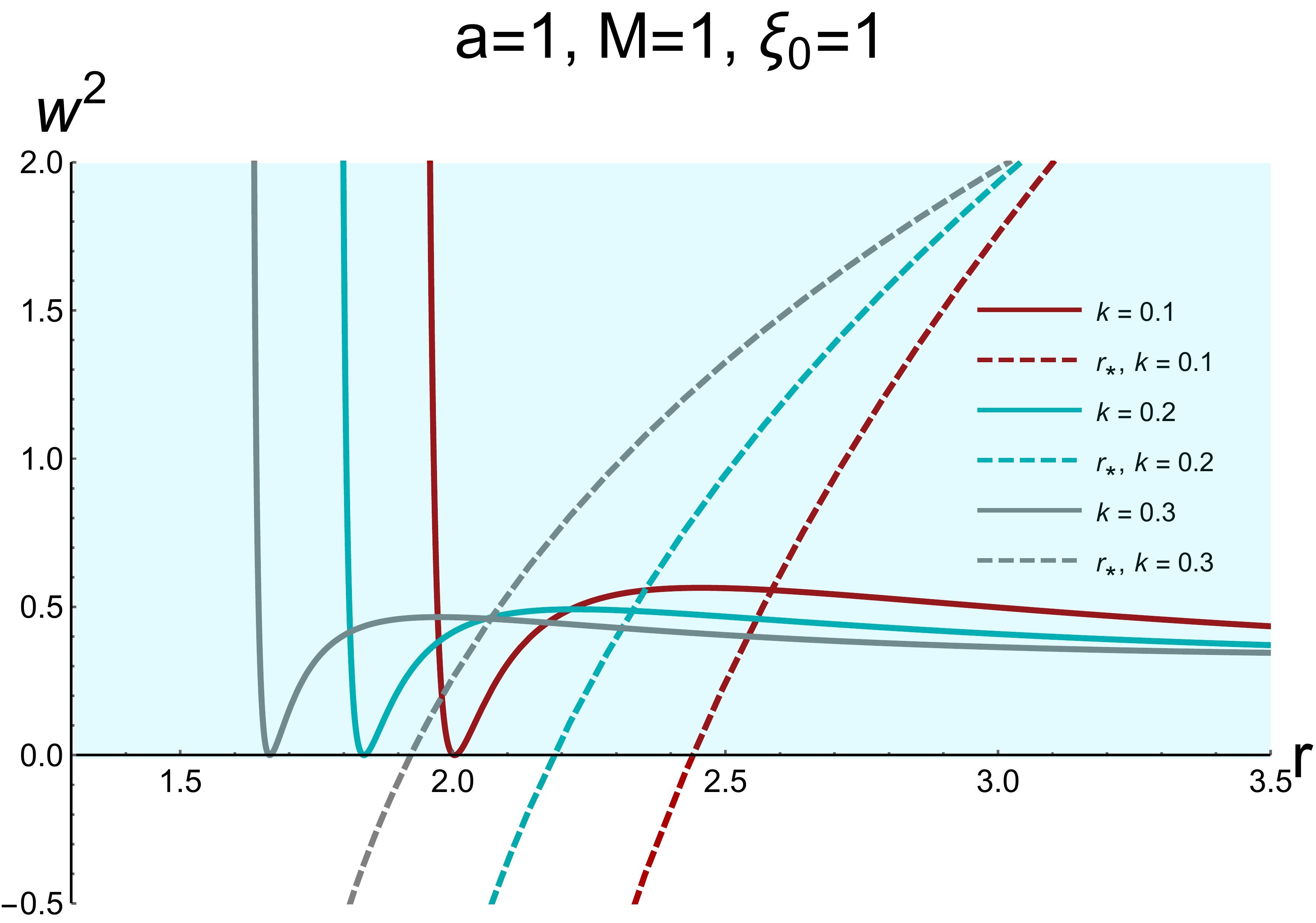}
    \\
        B\qquad \qquad \qquad \qquad \qquad C
  \caption{A) Trajectories in the AdS-Schwarzschild spacetime in the Schwarzschild coordinates corresponding to the observer with the fixed space coordinate $\xi_0$ in the  $(\eta,\xi)$  coordinates;  acceleration $ w ^ 2 $ vs $ r $ for trajectories in shown on A) for different values of $ k $;
 C) zoom of B).
 }
  \label{Fig:traj-R-AdS}
\end{figure}

\newpage
\subsubsection{Schwarzschild-dS in  L-coordinates}
  \begin{figure} [h!] \centering
\includegraphics[scale=0.25]{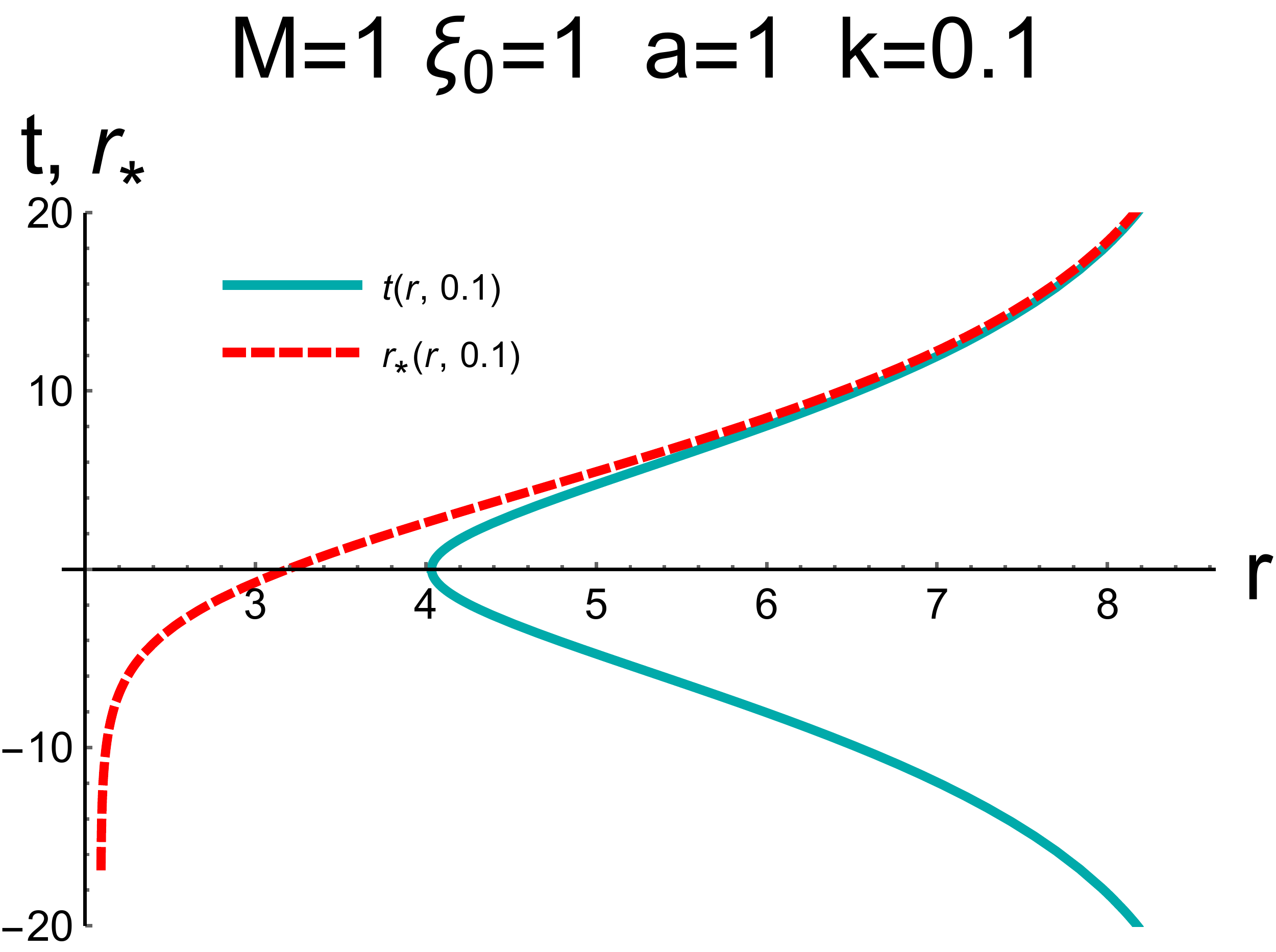}\qquad
\includegraphics[scale=0.25]{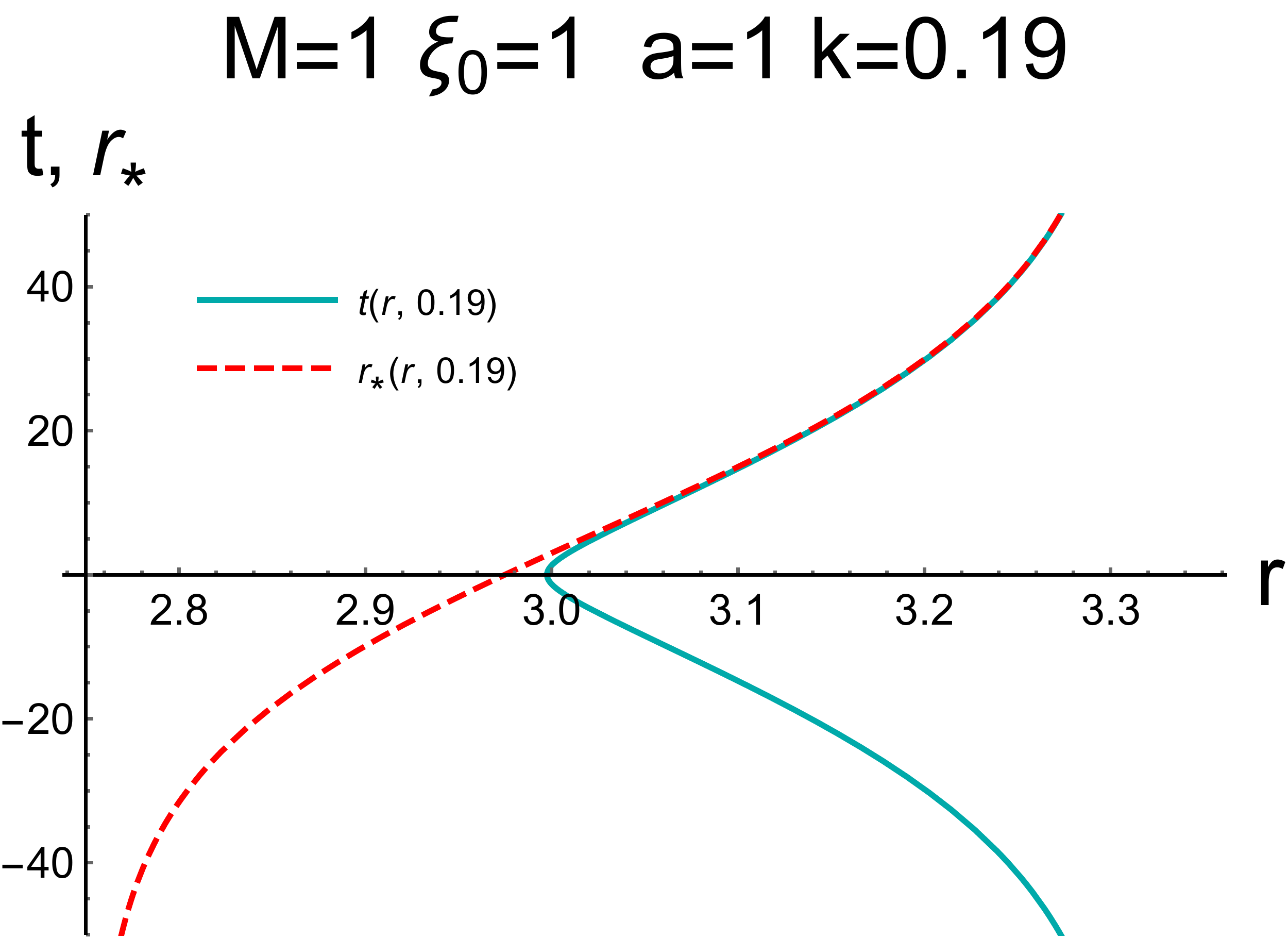}
  \caption{Trajectories in the dS-Schwarzschild spacetime in the Schwarzschild coordinates corresponding $\xi=\xi_0=1, M=1,a=1$ for $k=0.1$ (A) and  $k=0.19$ (B){\bf Math file: Geodefics-Rindler-short.nb+Rindler-Gen.nb}.}
  \label{Fig:traj-R-ds}
\end{figure}
\begin{figure}[h!] \centering
  \includegraphics[scale=0.3]{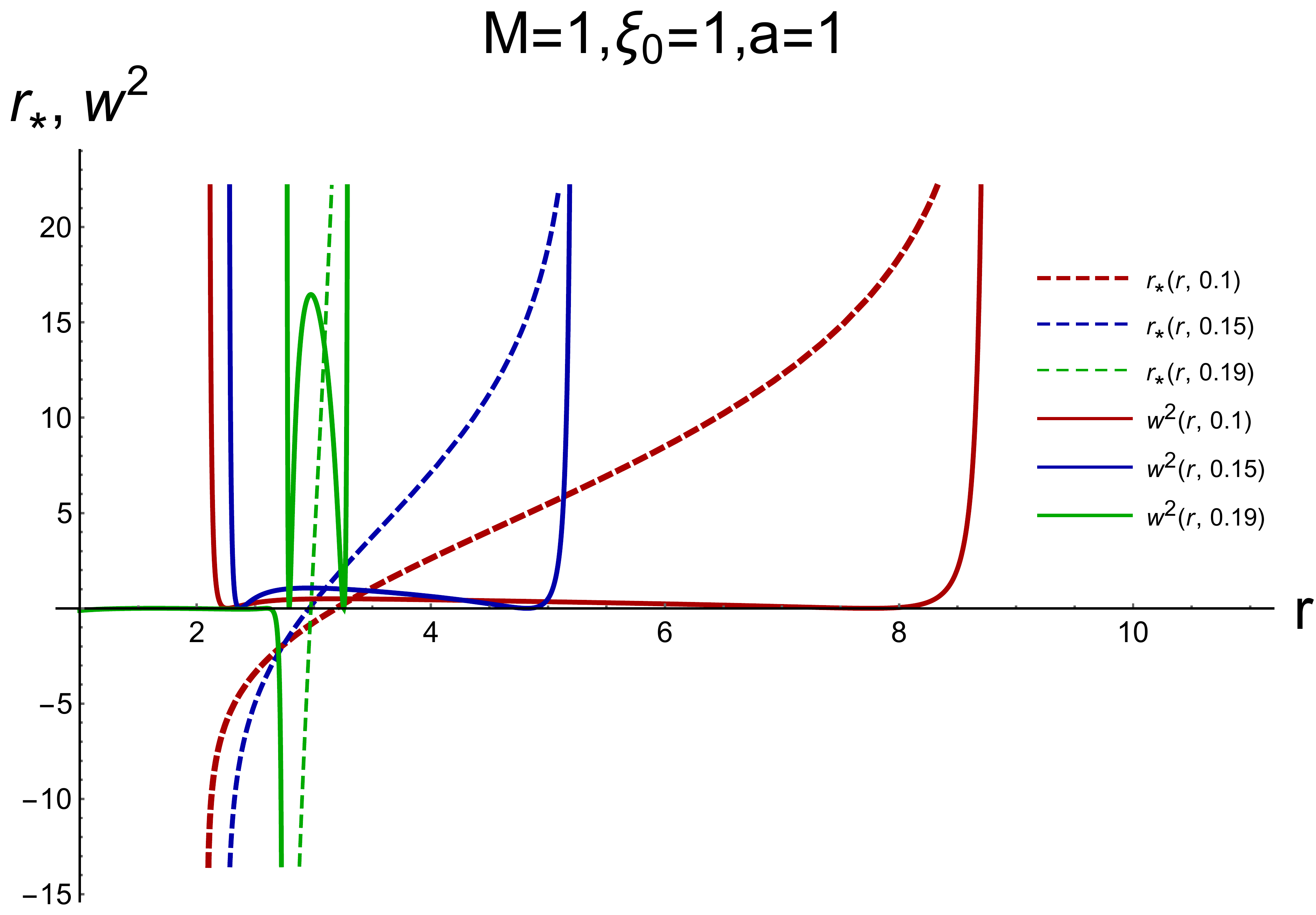}\\
  A\\$\,$\\
  \includegraphics[scale=0.17]{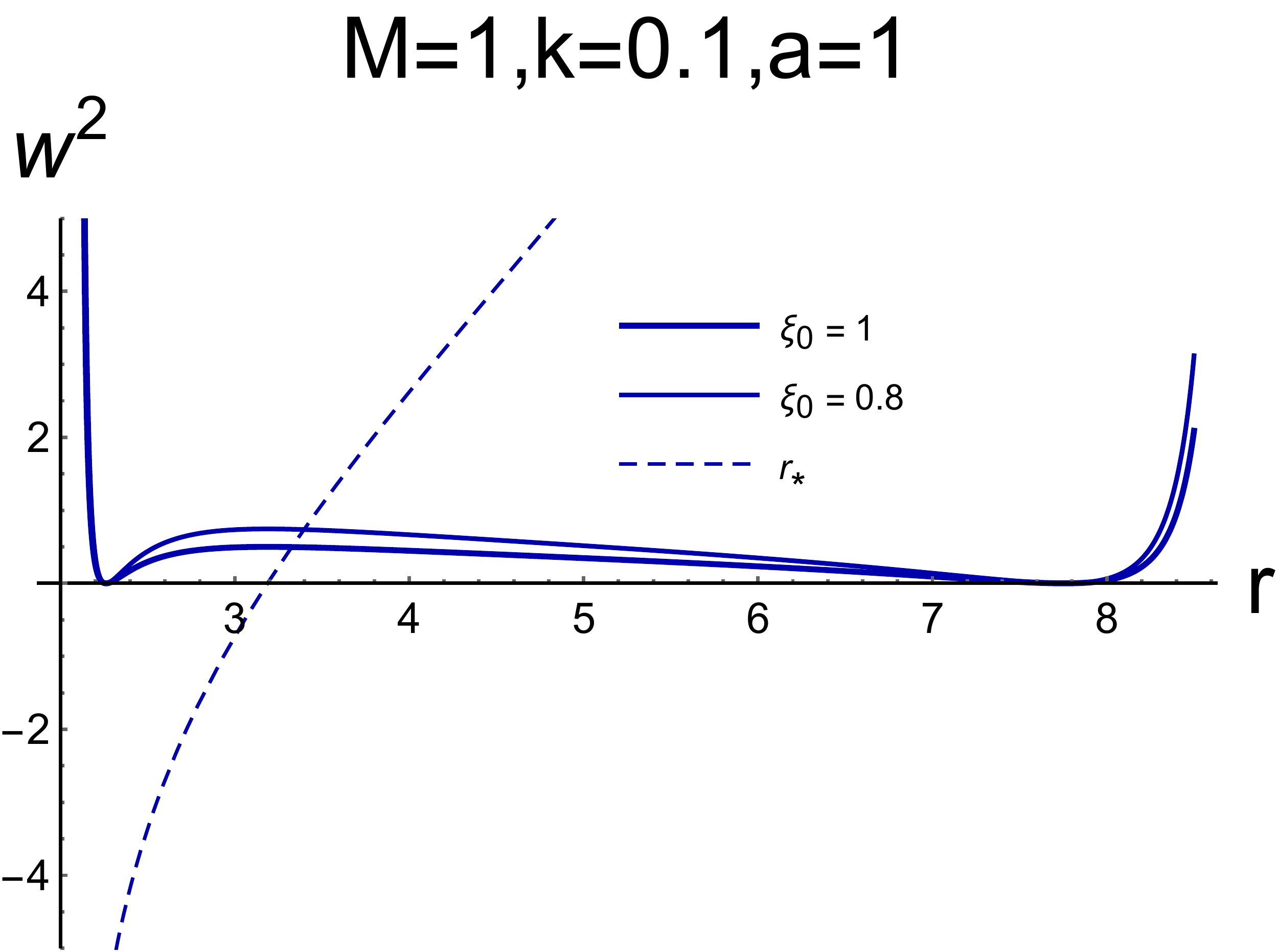}\qquad
  \includegraphics[scale=0.17]{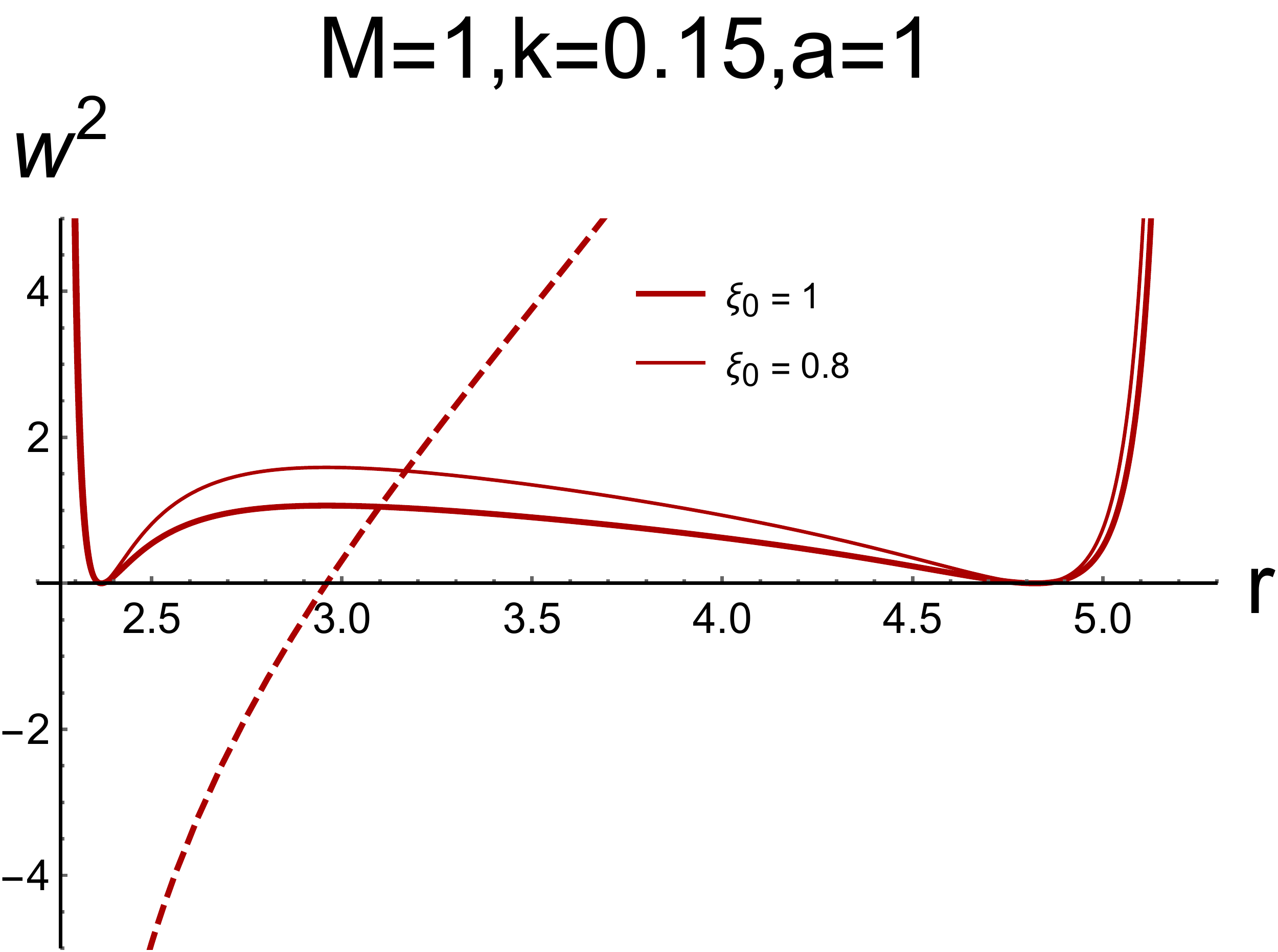}\qquad
    \includegraphics[scale=0.17]{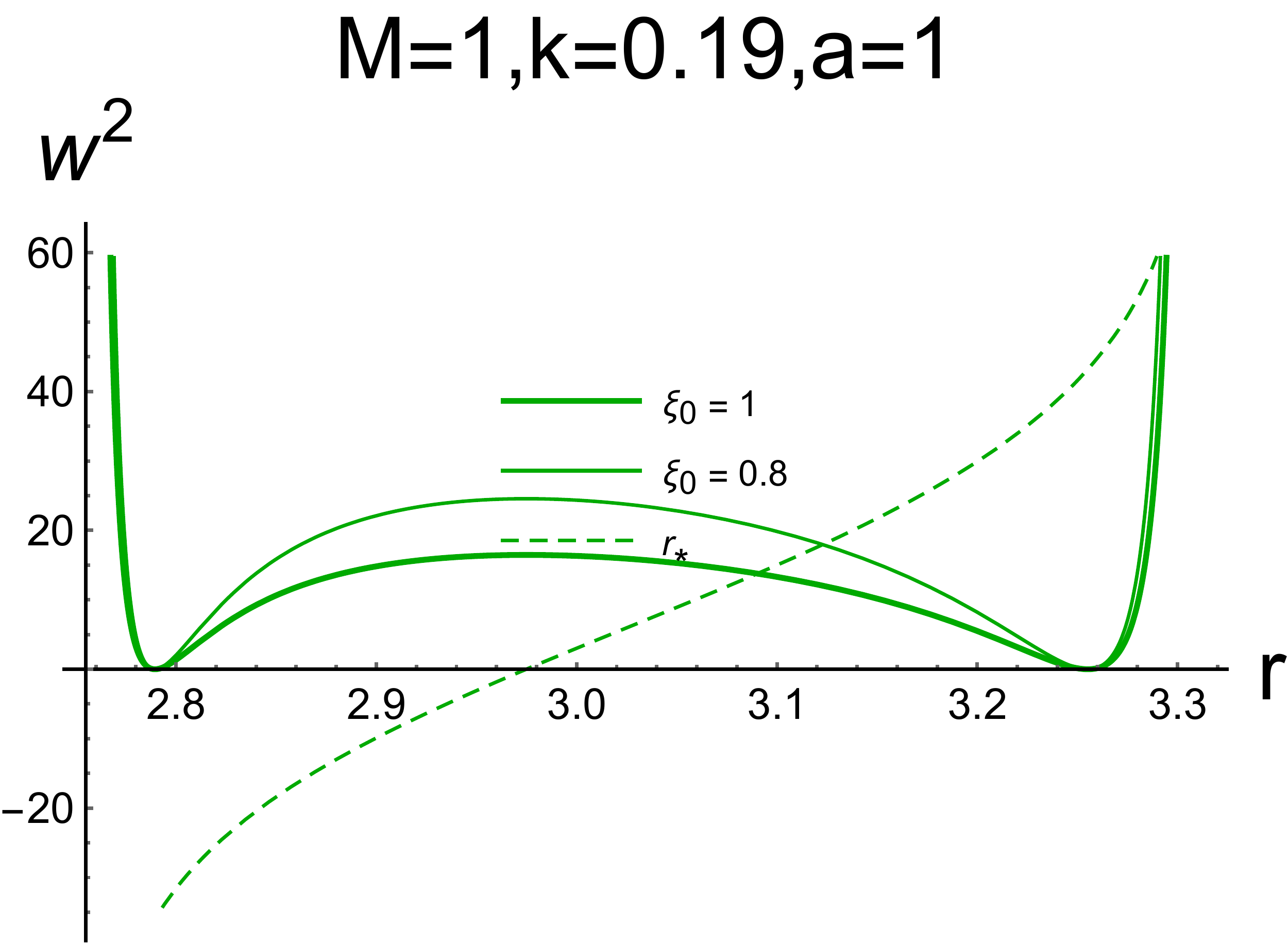}\\
    B\qquad \qquad \qquad \qquad \qquad C \qquad \qquad \qquad \qquad \qquad D
\caption{A) The accelerations along the trajectories  with $\xi=\xi_0=1$ for different $k=0.1$ (darker reds line),  $0.15$ (blue lines), $0.19$ (green lines). 
B), C) and D) are zooms of A)  as well as the same plots for $\xi_0=0.8$.
    }
  \label{Fig:dSa}
\end{figure}

Here we consider  Schwarzschild-dS metric in  L-coordinates. The geometry is characterized by formula 
\eqref{Sch-dS-rs} and \eqref{Sch-dS}
The dependences  of acceleration $w^2$ on $r$ are shown in Fig.\ref{Fig:dSa}.

\newpage
\subsection {Temperature in L-coordinates}\label{Sect: q-Rindler-T}
Here we show that the  accelerated observer moving along  special trajectories  defined by  requirement $\xi=\xi_0$
[[specified this condition in term of the original coordinates, the form of trajectory essentially depends on the blackening function $f(r)$]]  see  Hawking radiation with temperature, defined by only by parameter $a$.
\be
T=\frac{a}{2\pi}.\label{T-a}\ee
Indeed, comparing the solution  of the wave equation in $(u,v)$ and $ (\nu,\vartheta)$ coordinates  related as in \eqref{unn}
\bea
\partial_\nu\partial_ \vartheta\phi=0,\,\,\nu, \vartheta\in\mathbb{R}\label{plane-nu-weq}\\
\partial_v\partial_u\,\Phi=0,\quad u<0,\, v>0\label{fuv-weq}\eea
we get relation between corresponding creation and annihilation operators. For this purpose we write as usual, the representation 
for solutions of 2-dimensional wave equations as combinations of the left and right modes, 
$\phi(\nu,\vartheta)=\phi_R (\nu)+\phi_L (\vartheta)$, $
\Phi(u, v)=\Phi_R (u)+\Phi_L (v)$.
For the real right mode (for the left mode all consideration is similar) one has
\bea
\phi _R(\nu)&=&\int _0^\infty d\omega (f_\omega B_\omega+f_\omega ^*B^+_\omega),\quad 
f_\omega(\nu)=\frac{1}{\sqrt{4\pi \omega}}e^{-i\omega \nu},\label{phiR}
\eea
where $[B_\omega,B^+_{\omega^\prime}]=\delta(\omega-\omega^\prime)$
and 
\bea
\Phi _R( u)&=&\int _0^\infty d\mu \Big( \fB _\mu\ff_\mu(u) + \fB ^+_\mu\ff_\mu ^*(u)\Big),\quad 
\ff_\mu( u)=\frac{1}{\sqrt{4\pi \mu}}e^{-i\mu  u},
\eea
where
$
[\fB_\mu,\fB^+_{\mu^\prime}]=\delta(\mu-\mu^\prime)$.

Right (and left) modes in different coordinate system are related as
$
\phi _R(\nu)=\Phi _R( u)$ and therefore,
\bea
\int _0^\infty d\omega (f_\omega B_\omega+f_\omega ^*B^+_\omega)&=&\int_0^\infty d\mu (\ff_\mu \fB_\mu+\ff_\mu ^*\fB^+_\mu).
\label{PhiRQM-R}
\eea
Multiplying \eqref{PhiRQM} on  $f_{\omega'}(\nu)$ and integrate the first equation over $\RR$ one  gets 

 \bea 
 B_\omega=\int d\mu\Big(\beta ^*_{\omega\,\mu}\fB^+_{\mu}+\alpha^* _{\omega\,\mu}\fB_{\mu}\Big),\qquad  B^+_\omega&=&\int d\mu\Big(\beta _{\omega\,\mu}\fB_{\mu}+\alpha _{\omega\,\mu}\fB^+_{\mu}\Big),\label{BB}\\
\eea
where (compare with calculations in Sect.\ref{quasiK-Sch-T})
\bea
\beta _{\omega\,\mu}=\int _{\RR} \,\frac{d\nu}{2\pi }\sqrt{\frac{\omega}{ \mu}}e^{-i\omega \nu}e^{-i\mu u},
\qquad
\alpha _{\omega\,\mu}=\int _{\RR} \,\frac{d\nu}{2\pi }\sqrt{\frac{\omega}{ \mu}}e^{-i\omega \nu}e^{i\mu u}.
\label{gamma-R}
\eea

The EF (Eddington-Finkelstein) observer has the EF vacuum
\be
\fB_\omega|0_{EF} >=0,\ee
i.e. the state  $|0_{EF}\rangle$  does   not contain $\fB$ particles. But it contains  $B$ particles:
\begin{eqnarray}
&\,&\langle 0_{EF}|N_{\omega}(B)|0_{EF}\rangle \equiv\langle 0_{EF}|B_{\omega}^+B_{\omega}|0_{EF}\rangle
=\int_0^{\infty} d\mu \,|\beta_{\omega\mu}|^2.\label{beta-gamma-R}
\end{eqnarray}
The Bogoliubov coefficient $\beta_{\omega\nu}$ is given by \eqref{delta} with $u$ as in \eqref{unn},  so we have
\bea \beta _{\omega\,\mu}=\frac{1}{2\pi  a}\sqrt{\frac\omega\mu}\,e^{-\frac{\pi \omega}{2a}}\,\Big(\mu\Big)^{-i\frac{\omega}{a}} \Gamma (i\frac{\omega}{a}), \eea
and as in \eqref{delta2} we get he Planck distribution
\bea
|\beta _{\omega\,\mu}|^2&=&\frac{1}{2\pi a \mu}\frac{1}{e^{\frac{2\pi\omega }{a}}-1} \eea
 with the temperature \eqref{T-a}. 
So we have obtained that the temperature depends only on acceleration $a$, but the equation of the trajectory along which the observer is moving
depends on the metric through the coordinate $r^*$.

\section{Characteristic times in the black holes space times}\label{time-evap}
\subsection{Time of black hole evaporation}
The change of the black hole mass due to evaporation is described by equation
\be
\frac{d M(t)}{dt}=-L\ee
where $L$ is the luminosity,
$
L =C\, T^ 4 \cdot {\mbox {Area}}$ and
$C$ is a constant. In our case 
$
T=1/2\pi(b+4M),
$ and
 $\mbox {Area}=16\pi M^2
 $
so \be
L =\frac{C\,M^2}{\pi ^3 (b+4M)^4}\ee and we have the equation
\bea
\frac{d M}{dt}&=&-\frac{C\,M^2}{\pi ^3 (b+4M)^4},\quad M(0)=M_0\label{tM}
\eea
Therefore,
\bea
 \label{tMM}\\
\int_{M_0}^M\,\frac{(b+4M^{\prime} )^4}{M^{\prime 2}}\,d M^\prime&=&-\frac{C}{\pi ^3 }\,t, \label{tMM}
\eea
Taking the integral  we get
 \bea
    -\frac{b^4}{M}+8 b^3 \log (M)+96
   b^2 M+128 b M^2+\frac{256 M^3}{3}=\frac{C}{\pi ^3 }\,(t_0-t),\label{M-t}\eea
   where $t_0$ is 
   \be
  t_0=\frac{\pi ^3 }{C}\Big( -\frac{b^4}{M_0}+8 b^3 \log (M_0)+96
   b^2 M_0+128 b M_0^2+\frac{256 M_0^3}{3}\Big)\ee
 The evaporation time $t=t_{evap.time}$ is the time,  when $M(t_{evap.time})=\epsilon\to 0$. In our case the leading terms are the first two terms in the LHS of \eqref{M-t} and
   \be
    -\frac{b^4}{\epsilon}+8 b^3 \log (\epsilon)=\frac{C}{ \pi ^3 }\,(t_0-t_{evap.time}),\ee
   so we obtain
      \be
   t_{evap.time}\sim\frac{\pi ^3 }{C}(\frac{b^4}{\epsilon}-8b^3\log\epsilon) +t_0\to \infty.\ee
   Or, in other words
   \be
 M(t) =\frac{\pi ^3 }{C}\frac{b^4}{ t} \to 0 \qquad \mbox {when}\qquad t\to \infty\ee
   \\
   Note that the leading term is independent on the initial mass of the black hole.
   Note also that if we set $b=0$
   then the evaporation time  is
   \be
   t_{evap.time}=\frac{256 \pi ^3 }{3C}M_0^3.\ee
   
   \subsection{Small black holes and free falling observer}

Light  falling  on black hole in the Schwarzschild spacetime  is described by equation 
\bea
t&=& \int _{r_h} ^r\,\frac{dr}{(1-\frac{r_h}{r})},
\label{tr-LL1005}\eea
where $r_h=2M$. Here the integral is divergent at $r=r_h$ as
$r_h\ln (r-r_h)$. Usually one concludes an asymptotics  of approaching to the horizon is
\bea
r-r_h=const\, e^{-\frac{ t}{r_h}}.\label{r-rs}
\eea

We consider the question about the limit $r_h\to 0$ in more detailed. The solution of the equation 
\be
-dt=\frac{dr }{(1-\frac{r_h}{r})},\quad r(0)=r_0>r_h\ee   
is
\be
-t+r_0-r=2M\log (\frac{r-2M}{r_0-2M}).\label{t-r0}\ee
When $M\to 0$ one gets 
$
r=r_0-t$, as it should be.

From the other site if we rewrite  equation \eqref{t-r0} in the form 
\be
r=2M+(r_0-2M)e^{\frac{r_0-t-r}{2M}}\ee
then it is not obvious how to take the limit $M\to 0$.

 Let us discuss the leading term when the regularization parameters $B_!$
 and $B_2$ are introduced
 \bea
t=\int _{r_h+B_1}^r\, \frac{dr }{(1-\frac{r_h+B_2}{r})},\qquad B_1>B_2>0.
\eea
We have
\bea
r=r_h+B_2+(B_1-B_2)e^{\frac{-t+r-r_h-B_1}{r_h+B_2}}\label{r-rs-B}
\eea
and for $r$ near $r_h$ and large $t$ one obtains
\bea
r=r_h+B_2+(B_1-B_2)e^{-\frac{t}{r_h+B_2}}\label{r-r_s-1-B}
\eea

If $r_h$ is large we can set $B_2=0$ and we get the asymptotic formula \eqref{r-rs}. However for small black holes we can take the limit $r_h\to 0$ in 
\eqref{r-rs-B} to get 
\bea
r=B_2+(B_1-B_2)e^{-\frac{t}{B_2}}\label{r-rs-B0}
\eea
This consideration was a motivation to introduce the E-coordinates.

\section{Discussions and Conclusions}\label{Sec:Disc-Concl}
It is shown that the property to have a temperature distribution for quantum fields in classical gravitational background is not restricted to the cases of black holes or constant acceleration,    but is valid for any spherically symmetric metric written in thermal coordinates.

The Hawking temperature for  Schwarzschild black hole $T_H=1/8\pi M$ is singular in the limit of vanishing mass $M\to 0$. So, as the result of evaporation one gets an explosion of black hole. This is clearly unphysical since  the Schwarzschild metric in the original coordinates is regular when the black hole mass $M$ tends to zero. It is reduced to the Minkowski metric  and there are no reasons to believe that the temperature becomes  infinite. 

To improve the situation, new  coordinates, called thermal coordinates,  which depend on the black hole mass $ M $ and the parameter $ b $ that defines the semi-axis of a hyperbola along which an observer is moving
are used. Using the thermal coordinates the Schwarzschild black hole radiation is reconsidered and it is found that the Hawking formula for temperature is  valid only for large black holes while for small black holes the temperature is $T=1/2\pi (4M+b)$. 
The thermal coordinates are regular in the limit of vanishing black hole mass $ M $. In this limit, the Schwarzschild metric is reduced to the Minkowski metric, written in coordinates dual  to the Rindler coordinates.  
The thermal observer in Minkowski space
 sees  radiation with temperature $T=1/2\pi b$, similar to the Unruh effect but in our case the acceleration is not a constant.

 During evaporation, in the thermal coordinates with $b\neq 0$ the black hole mass is  decreasing inverse proportional to time and the black hole life time  is infinite. This is in a contrast with the case $b=0$ when the time life of black hole is finite. Sometimes the information paradox is formulated as follows. 
A collapsing black hole, described by a wave function, completely evaporates and leaves only radiation, described by a dense matrix.
So one gets a transformation of pure state to a mixed state, that contradicts to the unitary evolution in quantum mechanics.
Thus, a transformation of a pure state into a mixed one is occurred, which contradicts the unitary evolution in quantum mechanics, and means 
the loss of information. This formulation of the information lost paradox has a meaning only if the black hole time life is finite. In the case of using the 
E-coordinates the black hole life time is infinite. Therefore, formally speaking, the information loss paradox disappears. 

It would be interesting 
to estimate the entropy balance during the evaporation in the thermal coordinates.
It would also be interesting to study the fate of primordial black holes in the thermal coordinates and to use the thermal coordinates for investigation of massive fields of various spins in different dimensions in gravitational backgrounds.

\newpage
\section*{Acknowledgement}
We would like to thank V. Frolov, M. Katanaev, M. Kramtsov, K. Rannu, V. Sagbaev and  P. Slepov for useful discussions.This work is supported by the Russian Science Foundation (project 19-11-00320, Steklov Mathematical Institute).


\begin{thebibliography}{99}
\bibitem{Hawking:1974}
S.~W.~Hawking,
``Black hole explosions?''
Nature \textbf{248}, 30-31 (1974)
\bibitem{Hawking:1975} S.W. Hawking,  "Particle creation by black holes",
Comm. Math. Phys.
43 (1975) 199.

\bibitem{Hawking:1976} S.W. Hawking, 
"Breakdown of Predictability in Gravitational Collapse",
Phys. Rev. D \textbf{14}, 2460-2473 (1976)

\bibitem{Susskind:2005} 
L Susskind, J Lindesay,
 {\it   An introduction to black holes, information and the string theory revolution: The holographic universe},  2005
 
  \bibitem{tHooft:2012}
G.~'t Hooft,
{\it Introduction to general relativity}, 2012
 


\bibitem{HE} S.W. Hawking and G.F.R. Ellis, {\it The large scale structure
of space-time}, Cambridge University
Press (1973).
\bibitem{FN} V. Frolov, I. Novikov, {\it Black Hole Physics: Basic Concepts and New Developments}, Springer, 2012.

\bibitem{Wald} R.M. Wald, {\it General Relativity}, University of Chicago
Press (1984).


\bibitem{Ydri:2017} Badis Ydri, {\it Quantum Black Holes}, arXiv:1708.00748

\bibitem{Unruh:1976}
W.~G.~Unruh,
``Notes on black hole evaporation,''
Phys. Rev. D \textbf{14}, 870 (1976)
\bibitem{BD} N.D. Birell and P.C.W. Davies, {\it Quantum Fields in Curved
Space},
Cambridge University Press (1982).
\bibitem{Rindler:1966}
W.~Rindler,
``Kruskal Space and the Uniformly Accelerated Frame,''
Am. J. Phys. \textbf{34}, 1174 (1966)




\end{thebibliography}
\end{document}